\documentclass[11pt]{article}

\usepackage[margin=1in]{geometry}
\usepackage{microtype}
\usepackage{enumitem}
\usepackage{macros}
\usepackage{algorithm}
\usepackage{algpseudocode}
\usepackage{xcolor}
\usepackage{dsfont}
\usepackage{booktabs}

\usepackage{tcolorbox}

\newtcolorbox{questionbox}[1][Question]{
  colback=blue!3!white,
  colframe=blue!50!black,
  boxrule=0.5pt,
  arc=1mm
}

\title{Geometrically Constrained Learning I: \\ Agnostically Learning Adversarially Robust Classifiers}
\author{
Jane Lange\thanks{Supported by the National Science Foundation
under Grant No. DMS-2424441} \\
Institute for Advanced Study \\
\texttt{janeclange@gmail.com}
\and
Cassandra Marcussen\thanks{School of Engineering and Applied Sciences, Harvard University, Cambridge, Massachusetts, USA. Email: cmarcussen@g.harvard.edu. This material is based upon work supported by the Air Force Office of Scientific Research under award number FA9550-23-F-0014 in the amount of \$139,400 to Cassandra Marcussen. Supported in part by NSF Award 2152413 and a
Simons Investigator Award to Madhu Sudan.} \\ Harvard University \\ \texttt{cmarcussen@g.harvard.edu}
\and
Arsen Vasilyan \\
Aarhus University \\
\texttt{vasilyan@cs.au.dk}
}
\date{\today}

\begin{document}

\maketitle

\begin{abstract}
We study agnostic learning under the Gaussian distribution with guarantees on both classification error and robustness to 
adversarial perturbations.
 We say that a function has 
\emph{boundary profile} $K \ge 1$ if, for all $r > 0$, the probability that a 
point lies within Euclidean distance $r$ of the classification boundary is at most 
$Kr$. 
We give an algorithm that takes a robustness radius parameter $r$ and learns
a hypothesis with boundary volume $O(Kr)$ and classification error
$\mathsf{opt}_K + \varepsilon$, where $\mathsf{opt}_K$ is the minimum classification
error among functions with boundary profile at most \(K\).

For general $r, K,$ and $\varepsilon$, the algorithm runs in time 
$d^{\tilde{O}(K^2/\eps^2 + K^4r^2/ \varepsilon^4)}$.
In particular, in the $Kr = O(\varepsilon)$ regime, the algorithm runs in time 
$d^{\tilde{O}(K^2/\varepsilon^2)}$. Up to log factors in the exponent,
this matches the statistical query lower bound for agnostically
learning the $K$-profile class without recovering robustness.
For natural geometric classes, including convex sets and Boolean 
combinations of halfspaces, this matches the corresponding 
surface-area-based running-time bound for non-robust agnostic 
learning via $L_1$ polynomial regression, up to logarithmic factors in the exponent.

Our algorithm performs $L_1$ error minimization over polynomials,
with an auxiliary coordinate roughly standing in for distance to
the boundary. A convex constraint on the gradient of the 
polynomial, in this $(d+1)$-dimensional space, produces local
certificates of the range of the polynomial on balls of radius 
$r$. Our algorithm finds a polynomial such that these certified ranges are small, then obtains a robust classifier via
randomized rounding and partitioning of the domain.
\end{abstract}

\newpage
\section{Introduction}
\label{sec: intro}
\subsection{Background and results}
\label{sec:background-results}
The capabilities of modern Machine Learning systems have been progressing far faster than their reliability and security. Perhaps the most notorious security vulnerability inherent in ML systems is their susceptibility to backdoors. In their celebrated work, Goldwasser, Kim, Vaikuntanathan, and Zamir \cite{GKVZ2022} showed that it is possible for an adversary to plant an undetectable backdoor into a machine learning classifier, allowing the adversary to control the output of the classifier by adding a tiny change to the classifier’s input. Using cryptographic techniques, they showed that, despite there being a backdoor, it is cryptographically impossible for the user to even tell that the classifier was compromised.

How can we defend against such an attack? The work of Goldwasser, Shafer, Vafa, and Vaikuntanathan \cite{GSVV2025} introduces \emph{backdoor mitigation} — the notion of post-processing the classifier in order to make the backdoors disappear. Their work builds on classical TCS ideas of random self-reducibility to combine evaluations of the model on a number of correlated queries to neutralize the malicious backdoor-derived answers. This yields backdoor mitigation, for example, when the underlying uncorrupted classifier is a linear function or a low-degree polynomial. 

The backdoor mitigation theory of \cite{GSVV2025} builds a toolkit we can use for defending against an adversary who tries to attack a very well-behaved classifier by introducing a tiny change to it. However, how can this well-behaved classifier be efficiently trained to begin with?  This goal --- i.e., training classifiers for which tiny adversarial changes to the data point are unlikely to change the model prediction --- is studied by the area of adversarial robustness. Formally, this boils down to the goal of training a classifier with small boundary volume:
\begin{definition}
    The $r$\emph{-boundary volume} denotes the probability that a sample $x$ from the data distribution is within
distance $r$ of the classification boundary: 
\[\Bdry_r(f) = \Prx_{\bx  \sim \text{Data Distribution}} [\exists z: \ltwo{z} \le r, f(\bx) \ne f(\bx+z)].\]
\end{definition}
\noindent Achieving this goal for modern large-scale machine learning systems remains far from reach, and the theory needed for provable robustness guarantees is even further behind.
\begin{questionbox}
\textbf{Meta-Question:} How do we build a computational theory of efficient adversarial robustness? 
\end{questionbox}
One startling connection between adversarial robustness and Computational Learning Theory goes through the notion of \emph{Gaussian Surface Area}\footnote{This definition holds when $f$ has regular boundary. For general measurable functions, see \Cref{def:surface}.}
\[
\Gamma(f) := \limsup_{r \rightarrow 0 } \frac{\Bdry_r(f)}{2 r}
\]
This is a notion that has been studied extensively in probability theory, and the acclaimed work of Klivans, O'Donnell, and Servedio \cite{KOS2008} connected it to learning theory. They showed that for learning under Gaussian training data, the quantity $\Gamma(f)$ can be used as a \emph{complexity measure} that dictates the run-time sufficient to learn the function $f$.

Specifically, they show that a class of functions whose Gaussian Surface Area is bounded by $\Gamma$ can be learned in time $d^{O(\Gamma^2/\varepsilon^4)}$. They achieve this using the $L_1$ polynomial regression algorithm of Kalai, Klivans, Mansour and Servedio \cite{KKMS08}.  Klivans, O'Donnell, and Servedio also show that this algorithm applies in the challenging \emph{agnostic learning} setting where the data distribution $(x, y)$ cannot be perfectly described by the target function class $\mathcal{F}$ and the goal of the algorithm is to produce a classifier whose accuracy is $\epsilon$-competitive with respect to all classifiers in $\mathcal{F}$. The run-time analysis was recently refined to the optimal $d^{\tilde{O}(\Gamma^2/\varepsilon^2)}$ in the recent work of Pesenti, Slot, and Wiedmer \cite{PesentiSlotWiedmer2026}.

A simple corollary of this fact is that adversarially robust function classes can be learned efficiently, even in the agnostic setting. More formally, if our hypothesis class $\mathcal{F}$ has bounded boundary ratio, i.e., for all $f$ in $\mathcal{F}$ and $r$ we have
\[\Bdry_r(f) \leq Kr\] then it can be learned in time $d^{\tilde{O}(K^2/\eps^2)}$. 
Yet, as noted by Lange and Vasilyan \cite{LV26}, from the adversarial robustness perspective, this approach has a fatal flaw. The produced hypothesis $h$, although competitive with the best robust hypothesis in terms of \emph{accuracy}, is not itself guaranteed to be \emph{robust}. The reason is simple: the algorithm of \cite{KOS2008, KKMS08} is \emph{improper}, i.e., the hypothesis $h$ need not be in class $\mathcal{F}$. To be specific, it has the form $h(x)=\sign(P(x))$, where $P$ is a polynomial, and as observed in \cite{LV26}, such functions can have a very large boundary volume of $\Omega(1)$ for $r=\Theta(\eps^2/K^2)$.\footnote{Specifically, it returns a function of the form $\Ind[P(x) \geq 0]$ where $P$ is a polynomial of degree $\widetilde O(K^2/\varepsilon^2)$,
 which can in general have boundary profile $K^2/\varepsilon^2 \gg K$ (take, for example, 
 $f(x) = \sign(\prod_{i=1}^{K^2/\eps^2}x_i)$).} Can this limitation be overcome?
\begin{questionbox}
\textbf{Question 1: } Is there an inherent trade-off between accuracy and robustness under the Gaussian distribution? 
\end{questionbox}
\begin{questionbox}
\textbf{Question 2}: Are there hypothesis classes that consist of adversarially robust functions but can only be efficiently learned using non-robust hypotheses?
\end{questionbox}
\noindent In the setting above, even a simpler question has not been hitherto understood:
\begin{questionbox}
\textbf{Question 3: }
Can every robust function $f$ even be well-approximated by an \emph{easy-to-compute} adversarially robust function $f'$? What is the computing run-time needed?
\end{questionbox}
The work of Lange and Vasilyan \cite{LV26} show the lack of robustness can be overcome for a specific application of the framework of Klivans, O'Donnell, and Servedio \cite{KOS2008}, for learning halfspaces under a sub-Gaussian log-concave distribution. Here, we show it can be overcome for all adversarially robust functions using run-time $d^{\tilde{O}(K^2/\epsilon^4)}$ --- i.e., nearly matching the non-robust run-time of Klivans, O'Donnell, and Servedio \cite{KOS2008}. This resolves the questions above in the setting of Gaussian data distribution in the dependence on $K$, and up to the $\eps^2$ vs. $\eps^4$ gap in its dependence on the accuracy.

Formally, we consider the class ${\cal{C}}_K$, the class of Boolean functions $f$ 
satisfying $\Bdry_r(f) \le Kr$ for all $r$. Lange and Vasilyan \cite{LV26} considered linear classifiers, which are only a subset of ${\cal{C}}_{O(1)}$, while here we consider the class ${\cal{C}}_K$ in full generality. Under Gaussian data, we show the following:
\begin{theorem}[\Cref{thm:full}, informal]
\label{thm: main informal}
Let $\cal{D}$ be a distribution on $\R^d \times \{0,1\}$ whose data-points come from the standard Gaussian.
There is an algorithm that takes random example access to $\cal{D}$, as well as parameters $\eps, r$ in $(0,1)$, and outputs a hypothesis $h$ satisfying 
\[\Bdry_r(h) \le O(Kr), \qquad \Prx_{(\bx,\by) \sim \mathcal D}[h(\bx) \ne \by] \le \opt_K + \eps,\]
where $\opt_K$ denotes distance to the class ${\cal{C}}_K$.
Its running time is $d^{\tilde{O}(K^2/\eps^2 + K^4r^2/\eps^4)}$. In particular, when $Kr = O(\eps)$,
this running time is $d^{\tilde{O}(K^2/\eps^2)}$, and in general,
the running time is $d^{\tilde{O}(K^2/\eps^4)}$.\footnote{The notation $\tilde{O}$ suppresses log factors depending on $K$, $1/\varepsilon$, and $1/(Kr)$.}
\end{theorem}
\begin{remark}
    As we explain in Appendix \ref{apx:log-concave}, this theorem extends to the much more general class of isotropic strongly log-concave distributions, conditional on a polynomial approximation assumption for smooth functions under strongly log-concave distributions motivated by the recent announcement of \cite{BizeulKlartagAnnouncement2026}.
\end{remark}
The general running time bound in Theorem \ref{thm: main informal} follows from the fact that $r \le 1/K$, since otherwise the allowable boundary volume would be 1.
Another notion in the robust learning 
literature is the \emph{robust risk}, defined as 
\[\RobRisk(h) \coloneqq \Prx_{(\bx,\by) \sim D} [\exists z: \ltwo{\bz} \le r, h(\bx+\bz) \ne \by].\]
This is the total probability that an adversary can cause misclassification, whether that
misclassification comes from the hypothesis itself or the boundary.
Since this is at most the sum of boundary volume and the
classification error,
our algorithm also gives a guarantee on robust risk, obtained
by setting $\eps = Kr$:
\begin{corollary}
In time $d^{\tilde{O}(1/r^2)}$, this algorithm outputs a hypothesis satisfying 
\[\RobRisk(h) \le \opt_K + O(Kr).\]
\end{corollary}

\subsection{Some of the key technical insights}
\label{sec: key technical insights}
Although we would like to refer the reader to detailed recent surveys on adversarial robustness in Machine Learning \cite{peck2024introduction, costa2024survey} for a broad overview, one short summary is that there are two broad templates for achieving adversarial robustness:
\begin{itemize}
\item \textbf{Randomized Smoothing} \cite{CRK2019, lecuyer2019certified}.  The key idea is that adversarially fragile instances, such as the ones constructed in \cite{GKVZ2022}, are fragile because there are lots of tiny \textquotedblleft{}islands\textquotedblright{} in $\mathbb{R}^d$ on which the value of the classifier changes drastically. These islands are hard to find because their probability mass is small, but the adversary can nudge a data point into one of these islands by only perturbing it a little. This suggests a natural way to correct this issue --- query your function in a neighborhood of $x$ and average (or take a majority). This will average out these islands. 

This method was one of the key ingredients of Lange and Vasilyan \cite{LV26} as well. However, there is a problem with extending this method on its own beyond linear classifiers: smoothing at the rate necessary to remove the adversarial islands also introduces a lot of error when we try to move beyond linear classifiers. See section \ref{sec:previous-work} for more details and discussion on the work of Lange and Vasilyan \cite{LV26} and about the obstacles to extending their work for general functions with bounded boundary ratio.

\item \textbf{Adversarial Training} \cite{madry2018towards, goodfellow2015explaining} . Here, the core idea is also simple: when training a classifier, directly incentivize not only accuracy but also robustness. For example, for a datapoint $x$, one can try to make the training loss the sum of (a) if the prediction on $x$ is correct and (b) if the datapoint $x$ is $r$-far from the boundary. In practice, part (b) involves running some known attacks on the datapoint $x$, and penalizing if an attack succeeds and finds a small perturbation of $x$ that makes the value of the current classifier change.
\end{itemize}
Adversarial training is often considered less reliable because, end to end, it will protect us only from some anticipated set of attacks we tried during the training, but will not directly guarantee that the boundary volume is small. As a result, an adversary may successfully attack the classifier by using a novel attack not previously used during adversarial training.

A key idea for us is incorporating a theoretically secure version of adversarial training. For simplicity, assume we are in the realizable setting, so the unknown function $f$ we are trying to learn has a boundary ratio of $K$. A first attempt would try to combine the polynomial-regression method of Klivans, O'Donnell, and Servedio with the idea of adversarial training as follows:
\par\vspace{0.5em}
\begingroup
\setlength{\fboxsep}{8pt}
\noindent\fbox{\begin{minipage}{\dimexpr\linewidth-2\fboxsep-2\fboxrule\relax}
\setlength{\parindent}{0pt}

\textbf{Attempt 1:}\par\smallskip
Find a degree-$\operatorname{poly}(K/\epsilon)$ polynomial $P^{\text{robust}}$ for which
\begin{align}
\text{Accuracy:}\quad &
\mathbb{E}_{x\sim\mathrm{Gaussian}}
\bigl[|P^{\text{robust}}(x)-f(x)|\bigr]
\leq \epsilon, \notag\\[0.5em]
\text{Robustness:}\quad &
\mathbb{E}_{x\sim\mathrm{Gaussian}}\left[
  \max_{\|z\|\leq r} P^{\text{robust}}(x+z)
  - \min_{\|z\|\leq r} P^{\text{robust}}(x+z)
\right]
\leq O(Kr).
\label{eq:oscillation}
\end{align}
Pick a random $t$ in $[0,1]$ and output the hypothesis mapping $x$ to
$\Ind[P^{\text{robust}}(x)\geq t]$.

\end{minipage}}
\endgroup
\par\vspace{0.5em}
The optimization in Equation \ref{eq:oscillation} is a very natural approach, and the TRADES method \cite{zhang2019trades} is a close analogue of this in the applied literature.
\footnote{The work of \cite{zhang2019trades} also contains a theoretical component. However, in contrast with our work here, their theory addresses the statistical and not the computational aspects of the problem. }

It can be shown that polynomials $P^{\text{robust}}$ like this do exist, and once such a polynomial is found, the random rounding $\operatorname{sign}(P^{\text{robust}}(x)-t)$ would finish the job. The key issue is in the computational implementation. The quantities
\[
\max_{\|z\|\leq r} P^{\text{robust}}(x+z)
-\min_{\|z\|\leq r} P^{\text{robust}}(x+z)
\]
are NP-hard to compute in the worst case \cite{Nesterov2003}, making this approach a non-starter.

Instead of computing the extrema of our polynomial, we
introduce an auxiliary coordinate whose purpose is to produce
(possibly loose) bounds on the range.
This is an instance of \emph{functional lifting}, the
technique of introducing additional coordinates in order to
express geometric constraints as convex constraints in a
higher-dimensional space
\cite{PockCremersBischofChambolle2010, AlbertiBouchitteDalMaso2003}.
The goal is to learn the function
\[
\operatorname{B}_{P^{\text{robust}}}(x,a)=
\begin{cases}
\displaystyle \max_{\|z\|\leq |a|} P^{\text{robust}}(x+z) & \text{when } a\geq 0,\\[0.6em]
\displaystyle \min_{\|z\|\leq |a|} P^{\text{robust}}(x+z) & \text{when } a\leq 0.
\end{cases}
\]
Which we hope will approximate
\[
\operatorname{B}_f(x,a)=
\begin{cases}
\displaystyle \max_{\|z\|\leq |a|} f(x+z) & \text{when } a\geq 0,\\[0.6em]
\displaystyle \min_{\|z\|\leq |a|} f(x+z) & \text{when } a\leq 0.
\end{cases}
\]

But this goal may sound preposterous at first sight. The function $\operatorname{B}_f(x,a)$ encodes not only the value of $f$ at each point $x$ but also how far $x$ is from the boundary of $f$. But the training data has only information about which side of the boundary of $f$ each data point $x$ is. It has no information about how far each data point $x$ is from the boundary that this function describes! 

This is where the last key insight comes in. Although our samples do not directly reveal $\operatorname{B}_f(x,a)$ for non-zero values of $a$,
there are good approximations to $\operatorname{B}_f(x,a)$ that satisfy stringent \emph{self-consistency} conditions:
\begin{center}
\setlength{\fboxsep}{9pt}
\fbox{\begin{minipage}{\dimexpr\linewidth-2\fboxsep-2\fboxrule\relax}
\textbf{Key insight:} $\operatorname{B}_{P^{\text{robust}}}(x,a)$ needs to satisfy \emph{self-consistency conditions} that we can enforce in our optimization loop:
\[
\|\nabla_x\operatorname{B}_{P^{\text{robust}}}(x,a)\|_2
\leq \partial_a\operatorname{B}_{P^{\text{robust}}}(x,a).
\]
\end{minipage}
}
\end{center}
While self-consistency doesn't imply that $\operatorname{B}_{P^{\text{robust}}}(x,a)$ exactly computes the extrema on the $a$-ball, it does imply that it computes useful bounds on these extrema. 
We explain why this is the case in \Cref{sec:oscillation}.
We leverage these self-consistency checks in our optimization loop by searching for a polynomial $Q:\mathbb{R}^d\times\mathbb{R}\to\mathbb{R}$ that serves as a proxy for $\operatorname{B}_{P^{\text{robust}}}$ and, through that, is a proxy of $\operatorname{B}_f(x,a)$ and satisfies the following conditions for some very small $\beta$:
\par\vspace{0.5em}
\begingroup
\setlength{\fboxsep}{8pt}
\noindent\fbox{\begin{minipage}{\dimexpr\linewidth-2\fboxsep-2\fboxrule\relax}
\setlength{\parindent}{0pt}

\textbf{Blueprint of our approach:}\par\smallskip
Find a degree-$\operatorname{poly}(K/\epsilon)$ polynomial $Q$ for which
\begin{alignat*}{2}
&\text{\textbf{Accuracy:}}\qquad &&
\mathbb{E}_{x\sim\mathrm{Gaussian}}
\bigl[|Q(x,0)-f(x)|\bigr]
\leq \epsilon.
\\[0.5em]
&\text{\textbf{Robustness:}}\qquad &&
\mathbb{E}_{x\sim\mathrm{Gaussian}}
\bigl[Q(x,r)-Q(x,-r)\bigr]
\leq O(Kr).
\\[0.5em]
&\text{\textbf{Self-Consistency:}}\qquad &&
\mathbb{E}_{x,a\sim\mathrm{Gaussian}}\left[
\max\bigl(0,\|\nabla_xQ(x,a)\|_2-\partial_aQ(x,a)\bigr)^2
\right]
\leq \beta.
\end{alignat*}
Obtain the polynomial $P$ by applying Gaussian smoothing to $Q$.

Pick a random $t$ in $[0,1]$ and output the hypothesis mapping $x$ to
$\Ind(P(x, 0)\geq t)$.

\end{minipage}}
\endgroup
\par\vspace{0.5em}
In brief, we show that the self-consistency condition guarantees that the difference $Q(x,r)-Q(x,-r)$ is indeed a good proxy for
$\max_{\|z\|\leq r}Q(x+z)-\min_{\|z\|\leq r}Q(x+z)$,
after the polynomial $Q$ has been smoothed. See \Cref{sec:overview} for a more detailed overview of the
full learning algorithm and its analysis.

\subsection{Applications to concrete function classes}
Many natural geometric concept classes have small boundary ratio $K$, and are thus naturally adversarially robust. See \Cref{tab:boundary-profiles} for some examples of such classes.
We also include the run-time of our algorithm for those values of $K$, as well as
statistical query lower bounds for the agnostic learning of these function classes, which can be viewed as evidence that the run-times we give here are close to optimal.

\begin{table}[H]
\centering
\small
\setlength{\tabcolsep}{4pt}
\resizebox{\linewidth}{!}{
\begin{tabular}{@{}llll@{}}
\toprule
Function class
& Boundary Ratio 
& Our run-time
& \begin{tabular}[c]{@{}l@{}}
  SQ lower bound\\
  for agnostic learning
  \end{tabular} \\
\midrule

\begin{tabular}[c]{@{}l@{}}
Functions with\\
Boundary Ratio $K$
\end{tabular}
& $K$
& $d^{\tilde O(K^2/\eps^4)}$
& $d^{\Omega(K^2/\eps^2)}$
  \cite{DKPZ2021}$^\dagger$ \\
  
Euclidean balls
& $O(1)$
& $d^{\tilde O(1/\eps^4)}$
& $d^{\Omega(1/\eps^2)}$
  \cite{DKPZ2021}$^\dagger$ \\

Intersections/unions of $m$ halfspaces
& $O\!\left(\sqrt{\log m}\right)$
& $d^{\tilde O(\log m/\eps^4)}$
& $d^{\Omega\left(
    \frac{\log m}{\log\log m}+\eps^{-2}
  \right)}$
  \cite{HSSV2022} \\

Axis-aligned boxes
& $O\!\left(\sqrt{\log d}\right)$
& $d^{\tilde O(\log d/\eps^4)}$
& $d^{\Omega(\log(1/\eps))}$
  \cite{FeldmanKothari2015}$^\dagger$ \\

Boolean functions of $m$ halfspaces
& $O(m)$
& $d^{\tilde O(m^2/\eps^4)}$
& $d^{\Omega(m^2/\eps^2)}$
  \cite{DKPZ2021}$^\dagger$ \\

Boolean functions of $m$ balls
& $O(m)$
& $d^{\tilde O(m^2/\eps^4)}$
& $d^{\Omega(m^2/\eps^2)}$
  \cite{DKPZ2021}$^\dagger$ \\

Convex sets
& $O(d^{1/4})$
& $d^{\tilde O(\sqrt d/\eps^4)}$
& $2^{\Omega(\sqrt d+\log(d)/\eps^2)}$
  \cite{KOS2008,DKPZ2021}$^\dagger$ \\

\bottomrule
\end{tabular}
}
\caption{
Values of the Gaussian boundary ratio $K$ for various hypothesis classes, and our run-time for those values of $K$.
We are assuming $m\ge4$, $d\ge2$,
and the parameter ranges of the cited lower bounds.
Entries marked $\dagger$ follow by the reductions described in \Cref{apx:boundary-profile-bounds}.
See \Cref{apx:boundary-profile-bounds} for proofs of the
boundary-profile bounds.
}
\label{tab:boundary-profiles}
\end{table}

\subsection{Previous work
on computationally efficient adversarial robustness}
\label{sec:previous-work}
The work of~\cite{LV26} studies the same objective of
learning a classifier with near-optimal classification error and small
boundary volume, but $\opt$ is the minimum error among 
\textit{halfspaces}, rather than the minimum error among small-boundary
functions in general.
Under subgaussian isotropic
log-concave marginals, they obtain error $\operatorname{opt}+O(\varepsilon)$
and boundary volume $O(r+\varepsilon)$ in time
$d^{\widetilde O(1/\varepsilon^2)}$.
Their algorithm performs $L_1$ polynomial regression under noise-sensitivity-based
constraints, 
rounds the polynomial to an almost-robust function, then
uses a local corrector that converts the relaxed stability
guarantees into adversarial robustness. 

There is an inherent obstacle to extending their technique 
to learning small-boundary classes broader than halfspaces. 
They use two noise sensitivity constraints, one on the average
\emph{local noise sensitivity} at a larger scale $cr$
\[\Ex_{\bx} [\phi_f(\bx)] \le O(r), \qquad \phi_f(x) = \Prx_{\bz  \sim N(0,(cr)^2I_d)}[f(x) \ne f(x+\bz)]\]
and one on the \emph{isolation probability}
\[\mathsf{iso}(f) = \Prx_{\bx  \sim D}[\phi_f(\bx) > 0.8] \le \eps.\]
Their relaxed algorithm produces a function satisfying these constraints. Then, the local corrector flips
the label of isolated points. Their analysis uses a complementary
condition --- the condition $\phi_f(x) < 0.1$ --- as a certificate
that $x$ becomes robust after the local corrector is applied (i.e., that all disagreeing points in its $r$-ball must be isolated, and therefore flipped).
By Markov, the bound on $\E[\phi]$ implies that
$1 - O(r)$ fraction of points have such a certificate.

This is made possible by the fact 
that halfspaces have zero isolation. However, even
intersections of two halfspaces can have $O(r)$ isolation: consider a band defined by two parallel halfspaces that are at
distance $r/100$ from each other. For $r \gg \eps$, there is no 
$\eps$-approximation to this function with $\eps$ isolation, and
running the local corrector on this function would yield 
a constant function.

The exciting recently announced work of \cite{CMSYZ26} gives a different perspective
on computationally efficient adversarial robustness.
In their model, called \emph{testable sanitization}, the algorithm is
given access to a classifier $f$ that is claimed to have small boundary
volume, together with samples from the data distribution. The algorithm
is required to run fast and either (i) reject or (ii) produce a classifier
$f'$ that is close to $f$ and has small boundary volume. It must accept
whenever $f$ satisfies the promised conditions. Some of the key
differences between our work and theirs are
\begin{itemize}
    \item The model of \cite{CMSYZ26} is \emph{distribution-free},
    i.e., there is no assumption on the data distribution. This is more
    general than our distributional setting, which is specific to
    Gaussian distributions, with a conditional extension to isotropic
    strongly log-concave distributions.

    \item They have a deterioration in the boundary distance parameter
    while we do not. Qualitatively, the robustness radius guaranteed
    after correction is smaller by a factor of $\sqrt{d}$ than the
    radius in the input promise. They show that, in their general
    distribution-free model, this deterioration is inevitable up to
    logarithmic factors.

    \item Their algorithm can query $f$ at arbitrary points, and the
    classifier produced after correction retains this access.
    In our setting, the algorithm learns a predictor from random
    labeled examples.
\end{itemize}
We note that our algorithm incurs only a constant-factor deterioration
in the boundary-volume bound at the same radius, although we require
the target class to have a bounded boundary profile across all scales.
There is also an accuracy qualification in their results: their general
guarantee ties the permitted disagreement with $f$ to the boundary-volume
bound. Higher accuracy requires additional regularity assumptions and
a restriction on the robustness radius.

We would also like to point out that in our setting---the setting of
function learning rather than testable sanitization---learning the full
class of functions with bounded boundary profile without assumptions
on the data distribution can require exponentially many samples.
This can be shown by considering distributions that are uniform on
an exponentially large error-correcting code of constant distance,
whose points can be labeled arbitrarily while preserving robustness
in a neighborhood of each point.

\subsection{Related work}
\label{sec:related-work}
\paragraph{Proper agnostic learning.}
A proper learner automatically inherits any boundary-volume bound that
holds uniformly over its output class. The work of~\cite{DKKTZ2021} gave a proper agnostic learner for
halfspaces under Gaussian marginals. More recently, Tikhonov and
Vasilyan~\cite{TikhonovVasilyan2026} obtained proper agnostic learners for
Boolean functions of $m$ halfspaces and for intersections of $m$ halfspaces.
Their dimension exponents are respectively
$O(m^2\log(1/\varepsilon)/\varepsilon^2)$ and
$O(\log(m+1)\log(1/\varepsilon)/\varepsilon^2)$, with additional
terms depending on $m$ and $\varepsilon$ but not on $d$.
Thus robustness for these classes can also be obtained through proper
learning, but the accuracy competes with $\opt$ for these 
specific classes, which may be larger than $\opt_K$, the 
optimal error among members of the broader boundary-profile class.

\paragraph{Distribution-free adversarially robust learning.}
Much of the theoretical literature optimizes robust risk directly,
allowing perturbations from a specified family of sets.
As mentioned earlier, the robust risk model is incomparable to ours;
robust risk is at most the sum of error and boundary volume, but
it may be smaller when the two risk sets overlap.
Montasser, Hanneke, and Srebro~\cite{MHS2019} show that finite-VC classes
are robustly PAC learnable improperly, while proper robust learning can
fail. This is an information-theoretic result, rather than a general
efficient algorithm. The works of \cite{BLPR2019} and \cite{DNV2019} 
exhibit computational separations between ordinary and robust learning.
For halfspaces, the work of \cite{MGDS2020} studies efficient robust learning 
in realizable and
random-label-noise settings, while \cite{DKM2020} give algorithms and hardness 
results for proper robust learning with agnostic noise.
Jog~\cite{Jog2021} studies Gaussian surface-area bounds for parallel sets,
with applications to learning such sets and estimating robust risk.

\paragraph{Smoothing, gradient certificates, and polynomial optimization.}
Randomized smoothing converts a classifier's behavior under Gaussian noise
into certificates against worst-case Euclidean perturbations. Cohen, Rosenfeld, and Kolter~\cite{CRK2019} give a sharp analysis of this
connection. Gradient-based certificates and regularizers also have a long
history. Hein and Andriushchenko~\cite{HeinAndriushchenko2017} derive
instance-specific robustness guarantees and propose cross-Lipschitz
regularization. Awasthi, Dutta, and
Vijayaraghavan~\cite{ADV2019} connect provable adversarial robustness to
polynomial optimization.

\paragraph{Smoothed agnostic learning and lower bounds.}
The work of~\cite{CKKMS2026} introduces a smoothed agnostic benchmark
that evaluates functions under random Gaussian perturbations, and obtains
algorithms for low-intrinsic-dimensional classes with bounded Gaussian
surface area. 
Diakonikolas and Kane~\cite{DiakonikolasKane2026} establish statistical-query
lower bounds for that smoothed benchmark.
\paragraph{Backdoors, watermarks, and verifiable robustness.}
Backdoored classifiers exhibit behavior controlled by a secret trigger while 
retaining high accuracy on ordinary inputs. Goldwasser et al.~\cite{GKVZ2022}
show that such backdoors can be computationally undetectable, even when they 
introduce adversarial examples near every input. Related constructions use 
backdoors as watermarks to establish ownership of a model~\cite{AdiEtAl2018}.
These results motivate removing unwanted behavior without first identifying its
trigger, as studied in oblivious backdoor mitigation~\cite{GSVV2025} and in connections
between backdoor defense and learnability~\cite{CHLX2025}. Our algorithm admits a 
complementary interpretation as a robustness-preserving distillation procedure. Given 
query access to an unknown classifier supplied by an untrusted party, we can label
fresh Gaussian examples using that classifier and learn a replacement with boundary
volume $O(Kr)$. If the supplied classifier is close in $L_2$ to a 
function with boundary profile $K$, the replacement approximately preserves its
predictions, while every perturbation of norm at most $r$ preserves the
replacement's prediction outside a set of Gaussian mass $O(Kr)$. This guarantee
holds regardless of whether the supplied classifier is actually robust, and limits
the success of backdoors without requiring their detection. It does
not imply removal of arbitrary watermarks or correct predictions on every trigger
input. 

Our explicit local range certificates also suggest a \emph{verifiably robust} 
variant, following \cite{LV26}, in which a user checks local constancy of the
replacement rather than trusting the supplier's robustness claim. As in that work,
deterministic soundness must be distinguished from guarantees that hold with high
probability over trusted randomness used in constructing the certificates.

\subsection{AI methodology}

We used OpenAI's GPT-6 Sol, GPT-6 Astra, and GPT-5.6 Sol to assist with developing and checking arguments, drafting and revising the 
exposition, and searching for related work.
In particular, the functional lifting method and the distribution-free rounding argument were proposed by GPT-5.6 Sol.
We also used Claude Fable 5.1 to assist with writing.
The tools materially affected all sections of the paper except for \Cref{sec:background-results,sec:overview}. 
The authors take full responsibility for all results,
proofs, references, and claims of originality.

\section{Technical overview}
\label{sec:overview}
 In \Cref{sec: key technical insights}, we gave an informal overview of some of the key ideas that went into our approach.
We now go one level deeper and develop the ideas in a little bit more detail.
We begin with the guarantees needed for rounding, then describe
the convex program and show that it contains a good low-degree
approximator. Finally, we return to the rounding step to obtain
the additive agnostic error guarantee.

\subsection{Rounding}
\label{sec:overview-rounding}
We begin with the randomized rounding strategy of \cite{KKMS08}.
Consider $p:\R^d\to[0,1]$.
For $\theta\in(0,1)$, define the threshold function
$g_\theta(x)=\Ind[p(x)\ge\theta]$.

For a fixed labeled example $(x,y)$, the fraction of thresholds
$\theta\in(0,1)$ for which $g_\theta(x)\ne y$ is exactly
$|p(x)-y|$. Averaging over examples gives
\[
 \Ex_{\btheta \sim\mathrm{Unif}(0,1)}
       \Prx_{(\bx,\by)\sim\mathcal D}[g_{\btheta} (\bx)\ne \by]
 =\Ex_{(\bx,\by)\sim\mathcal D}|p(\bx)-\by|
\]
Thus an $L_1$ prediction-loss bound controls the expected
classification error after rounding.

Boundary volume has an analogous interpretation. Define the
\emph{oscillation} of a function $f$ on a ball by
\[
 \mathsf{osc}_{B(x,r)}f
 =\sup_{z\in B(x,r)}f(z)-\inf_{z\in B(x,r)}f(z).
\]
For a fixed center $x$, the thresholds that allow $g_\theta$ to change within $B(x,r)$ lie between the minimum and maximum of
$p$ on the ball. The length of this interval is therefore
$\mathsf{osc}_{B(x,r)}p$. Averaging over centers gives
\[
 \Ex_{\btheta \sim\mathrm{Unif}(0,1)}\Bdry_r(g_{\btheta} )
 =\Ex_{\bx \sim\gamma}\mathsf{osc}_{B(\bx,r)}p.
\]

Consequently, bounds on $L_1$ prediction loss and average
$r$-ball oscillation give bounds on both error and boundary volume
in expectation over thresholds, though they need not be simultaneously
attained by any single threshold.

In the realizable setting, where $\opt=0$, or the semi-agnostic
setting, where a constant-factor loss in $\opt$ is allowed,
Markov's inequality suffices: at least $4/5$ of thresholds have
both error and boundary volume at most ten times their respective
expectations. This gives error $O(\opt+\eps)$ and boundary volume
$O(Kr)$. For the agnostic guarantee $\opt+O(\eps)$, the rounding
step needs more care, as discussed in
\Cref{sec:intro-round-agnostic}. The required properties of $p$
remain the same: small $L_1$ prediction loss and small average
$r$-ball oscillation.

\subsection{A polynomial with oscillation bounds}
\label{sec:oscillation}

Having observed that average oscillation supplies a bound on
boundary volume of the rounded polynomial, we would like to
learn under an oscillation constraint.
Since oscillation is convex, we could impose the constraint
\[
\E_{\bx}[\mathsf{osc}_{B(\bx,r)}p]\le O(Kr),
\]
which, if satisfied, would imply the desired rounding bounds.
Clipping to $[0,1]$ can only reduce oscillation and $L_1$ error,
so this constraint suffices even when $p$ takes values outside
$[0,1]$. However, as discussed in \Cref{sec: key technical insights},
it is intractable to evaluate whether or not a polynomial
satisfies the constraint.

A first try at circumventing this problem could be a Lipschitz
constraint: $K$-Lipschitzness would imply that the $r$-ball
oscillation is $\le 2Kr$ everywhere.
The problem with this approach is that the $L_1$ error would
be large. To approximate a Boolean function, the polynomial
needs to have a sharp transition from the negative side to
the positive side, which a fixed global Lipschitz bound
does not allow.
We would like to control average oscillation while allowing
the predictor to have a sharp transition near the
classification boundary.
Thus instead of using a fixed Lipschitz bound in $\R^d$,
we will control the rate of change in $\R^d$ \emph{relative}
to the rate of change in another coordinate, so that the
other coordinate can be used to certify a local oscillation
bound.

Using the auxiliary certificate variable $a$ introduced in
\Cref{sec: key technical insights}, we will learn a $(d+1)$-dimensional
polynomial $Q(x,a):\R^{d+1}\to\R$ by the following convex program:
\begin{itemize}
\item Minimize:
$\E_{(\bx,\by)\sim\mathcal D}[|Q(\bx,0)-\by|]$;
\item Signed width constraint:
$\E_{\bx}[Q(\bx,r)-Q(\bx,-r)]\le O(Kr)$;
\item Certificate validity constraint:
$\forall x,a:\ \ltwo{\nabla_x Q(x,a)}
-\partial_a Q(x,a)\le0$.
\end{itemize}
In fact the validity constraint will be relaxed; the next
subsection will explain why and how to relax the constraint.
For now, imagine that this constraint is exactly satisfied.

The $d$-dimensional slice $Q(x,0)$ is the predictor; we minimize
the $L_1$ error of this slice.
We call $Q(x,r)-Q(x,-r)$ the \emph{signed width} at $x$.
When the validity constraint holds, the signed width is an
upper bound on $\mathsf{osc}_{B(x,r)}Q(\cdot,0)$.
Indeed, consider some $y$ with $\ltwo{y-x}\le r$, and integrate
along the path from $(x,r)$ to $(y,0)$.
Using the parametrization
\[
v(t)=\left(x+\frac{t}{r}(y-x),r-t\right),
\qquad t\in[0,r],
\]
we have
\begin{align*}
Q(y,0)-Q(x,r)
&=\int_0^r
\left[
\left\langle\nabla_x Q(v(t)),\frac{y-x}{r}\right\rangle
-\partial_a Q(v(t))
\right]dt\\
&\le\int_0^r
\left[
\ltwo{\nabla_x Q(v(t))}-\partial_a Q(v(t))
\right]dt\\
&\le0.
\end{align*}
Symmetrically, integrating along the path from $(x,-r)$
to $(y,0)$ gives $Q(y,0)\ge Q(x,-r)$.
Thus
\[
Q(x,-r)\le Q(y,0)\le Q(x,r)
\qquad\text{for every }y\in B(x,r).
\]
In other words, $[Q(x,-r),Q(x,r)]$ is an interval guaranteed
to contain $Q(y,0)$ for all $y\in B(x,r)$, which we call the
\emph{danger interval} for $x$.
Not only is the size of the interval a bound on the oscillation
on this ball, but we can even evaluate whether or not a
specific threshold is certified to be good for this ball
(it is good when it is outside of this interval).

\subsection{Relaxing the validity constraint}
This convex program, as written in the above section, might not contain a polynomial 
of error $\opt+\eps$. At least,
our analysis does not give one. To prove that a good approximator exists in the feasible
region, we need to add a little bit of slack to this constraint. The relaxed constraint is
\begin{equation}
\label{eq:defect-constraint}
\E_{\bx,\ba} [(\ltwo{\nabla_x Q (\bx,\ba)} - \partial_a Q(\bx,\ba))_+^2] \ll \exp(-r^2/{\sigma}^2),
\end{equation}
where ${\sigma}$ is the {standard deviation} of an Ornstein-Uhlenbeck smoothing transformation to be
performed on $Q$ (see \Cref{sec:OU-smoothing} for details
on the OU operator).
The error objective and the signed width constraint will actually be applied to the
smoothed polynomial $P \coloneqq {T_\sigma} Q$, which is the actual output of
the polynomial learning program.

We show that this exponentially small squared ``certificate defect'' of $Q$ implies
a useful nearly-pointwise bound on $\ltwo{\nabla_x P (x,a)} - \partial_a P(x,a)$, the
defect of the smoothed output. The key is to show that the smoothed squared
defect at a point, $[{T_\sigma}(\defect_Q)^2](x,0)$, controls the defect of $P={T_\sigma}Q$ uniformly on the whole
$r$-ball around $x$, and for every $|a| \le r$. The precise inequality is
\[{T_\sigma} \defect_{Q}(z,a) \le \sqrt{[{T_\sigma} (\defect_{Q}^2)](x,0)} \cdot \exp\Paren{\frac{\ltwo{z-x}^2 + a^2}{2{\sigma}^2}}.\]
This is sometimes referred to as the dimension-free power-Harnack inequality, and 
combined with \Cref{eq:defect-constraint} and a transfer of the squared-defect bound to $\E_{\bx}[{T_\sigma} \defect_Q^2](\bx,0)$,
it allows us to give valid danger intervals
for many points.
Aside from an ``exception set'' of very small mass (points where $[{T_\sigma} \defect_Q^2](x,0)$ is much higher than its expectation), all points are
surrounded by an $r$-ball of defect at most 1. This gives a valid certificate for the 
expanded danger interval $[P(x,-r)-r, P(x,r)+r]$, obtained again by integrating on the path
from $(x,-r)$ to $(y,0)$ and from $(y,0)$ to $(x,r)$. Points in the exception set have
invalid certificates, but the set is small enough to contribute little to the boundary volume.

\subsection{Showing that a feasible polynomial exists}
\label{sec:overview-witness}
Now it remains to show that this relaxed program is indeed 
guaranteed to contain a polynomial with error $\opt + \eps$.
For this, it suffices to show that each function $f \in \CK$ has
some witness, a polynomial that is $\eps$-close to $f$ and satisfies the constraints.
To do so, we will let the extra coordinate describe 
distance to the boundary, and explain why this provides a witness.

Consider, for a Boolean function $f$ with boundary profile
$K$, a signed 
distance-to-boundary function 
\[\sdist_f(x) = \begin{cases} \mathsf{dist}(x, f^{-1}(0)) \quad | \quad f(x) = 1 \\
-\mathsf{dist}(x, f^{-1}(1)) \quad | \quad f(x) = 0  \end{cases}. \]
The witness comes from the \emph{subgraph lift} of this function.
Consider a $(d+1)$-dimensional function indexed by $x$ and an offset $a$: 
\[B_f(x,a) = \Ind[\sdist_f(x) \ge -a].\]
Positive offsets expand the accepted set $f^{-1}(1)$, while negative
offsets erode it.
It satisfies $B_f(x,0) = f(x)$ and $\E_{\bx} [B_f(\bx,r)-B_f(\bx,-r)] \le Kr$, since
this expectation is exactly the $r$-boundary volume of $f$. 
It satisfies a finite-difference analogue of the validity constraint:
\[B_f(x+z, a) - B_f(x,a+u) \le 0, \qquad \ltwo{z} \le u,\]
but it is not everywhere differentiable, and it is not a low-degree polynomial.

Applying OU smoothing to $B_f$ fixes both of these problems: ${T_\sigma} B_f$ satisfies the
validity constraint, and ${T_\sigma} B_f$ and its derivatives have geometrically decreasing
Hermite tails. Thus, truncating at degree $k$ costs about ${\exp(-\Omega(\sigma^2 k))}$ in both
the error $\ltwo{({T_\sigma} B_f)^{\le k}(\cdot,0) - f}$ and the squared certificate defect. The defect
term dominates here; we need to set $k$ to be about ${r^2/\sigma^4}$
in order to guarantee that \Cref{eq:defect-constraint} is satisfied. Thus, we would
like to take ${\sigma}$ to be as large as possible; that is, to smooth at the largest scale the
error budget will allow.

For a function of boundary profile $K$, smoothing {with standard deviation $\sigma$} introduces
error about ${K\sigma}$. Thus, to approximate to error $\eps$, ${\sigma}$ needs to be at most
${\eps/K}$. Combining the parameters yields the claimed degree bound for the
program-feasibility witness of $f$.

\subsection{Rounding in the agnostic setting}
\label{sec:intro-round-agnostic}
We now return to rounding. To preserve the additive agnostic
error guarantee, we use a mixture of thresholds rather than
a single threshold, then turn that mixture into a deterministic
classifier by partitioning the domain. The partition must
reproduce the mixture's error and boundary-volume guarantees
while introducing little additional boundary. Letting $E_\theta$ denote the 
set of points misclassified by rounding at $\theta$, and $C_\theta$
the set of possible boundary points (points such that $\theta$ falls inside their danger intervals), we
ensure that 
\[\Pr[x \in E_\theta \land x \in A_i] \approx w_i \cdot \Pr[x \in E_\theta] \qquad \Pr[x \in C_\theta \land x \in A_i] \approx w_i \cdot \Pr[x \in C_\theta] \]
where $A_i$ is the $i_{th}$ part of the partition and $w_i$ is the 
$i_{th}$ mixing weight. In other words, the partition behaves like 
the mixture.

To avoid introducing too much boundary volume between parts of the 
partition, it is important that the mixture have a constant number of components.
The rounding argument of \cite{LV26} establishes existence of such
a mixture and gives an algorithm for finding it and turning it 
into a partition, which requires some 
assumptions on the distributions of the error and boundary sets.
We could use the same method here, but we give a simpler 
distribution-free partitioning algorithm. This doesn't eliminate any 
distributional assumptions from the full learner, but it may be of 
independent interest.

The new partitioning ingredient is a consequence of classic results in
\emph{consensus splitting}, also called cake cutting or necklace 
splitting. The classic necklace splitting result of \cite{AW86} says
that for any $q$-coloring of an interval, there is a partition of the
interval using $q$ cuts such that each part contains half the points
of each color. A result of \cite{StromquistWoodall1985} allows splitting 
into two parts of arbitrary weights, and a straightforward corollary
allows splitting into $k$ parts of arbitrary weights.
This is the basis of our partitioning algorithm: 
we project onto a line and split the sets $E_\theta$ and $C_\theta$ for each of the thresholds $\theta$ in our mixture, according to the mixing weights $\vec{w}$.
We use this existence result, together with VC uniform convergence, to
implement the division empirically from samples.

\section{Preliminaries}
Throughout the paper, $\gamma$ denotes the
standard Gaussian distribution $\mathcal N(0,I_d)$ on $\R^d$.
We retain explicit subscripts for other dimensions: in particular,
$\gamma_{d+1}=\mathcal N(0,I_{d+1})$ and
$\gamma_1=\mathcal N(0,1)$. More generally, $\gamma_n$ denotes the
standard Gaussian distribution on $\R^n$ when the dimension is specified.
We use bold symbols (e.g. $\bx$) to denote
random variables.
{When two samplings are listed under an expectation sign,
they are independent unless stated otherwise.}

We will often work with $(d+1)$-dimensional variables, decomposed as $(x,a) \in \R^d \times \R$.
We refer to $x$ as the \emph{spatial coordinates} and $a$ as the \emph{offset coordinate}.

\subsection{Surface area and boundary volume}
\begin{definition}[Gaussian surface area]
\label{def:surface}
We use the Minkowski surface area, as in \cite{KOS2008}.
For a Borel set $A$, the surface area $\Gamma_\gauss(A)$ is defined
\[\Gamma_\gauss(A) = \underset{\delta \to 0}{\lim \inf} \frac{\gauss(A_\delta \setminus A)}{\delta},\]
where $A_\delta$ denotes 
\[\{x: \dist(x,A) \le \delta\}.\]
\end{definition}

\begin{fact}
\label{fact:bdry-to-surface-area}
For all $f \in \CK$, $\Gamma_\gauss(f^{-1}(1)) \le K$.
\end{fact}
\begin{proof}
By the boundary condition, we have $\gauss(A_\delta \setminus A) \le K\delta$ for all $\delta$.
By taking the limit as $\delta \to 0$, we obtain $\Gamma_\gauss(f^{-1}(1)) \le K$.
\end{proof}

\begin{fact}[Corollary 14 of \cite{KOS2008}]
\label{fact:surface-area-to-NS}
Let $f$ be a function satisfying $\Gamma_\gauss(f^{-1}(1)) \le K$ and let $\bx,\bz  \sim \gauss$ be independent.
For ${\sigma\in(0,1/\sqrt2]}$, write ${\rho_\sigma=\sqrt{1-\sigma^2}}$, and set
\[{\bx_\sigma = \rho_\sigma \bx+ \sigma \bz.}\]
Then there is a universal constant $C$ such that
\[\Pr[f(\bx) \ne f({\bx_\sigma})] \le CK{\sigma}.\]
\end{fact}

\subsection{Hermite expansion}
\label{sec:Hermite}
We use the normalized probabilists' Hermite polynomials
\[
 H_j(t)=\frac{(-1)^j}{\sqrt{j!}}e^{t^2/2}
                  \frac{d^j}{dt^j}e^{-t^2/2},
 \qquad j\ge0.
\]
For a multi-index $\nu\in\mathbb Z_{\ge0}^n$, write
$|\nu|=\sum_i\nu_i$ and $H_\nu(z)=\prod_{i=1}^nH_{\nu_i}(z_i)$.
These polynomials form an orthonormal basis of $L_2(\gamma_n)$.
For $F\in L_2(\gamma_n)$, we write
\[
 \widehat F(\nu)=\Ex_{\bz\sim\gamma_n}F(\bz)H_\nu(\bz),
 \qquad F_j=\sum_{|\nu|=j}\widehat F(\nu)H_\nu,
 \qquad F=\sum_{j\ge0}F_j,
\]
where the last sum converges in $L_2$. Parseval's identity gives
$\|F\|_2^2=\sum_j\|F_j\|_2^2=\sum_\nu\widehat F(\nu)^2$.
We use $F^{\le k}=\sum_{j=0}^kF_j$ for the degree-$k$ truncation
and $F^{>k}=F-F^{\le k}$ for its tail. In particular,
$\|F^{\le k}\|_2\le\|F\|_2$.
The space of degree-$\le k$ polynomials on $\R^n$ has dimension
$\binom{n+k}{k}$, and the Gaussian $L_2$ norm of a polynomial is
the Euclidean norm of its Hermite coefficient vector.

The derivative identity
$\partial_{z_i}H_\nu=\sqrt{\nu_i}H_{\nu-e_i}$ (with derivative
zero if $\nu_i=0$) implies, for every polynomial $Q$,
\[
 \Ex_{\bz\sim\gamma_n}\|\nabla Q(\bz)\|_2^2
 =\sum_j j\|Q_j\|_2^2.
\]
For the product basis on $\R^d\times\R$, spatial averaging removes
every term with nonzero degree in a spatial coordinate:
\[
 \Ex_{\bx\sim\gamma}H_{(\mu,j)}(\bx,a)
 =\begin{cases}H_j(a),&\mu=0,\\0,&\mu\ne0.\end{cases}
\]
\subsection{The Ornstein--Uhlenbeck semigroup}
\label{sec:OU-smoothing}
{We adopt the convention of indexing the OU operator by the standard
 deviation of smoothing.\footnote{{Another convention indexes the
 operator by time $s$, with Gaussian kernel
 $e^{-s}z+\sqrt{1-e^{-2s}}\,\bg$ and additive composition in $s$.
 Our parameter is the standard deviation
 $\sigma=\sqrt{1-e^{-2s}}$.}}}
{For $0<\sigma<1$, write $\rho_\sigma=\sqrt{1-\sigma^2}$ and define}
\[
 ({T_\sigma}F)(z)
 =\Ex_{\bg \sim\gamma_n}
       F({\rho_\sigma}z+{\sigma}\bg),
 \qquad z\in\mathbb R^n.
\]
{The smoothing parameter used by the learner satisfies $\sigma<1/2$;
compositions of operators may have larger standard deviations.} We use this operator with $n=d+1$, writing $z=(x,a)$.
The following standard properties will be used throughout.

\begin{enumerate}
\item \textbf{Hermite coefficients.}
The OU operator defined above acts diagonally in Hermite basis:
\[
 {T_\sigma}F=\sum_{j\ge0}{(1-\sigma^2)^{j/2}}F_j,
 \qquad \|{T_\sigma}F\|_2^2=\sum_{j\ge0}{(1-\sigma^2)^j}\|F_j\|_2^2.
\]
Thus smoothing preserves the degree bound and multiplies each
degree-$j$ coefficient by ${(1-\sigma^2)^{j/2}}$. On the degree-$\le k$
polynomial space we also use its inverse ${T_\sigma^{-1}}$, which multiplies
those coefficients by ${(1-\sigma^2)^{-j/2}}$ and has $L_2$ operator norm at most
${(1-\sigma^2)^{-k/2}}$.
\item \textbf{Semigroup and stationarity.}
For ${\sigma,\eta\in[0,1)}$, with $T_0$ the identity,
\[
 {T_\sigma T_\eta
 =T_{\sqrt{\sigma^2+\eta^2-\sigma^2\eta^2}}.}
\]
Moreover, if $\bz,\bg \sim\gamma_n$ are independent, then
${\rho_\sigma}\bz+{\sigma}\bg \sim\gamma_n$. Consequently,
\[
 \mathbb E_{\gamma_n}{T_\sigma}F=\mathbb E_{\gamma_n}F,
 \qquad
 \|{T_\sigma}F\|_{L_p(\gamma_n)}
 \le \|F\|_{L_p(\gamma_n)}
 \quad(1\le p<\infty).
\]

\item \textbf{Positivity and smoothing.}
The operator ${T_\sigma}$ preserves pointwise inequalities and constant
functions. In particular, $0\le F\le1$ implies $0\le {T_\sigma}F\le1$.
If $F$ is bounded and measurable, then ${T_\sigma}F$ is
$C^\infty$ for every ${0<\sigma<1}$. Its first derivatives satisfy
\[
 \partial_{z_i}{T_\sigma}F(z)
 =\frac{{\rho_\sigma}}{{\sigma}}
   \mathbb E_{\bg} \!\left[
       \bg_iF({\rho_\sigma}z+{\sigma}\bg)
   \right].
\]
This formula follows by differentiating the Gaussian kernel
and does not require $F$ to be differentiable.

\item \textbf{Gradient commutation.}
If $F$ is Lipschitz, then
\[
 \nabla {T_\sigma}F={\rho_\sigma}{T_\sigma}(\nabla F),
\]
where $\nabla F$ is its almost-everywhere gradient and ${T_\sigma}$
acts coordinatewise on vector-valued functions.

\item \textbf{Averaging over the spatial coordinates.}
For bounded measurable $F:\mathbb R^{d+1}\to\mathbb R$, define
$m_F(a)=\Ex_{\bx \sim\gamma}F(\bx,a)$.
Spatial stationarity gives
\[
 m_{{T_\sigma}F}(a)
 =\Ex_{\bg \sim\gamma_1}
       m_F({\rho_\sigma}a+{\sigma}\bg).
\]
In particular, averaging over $x$ reduces OU smoothing to
one-dimensional smoothing in the offset coordinate.
\end{enumerate}

\section{Approximation by a low-degree polynomial}
In this section, we will define the convex program, then 
show that every function $f$ in the class $\CK$ corresponds to a good solution.

Fix a degree bound $k$, a {smoothing standard deviation $0<\sigma<1/2$}, and a
squared-defect budget $\beta>0$.
For a polynomial $Q:\mathbb R^{d+1}\to\mathbb R$, write $P={T_\sigma}Q$,
and define its certificate defect by
\[
 \defect_Q(x,a)
 =\bigl(\|\nabla_xQ(x,a)\|_2-\partial_aQ(x,a)\bigr)_+,
 \qquad (u)_+=\max\{u,0\}.
\]
Write 
\[
m_P(a) \coloneqq \Ex_{\bx  \sim \gauss} [P(\bx,a)].
\]
We call $P(x,r)-P(x,-r)$ the \emph{signed width} at $x$;
its expectation is $m_P(r)-m_P(-r)$.
\Cref{alg:learn-polynomial} describes the polynomial-learning
subroutine used by \Cref{alg:full}. We state its convex program
in terms of exact population statistics; its sample implementation
in exact real arithmetic is given by
\Cref{lem:empirical-implementation}.

\begin{algorithm}[H]
\caption{\textsc{LearnPolynomial}$(K,r,\varepsilon,\delta,\mathcal D)$}
\label{alg:learn-polynomial}
\begin{algorithmic}[1]
\Require $K\ge1$; $0<r\le1/K$; accuracy
         $\varepsilon\in(0,1/4)$; failure probability
         $\delta\in(0,1/4)$; sample access to $\mathcal D$.
\Ensure A smoothed polynomial $P:\R^{d+1}\to\R$ of degree at most $k$.
\State Set ${\sigma},\beta,k$ as in \Cref{par:parameters}.
\State Optimize over degree-$\le k$ polynomials $Q$, writing $P={T_\sigma}Q$:
\begin{equation}
\label{eq:polynomial-learning-program}
\begin{aligned}
 \underset{Q}{\operatorname{minimize}}
 \quad&
 \Ex_{(\bx,\by)\sim\mathcal D}
       |P(\bx,0)-\by|
 \\[2pt]
 \text{subject to}\quad
&\text{(a) }\|Q\|_{2}\le1,
 \\
&\text{(b) }m_P(r)-m_P(-r)\le C_0Kr,
 \\
&\text{(c) }\Ex_{(\bx,\ba)\sim\gamma_{d+1}}
       \defect_Q(\bx,\ba)^2\le\beta,
\end{aligned}
\end{equation}
\State \Return $P={T_\sigma}Q$.
\end{algorithmic}
\end{algorithm}
Here $C_0$ is a sufficiently large universal constant.
The optimization variables are the coefficients of the auxiliary
polynomial $Q$. The subroutine returns $P={T_\sigma}Q$, whose coefficients
depend linearly on those of $Q$, as in
\Cref{sec:OU-smoothing}.

The objective and constraint (a) are convex, and constraint (b) is linear. 
For each $(x,a)$, the function $\defect_Q(x,a)$ is nonnegative
and convex in the coefficients of $Q$, so its square is
also convex. Thus constraint (c) is convex.

\subsection{The signed distance function and offset lift}
\label{sec:offset-lift}
We aim to construct, for each $f \in \CK$,
a witness that the convex program defined in \Cref{eq:polynomial-learning-program} can find a 
polynomial close to $f$.
The starting point for this witness will be what we call an \emph{offset lift}: 
a function that takes an input $x$ and an offset $a$, and moves the boundary
of $f$ out by a distance $a$ (or in by a distance $|a|$,
if $a$ is negative). 

More formally, for a Boolean function $f \in \CK$, we define its \emph{signed distance function}:
\[\sdist_f(x) = \begin{cases}\dist(x, f^{-1}(0)) \quad f(x) = 1\\
-\dist(x,f^{-1}(1)) \quad f(x) = 0 \end{cases}.\]
Using the signed distance, we define the offset lift:
\[B_f(x,a) = \Ind[\sdist_f(x) \ge -a].\]
Positive offsets expand the accepted set, and negative offsets erode it.
Observe that this function recovers $f$ in that $B_f(x,0) = f(x)$ almost surely.
It also recovers a finite-difference analogue of the 
certificate validity constraint: it satisfies
\[
 B_f(x,a)\le B_f(z,a+u)
 \qquad\text{whenever }u\ge\|x-z\|_2,
\]
which makes $[B_f(x,-r), B_f(x,r)]$ a certified bound on the range of $B_f(\cdot,0)$ in the $r$-ball
around $x$.
Thus, it also satisfies the expected signed width bound:
\[\Ex_{\bx \sim \gamma}[B_f(\bx, r)] - \Ex_{\bx \sim \gamma}[B_f(\bx, -r)] = \Bdry_r(f) \le Kr.\]
The witness that the convex program has a good solution will be a degree-$k$
truncation of a smoothed version of this function: 
\[{Q^\star} = ({T_\sigma} B_f)^{\le k}, \qquad {P^\star} = {T_\sigma} {Q^\star},\]
for a parameter ${\sigma}$ to be set later.
\subsection{Smoothing the offset lift}
We will show that before truncation, smoothing 
preserves satisfaction of the program's constraints
and does not introduce too much error.
We will handle the error and defect first (the objective and constraint (c)), then bound
the expected signed width (constraint (b)).
First, we will show that before truncation, ${T_\sigma} B_f$ has zero defect, and that its error on the $a=0$ slice scales with $K{\sigma}$.
Later we will set ${\sigma}$ so that this is at most $\eps$.
\begin{lemma}[Smoothing $B_f$ preserves error and zero defect]
\label{lem:smoothed-hard-lift}
Let $f\in\mathcal C_K$ and fix ${\sigma\in(0,1/\sqrt2]}$. Write
\[
 V={T_\sigma}B_f,\qquad {\rho_\sigma=\sqrt{1-\sigma^2}}.
\]
Then $V$ takes values in $[0,1]$ and satisfies the certificate validity constraint
\[
 \|\nabla_xV(x,a)\|_2-\partial_aV(x,a)\le0
 \qquad\text{for every }(x,a)\in\mathbb R^{d+1}.
\]
Moreover, for an absolute constant $C$,
\[
 \Ex_{\bx \sim\gamma}|V(\bx,0)-f(\bx)|
 \le CK{\sigma}.
\]
\end{lemma}

\begin{proof}
First observe that the offset lift satisfies
\[
 B_f(x,a)\le B_f(z,a+u)
 \qquad\text{whenever }u\ge\|x-z\|_2.
\]
We show the same holds for $V$. Couple the Gaussian perturbations
at the two points to obtain
\[
\begin{aligned}
 V(x,a)-V(z,a+u)
 =\Ex_{\by \sim\gamma,\,\bg \sim\gamma_1}\bigl[
 &B_f({\rho_\sigma}x+{\sigma}\by,{\rho_\sigma}a+{\sigma}\bg)\\
 &-B_f({\rho_\sigma}z+{\sigma}\by,{\rho_\sigma}(a+u)+{\sigma}\bg)\bigr].
\end{aligned}
\]
The spatial separation and offset increment inside the expectation
are ${\rho_\sigma}\|x-z\|_2$ and ${\rho_\sigma}u$, respectively.
Thus the integrand is nonpositive whenever $u\ge\|x-z\|_2$, giving
\[
 V(x,a)\le V(z,a+u)
 \qquad\text{whenever }u\ge\|x-z\|_2.
\]
In particular, for every unit vector $v$ and every $u>0$,
\[
 V(x+uv,a-u)\le V(x,a).
\]
The boundedness of $B_f$ implies that $V$ is smooth (see \Cref{sec:OU-smoothing}).
Differentiating at $u=0$ and maximizing over $v$ gives
\[
 \|\nabla_xV(x,a)\|_2-\partial_aV(x,a)\le0.
\]
Also, $0\le V\le1$ because $B_f$ is $\{0,1\}$-valued.

For the approximation bound, let $\bx,\bz \sim\gamma$ and
$\bg \sim\gamma_1$ be independent, and set
$\bw={\rho_\sigma}\bx+{\sigma}\bz $.
By the definition of the offset lift,
\[
 |B_f(\bw,{\sigma}\bg)-f(\bw)|
 \le \Ind[\bw \in\operatorname{Bdry}_{{\sigma}|\bg|}(f)].
\]
Since $\bw \sim\gamma$ independently of $\bg $, the boundary-profile
assumption and the Gaussian noise-stability bound
from \Cref{fact:bdry-to-surface-area,fact:surface-area-to-NS} imply
\[
\begin{aligned}
 \mathbb E|V(\bx,0)-f(\bx)|
 &\le
 \mathbb E|B_f(\bw,{\sigma}\bg)-f(\bw)|
 +\mathbb E|f(\bw)-f(\bx)|\\
 &\le K{\sigma}\mathbb E|\bg|+CK{\sigma}
 \le C'K{\sigma},
\end{aligned}
\]
for another absolute constant $C'$.
\end{proof}
We will now show that ${T_\sigma}B_f$ also preserves the expected signed width bound up to a constant factor.
\begin{lemma}[Smoothing preserves the expected signed width bound]
\label{lem:smoothing-signed-width}
Let $f \in \CK$, ${\sigma\in(0,1/\sqrt2]}$, and $V = {T_\sigma} B_f$, and let
\[m_V(a) \coloneqq \Ex_{\bx  \sim \gauss}[V(\bx,a)].\]
Then 
\[m_V(r)-m_V(-r) \le O(Kr).\]
\end{lemma}

\begin{proof}
{Write $\rho_\sigma=\sqrt{1-\sigma^2}$.}
By spatial stationarity, for independent $\bx \sim\gamma$ and
$\bg \sim\gamma_1$,
\[
 m_V(a)=\mathbb E_{\bx,\bg}B_f(\bx,{\rho_\sigma}a+{\sigma}\bg)
 =\mathbb E_{\bx} \Phi\left(
       \frac{\sdist_f(\bx)+{\rho_\sigma}a}{{\sigma}}\right),
\]
where $\Phi$ is the standard Gaussian CDF. Hence
\begin{equation}
\label{eq:m-deriv}
 m_V'(a)=\frac{{\rho_\sigma}}{{\sigma}}
 \mathbb E_{\bx} \varphi\left(
       \frac{\sdist_f(\bx)+{\rho_\sigma}a}{{\sigma}}\right)\ge0,
\end{equation}
where $\varphi$ is the standard Gaussian density.

First suppose $r\le{\sigma}$. For $|a|\le r$, the inequality
$(u+v)^2\ge u^2/2-v^2$ gives
\[
 \varphi\left(\frac{\sdist_f(\bx)+{\rho_\sigma}a}{{\sigma}}\right)
 \le C\exp\left(-\frac{\sdist_f(\bx)^2}{4{\sigma}^2}\right).
\]
The boundary-profile assumption gives
$\Pr[|\sdist_f(\bx)|\le u]\le Ku$. Therefore, by Tonelli's theorem,
\[
\begin{aligned}
 \mathbb E_{\bx} \exp\left(-\frac{\sdist_f(\bx)^2}{4{\sigma}^2}\right)
 &=\int_0^\infty\frac{u}{2{\sigma}^2}
       e^{-u^2/(4{\sigma}^2)}
       \Pr[|\sdist_f(\bx)|\le u]\,du\\
 &\le K\int_0^\infty\frac{u^2}{2{\sigma}^2}
       e^{-u^2/(4{\sigma}^2)}\,du
 \le C'K{\sigma}.
\end{aligned}
\]
Substituting into \Cref{eq:m-deriv} gives $m_V'(a)\le C''K$
on $[-r,r]$. Integration yields $m_V(r)-m_V(-r)\le CKr$.

If $r>{\sigma}$, then, up to null endpoints,
\[
\begin{aligned}
 m_V(r)-m_V(-r)
 &=\mathbb E_{\bx,\bg}\bigl[
 B_f(\bx,{\rho_\sigma}r+{\sigma}\bg)-B_f(\bx,-{\rho_\sigma}r+{\sigma}\bg)\bigr]\\
 &=\Pr[-{\rho_\sigma}r-{\sigma}\bg \le\sdist_f(\bx)<{\rho_\sigma}r-{\sigma}\bg]\\
 &\le\mathbb E_{\bg} \Pr_{\bx} [
       |\sdist_f(\bx)|\le{\rho_\sigma}r+{\sigma}|\bg|]\\
 &\le K({\rho_\sigma}r+{\sigma}\mathbb E|\bg|)
 \le CKr.
\end{aligned}
\]
This completes the proof.
\end{proof}
\subsection{Truncating the smoothed offset lift}
Fix $f\in\CK$. We will choose the {smoothing standard deviation $\sigma$} and degree
bound $k$ so that $Q^\star=({T_\sigma}B_f)^{\le k}$ is a feasible witness.
Besides satisfying the norm constraint, truncation must preserve
three properties: small $L_1$ approximation error on the prediction
slice, a small expected signed width,
and a sufficiently small squared certificate defect.

{The smoothing standard deviation is set by the error budget.} The defect
requirement then determines what Hermite degrees we must retain:
the squared defect must be exponentially small in ${r^2/\sigma^2}$ to
control it uniformly on perturbation balls outside a small
exceptional set (see \Cref{thm:output-guarantees}\textup{(i)--(ii)}). We first prove the required defect and norm
bounds, then verify the error and expected signed width bounds.
In particular, writing $P^\star={T_\sigma}Q^\star$, we will obtain
$\Ex_{\bx} |P^\star(\bx,0)-f(\bx)|\le\eps$. The triangle inequality gives
\[
 \Ex_{(\bx,\by)\sim\mathcal D}|P^\star(\bx,0)-\by|
 \le \Prx_{(\bx,\by)\sim\mathcal D}[f(\bx)\ne \by]+\eps.
\]
Thus the program's optimal loss is at most $\opt_K+\eps$.

\paragraph{Parameters.}
\label{par:parameters}
Assume $K\ge1$ and $0<r\le1/K$. Set
\[
 {\sigma=c_\sigma\frac{\varepsilon}{K}},
 \qquad
 {\beta=c_{\beta} Kr\,\sigma^2
       \exp\left(-C_{\beta}\frac{r^2}{\sigma^2}\right)},
\]
where ${0<c_\sigma\le1}$ is sufficiently small,
$0<c_{\beta}\le1/4$, and $C_{\beta}\ge4$ are universal constants.
{This choice gives}
\[
 \beta\le\frac{Kr}{4}{\sigma}^2
              \exp\left(-\frac{2r^2}{{\sigma}^2}\right).
\]
Thus \Cref{thm:output-guarantees}\textup{(i)--(ii)}, with
${\alpha_{\mathrm{Exceptional\ Set}}}=Kr/4$, ensures that the exceptional set after smoothing
has Gaussian mass at most $Kr/4$.
The truncation degree will then be 
\[
{k=\left\lceil\frac2{\sigma^2}\log\frac8{\sigma^2\beta}\right\rceil=
O\!\left(
\frac1{\sigma^2}\left[
1+\frac{r^2}{\sigma^2}
+\log\frac1{Kr\,\sigma^4}
\right]\right).}
\]
Substituting in ${\sigma=c_\sigma\eps/K}$, this yields
\[
k=\tilde{O}\Paren{\frac{K^2}{\eps^2} + \frac{K^4r^2}{\eps^4}}.
\]
Since $r$ is always at most $1/K$ (otherwise the allowable boundary volume would be 1, 
making the boundary guarantee trivial), this degree is always at most $\tilde{O}(K^2/\eps^4)$.
When $Kr = O(\eps)$, $k$ becomes $\tilde{O}(K^2/\eps^2)$.
First we will show that truncating at this degree $k$ 
gives the desired exponentially small squared defect.
\begin{lemma}[Truncation with small squared defect]
\label{lem:truncation-defect}
Let $f\in\mathcal C_K$, let ${\sigma\in(0,1/2)}$, and let
$\beta\in(0,1]$. Write
\[
 V={T_\sigma}B_f,
 \qquad
 {Q^\star}=V^{\le k},
\]
using the truncation notation from \Cref{sec:Hermite}. If
\[
 k\ge
 {\left\lceil\frac2{\sigma^2}\log\frac{8}{\sigma^2\beta}\right\rceil},
\]
then $\|{Q^\star}\|_{2}\le1$ and
\[
 \mathbb \Ex_{(\bx,\ba)\sim\gamma_{d+1}}
 \left[
   \bigl(\|\nabla_x{Q^\star}(\bx,\ba)\|_2
                -\partial_a{Q^\star}(\bx,\ba)\bigr)_+^2
 \right]
 \le\frac{\beta}{2}.
\]
\end{lemma}

\begin{proof}
Since $B_f$ is Boolean, $\|B_f\|_2\le1$.
The truncation and smoothing identities in \Cref{sec:Hermite}
therefore give $\|{Q^\star}\|_2\le1$.
Write $B_j=(B_f)_j$ and $R=V-{Q^\star}$. The gradient identity gives
\[
 \mathbb E\|\nabla R\|_2^2
 =\sum_{j>k}j{(1-\sigma^2)^j}\|B_j\|_2^2
 \le\sup_{j>k}j{e^{-\sigma^2j}}.
\]
Using ${u e^{-\sigma^2u/2}\le2/(e\sigma^2)}$ for $u\ge0$, we obtain
\[
 \mathbb E\|\nabla V^{> k}\|_2^2
 \le{\frac2{e\sigma^2}e^{-\sigma^2(k+1)/2}}.
\]
By \Cref{lem:smoothed-hard-lift}, $V$ has zero defect everywhere. Hence, pointwise,
\[
\begin{aligned}
 \bigl(\|\nabla_x{Q^\star}\|_2-\partial_a{Q^\star}\bigr)_+
 &\le \|\nabla_x R\|_2+|\partial_aR|\\
 &\le\sqrt2\,\|\nabla R\|_2.
\end{aligned}
\]
Squaring and taking expectations therefore gives
\[
 \E[\defect_{{Q^\star}}(\bx,\ba)^2]
 \le{\frac4{e\sigma^2}e^{-\sigma^2(k+1)/2}}
 \le\frac{\beta}{2},
\]
where the last inequality follows from the prescribed
choice of $k$.
\end{proof}

Now we will show that truncation does not introduce too
much error or too much expected signed width.
\begin{lemma}[Error and expected signed width bounds for the truncated witness]
\label{lem:witness-slices}
Let $K\ge1$, $f\in\mathcal C_K$, ${\sigma\in(0,1/2)}$, and
$0<r\le1/K$. {Suppose}
\[
 0<\beta\le Kr\,{\sigma}.
\]
Set
\[
 {Q^\star}=({T_\sigma}B_f)^{\le k},
 \qquad
 {P^\star}={T_\sigma}{Q^\star},
 \qquad
 k\ge{\left\lceil\frac2{\sigma^2}\log\frac8{\sigma^2\beta}\right\rceil}.
\]
Then
\[
 \Ex_{\bx \sim\gamma}|{P^\star}(\bx,0)-f(\bx)|
 \le CK{\sigma},
\]
and
\[
 m_{{P^\star}}(r)-m_{{P^\star}}(-r)\le CKr,
 \qquad
 m_{{P^\star}}(a)=\Ex_{\bx \sim\gamma}{P^\star}(\bx,a).
\]
In particular, taking ${\sigma=c_\sigma\varepsilon/K}$ gives
prediction error $\varepsilon$.
\end{lemma}

\begin{proof}
Let
\[
 V={T_\sigma}B_f,\qquad
 R={Q^\star}-V,\qquad
 W={T_\sigma}V={T_{\sigma\sqrt{2-\sigma^2}}}B_f.
\]
Since $\|B_f\|_2\le1$, the Hermite expansion gives
\[
 \|R\|_2\le{(1-\sigma^2)^{(k+1)/2}
 \le e^{-\sigma^2(k+1)/2}
 \le\frac{\sigma^2\beta}{8}}.
\]
We first show that smoothing this tail controls its values
on the required slices $a=0$, $a=-r$, and $a=r$.

For fixed $a$, Jensen's inequality and OU stationarity
give
\[
 \mathbb E_{\bx} |{T_\sigma}R(\bx,a)|
 \le\mathbb E_{\bz,\ba}|R(\bz,\ba)|,
\]
where $\bz \sim\gamma$ and
$\ba \sim N({\rho_\sigma} a,{\sigma}^2)$ are independent.
We can perform a change of measure to express this as an
expectation over $\bz $ and $\gauss_1$:
\[\Ex_{\bz,\ba}|R(\bz,\ba)| = \Ex_{{\bz\sim\gauss,\,\bg\sim\gauss_1}}\Brac{|R(\bz,\bg)|\cdot \tfrac{d\ba}{d\gauss_1}}.\]
The density ratio between Gaussians is
\[
\frac{d\ba}{d\gauss_1}(u)=
\frac1{{\sigma}}
\exp\!\left(
\frac{u^2}{2}
-\frac{(u- {\rho_\sigma}a)^2}{2{\sigma}^2}
\right) \le \frac{e^{a^2/2}}{{\sigma}}.
\]
Cauchy--Schwarz then implies
\begin{equation}
\label{eq:slice-estimate}
 \mathbb E_{\bx} |{T_\sigma}R(\bx,a)|
 \le {\sigma}^{-1/2}e^{a^2/4}\|R\|_2
 \le {\frac{\sigma^{3/2}\beta}{8}}e^{a^2/4}.
 \end{equation}
Since ${P^\star}-W={T_\sigma}R$, this bounds the truncation error
at each offset.

{For the $a=0$ slice, \Cref{lem:smoothed-hard-lift}
applied with standard deviation $\sigma\sqrt{2-\sigma^2}\le\sqrt2\,\sigma$
(which lies in $(0,1/\sqrt2)$ because $\sigma<1/2$) gives}
\[
 \mathbb E|W(\bx,0)-f(\bx)|\le CK{\sigma}.
\]
Therefore
\[
\begin{aligned}
 \mathbb E|{P^\star}(\bx,0)-f(\bx)|
 &\le CK{\sigma}
       +{\frac{\sigma^{3/2}\beta}{8}}\\
 &\le CK{\sigma}
       +{\frac18Kr\,\sigma^{5/2}}
 \le CK{\sigma},
\end{aligned}
\]
using $Kr\le1$, ${\sigma}\le1$, and $K\ge1$.

For the bound on $m_{{P^\star}}(r)-m_{{P^\star}}(-r)$, we have from \Cref{lem:smoothing-signed-width}
\[
 m_W(r)-m_W(-r)\le CKr
\]
for a constant $C$.
Applying \Cref{eq:slice-estimate} at $a=-r$ and $a=r$ yields
\[
\begin{aligned}
 m_{{P^\star}}(r)-m_{{P^\star}}(-r)
 &\le CKr+
       {\frac{\sigma^{3/2}\beta}{4}}e^{r^2/4}\\
 &\le CKr+
       {\frac14Kr\,\sigma^{5/2}}\,e^{1/4}
 \le C'Kr,
\end{aligned}
\]
where we used $r\le1/K\le1$.
\end{proof}

\section{Guarantees on the program's output}
\label{sec:output-guarantees}

This section establishes the analytic certificates used in rounding.
Here $f$ denotes a polynomial on $\R^{d+1}$. In the learning algorithm
we apply the results with $f=Q$, where $P={T_\sigma}Q$ is returned by
\textsc{LearnPolynomial} and $Q$ is its auxiliary polynomial.

\subsection{Sandwiching a function via its defect}

Throughout, $d\ge1$ is an integer and $f:\R^d\times\R\to\R$ is a polynomial.
We write $\nabla_x f$ for the gradient of $f$ in the $x$ variables,
$\partial_a f$ for its derivative in the $a$ variable,
and $\|\cdot\|_2$ for the Euclidean norm on $\R^d$.
For a point $(y,a)\in\R^d\times\R$, define
\[
  \defect_f(y,a) := \max\bigl\{0,\ \|\nabla_x f(y,a)\|_2 - \partial_a f(y,a)\bigr\}.
\]
Here, we give proofs of the following two comparison inequalities,
which control the value of $f$ at a point $(z,0)$ by its values at $(x,\pm r)$ for any $r\ge\|z-x\|_2$,
and of \Cref{thm:output-guarantees} below, our main result concerning the Gaussian noise operator,
together with \Cref{prop:defect,prop:ball,prop:avg,prop:width} on which it rests.

The defect measures by how much $\partial_a f$ fails to dominate $\|\nabla_x f\|_2$.
When it vanishes, moving the $a$ variable forward increases $f$ at least as fast as moving the $x$ variable
in any direction can decrease it, so the $a$ variable acts as a budget that pays for displacements in $x$.
This intuition is formalized in the following proposition.

\begin{proposition}
\label{prop:comparison}
Suppose that
\[
  \defect_f(y,a) = 0
  \qquad\text{for every } (y,a)\in\R^d\times\R.
\]
Then, for any $x,z\in\R^d$ and any $r\ge\|z-x\|_2$, it holds that
\[
  f(x,-r) \le f(z,0) \le f(x,r).
\]
\end{proposition}

We also need the following robust version, which holds when the defect is close to zero
(rather than exactly zero).

\begin{proposition}
\label{prop:robust}
Fix $x,z\in\R^d$, let $r\ge\|z-x\|_2$, and let $\beta\ge0$.
Suppose that
\[
  \defect_f(y,a) \le \beta
\]
for every $y$ on the line segment joining $x$ to $z$ and every $a\in[-r,r]$.
Then
\[
  f(x,-r) - \beta r \le f(z,0) \le f(x,r) + \beta r.
\]
\end{proposition}

\Cref{prop:comparison} is the special case $\beta=0$ of \Cref{prop:robust}.
We nevertheless prove \Cref{prop:comparison} first,
since the proof of \Cref{prop:robust} builds on the derivative bounds established there.

\subsection{The Gaussian noise operator and the main theorem}

Recall the OU {smoothing operator} ${T_\sigma}$ from \Cref{sec:OU-smoothing},
which adds independent Gaussian noise to both the $x$ and the $a$
variable. Since ${T_\sigma}$ preserves polynomials, ${T_\sigma}f$ is a polynomial
and $\defect_{{T_\sigma}f}$ is well-defined.

Our main result says that if the defect of a polynomial $f$ is small on average,
then the comparison inequality of \Cref{prop:robust} holds for ${T_\sigma} f$ around most points,
and the expected negative width $\Brac{({T_\sigma} f)(x,r)-({T_\sigma} f)(x,-r)}_-$ is small.

\begin{theorem}[Program output guarantees]
\label{thm:output-guarantees}
Let $r\in[0,1]$ and ${\alpha_{\mathrm{Exceptional\ Set}}}\in(0,1]$. Suppose that
\[
  \Ex_{(\bx,\ba)\sim\mathcal N(0,I_{d+1})}\bigl[\defect_f(\bx,\ba)^2\bigr]
  \le {\alpha_{\mathrm{Exceptional\ Set}}}\,{\sigma}^2\, e^{-2r^2/{\sigma}^2}.
\]
Then the following hold.
\begin{enumerate}
  \item[\textup{(i)}]
    $\displaystyle
    \Ex_{\bx \sim\gauss}\Bigl[{T_\sigma}\bigl(\defect_f^2\bigr)(\bx,0)\Bigr]
    \le {\alpha_{\mathrm{Exceptional\ Set}}}\, e^{-2r^2/{\sigma}^2}$.
  \item[\textup{(ii)}]
    For every $x\in\R^d$ with ${T_\sigma}(\defect_f^2)(x,0) \le e^{-2r^2/{\sigma}^2}$
    and every $z\in\R^d$ with $\|z-x\|_2\le r$,
    \[
      ({T_\sigma} f)(x,-r) - r
      \le ({T_\sigma} f)(z,0)
      \le ({T_\sigma} f)(x,r) + r .
    \]
  \item[\textup{(iii)}]
    For $x\in\R^d$, let $W(x) := ({T_\sigma} f)(x,r) - ({T_\sigma} f)(x,-r)$.
    If $K\ge0$ and $\Ex_{\bx \sim\gauss}[W(\bx)]\le Kr$, then
    \[
      \Ex_{\bx \sim\gauss}\bigl[\max\{0,W(\bx)\}\bigr]
      \le (K + 4)\, r .
    \]
\end{enumerate}
\end{theorem}

We will break the proof of \Cref{thm:output-guarantees} into a number of propositions,
stated next, which we then combine in \Cref{sec:main} to prove the theorem.

The hypothesis of \Cref{thm:output-guarantees} controls the defect of $f$ only on average,
whereas \Cref{prop:robust} needs a pointwise bound along a segment, and an average cannot provide one.
This is why \Cref{thm:output-guarantees} is about ${T_\sigma} f$ rather than $f$:
averaging over Gaussian noise turns average information into pointwise information,
as \Cref{prop:ball,prop:avg} below make precise.
The first step is to check that smoothing does not create defect.

\begin{proposition}[Noise does not increase the defect]
\label{prop:defect}
For every $(x,a)\in\R^d\times\R$,
\[
  \defect_{{T_\sigma} f}(x,a)
  \le \bigl({T_\sigma} \defect_f\bigr)(x,a).
\]
\end{proposition}

Recall that in \Cref{thm:output-guarantees} we want the bound
\[
  \Ex_{\bx \sim\gauss}\Bigl[{T_\sigma}\bigl(\defect_f^2\bigr)(\bx,0)\Bigr] \le {\alpha_{\mathrm{Exceptional\ Set}}}\, e^{-2r^2/{\sigma}^2}
\]
to ensure that $\defect_{{T_\sigma} f}(z,a)$ is small when $z$ is $r$-close to $x$.
This uses the fact that ${T_\sigma}$ makes functions smooth,
which is formalized in the Harnack inequality of \Cref{fact:harnack} and lets us show the following.

\begin{proposition}[Certifying smallness of the defect on a ball]
\label{prop:ball}
Let $x\in\R^d$ and $r\ge0$. Suppose that
\[
  {T_\sigma}\bigl(\defect_f^2\bigr)(x,0) \le e^{-2r^2/{\sigma}^2}.
\]
Then the following hold.
\begin{enumerate}
  \item[\textup{(i)}]
    $\defect_{{T_\sigma} f}(z,a) \le 1$ for every $z\in\R^d$ with $\|z-x\|_2\le r$ and every $a\in[-r,r]$.
  \item[\textup{(ii)}]
    For every $z\in\R^d$ with $\|z-x\|_2\le r$,
    \[
      ({T_\sigma} f)(x,-r) - r
      \le ({T_\sigma} f)(z,0)
      \le ({T_\sigma} f)(x,r) + r .
    \]
\end{enumerate}
\end{proposition}

\Cref{prop:ball} needs a bound at the specific point $(x,0)$, while the hypothesis of \Cref{thm:output-guarantees}
is an average over $(\bx,\ba)\sim\mathcal N(0,I_{d+1})$, in particular an average over $\ba \sim\mathcal N(0,1)$.
The next proposition bridges the two, again relying on the smoothing properties of the noise operator ${T_\sigma}$.

\begin{proposition}[Certificates exist for most points]
\label{prop:avg}
It holds that
\[
  \Ex_{\bx \sim\gauss}\Bigl[{T_\sigma}\bigl(\defect_f^2\bigr)(\bx,0)\Bigr]
  \le \frac{1}{{\sigma}}\,\Ex_{(\bx,\ba)\sim\mathcal N(0,I_{d+1})}\bigl[\defect_f(\bx,\ba)^2\bigr].
\]
\end{proposition}

Recall that when the defect vanishes identically, \Cref{prop:comparison} with $z=x$ gives
$f(x,r)\ge f(x,-r)$ for every $x$: the function never decreases as $a$ moves from $-r$ to $r$.
When the defect is merely small on average, we cannot expect this at every $x$,
but after smoothing by ${T_\sigma}$ we can show it is close to true on average:
the amount by which $({T_\sigma} f)(x,r)$ falls below $({T_\sigma} f)(x,-r)$, averaged over $x$,
is controlled by the average defect.
This is the content of the last proposition, which is the last ingredient we use to prove \Cref{thm:output-guarantees}.

\begin{proposition}[Small average defect controls the negative part of the signed width]
\label{prop:width}
Let $r\in[0,1]$.
For $x\in\R^d$, define the signed width
\[
  W(x) := ({T_\sigma} f)(x,r) - ({T_\sigma} f)(x,-r).
\]
Then the following hold.
\begin{enumerate}
  \item[\textup{(i)}]
    $\displaystyle
    \Ex_{\bx \sim\gauss}\bigl[\max\{0,-W(\bx)\}\bigr]
    \le \frac{4r}{{\sigma}}\,
        \sqrt{\Ex_{(\bx,\ba)\sim\mathcal N(0,I_{d+1})}\bigl[\defect_f(\bx,\ba)^2\bigr]}$.
  \item[\textup{(ii)}]
    Consequently, if $K\ge0$ and $\Ex_{\bx \sim\gauss}[W(\bx)]\le Kr$, then
    \[
      \Ex_{\bx \sim\gauss}\bigl[\max\{0,W(\bx)\}\bigr]
      \le Kr + \frac{4r}{{\sigma}}\,
          \sqrt{\Ex_{(\bx,\ba)\sim\mathcal N(0,I_{d+1})}\bigl[\defect_f(\bx,\ba)^2\bigr]} .
    \]
\end{enumerate}
\end{proposition}

\subsection{A Harnack inequality for the Gaussian noise operator}

We use the OU {smoothing operator} from \Cref{sec:OU-smoothing} in arbitrary
dimension $n\ge1$. Our application takes $n=d+1$, identifying
$\R^{d+1}$ with $\R^d\times\R$.
We use the following inequality for ${T_\sigma}$; for completeness, we include its proof in \Cref{app:harnack}.

\begin{fact}[Harnack inequality for the Gaussian noise operator {\cite[Lemma~2.1]{Wang}}]
\label{fact:harnack}
Let $n\ge1$ and let $g:\R^n\to[0,\infty)$ be measurable. Then for all $x,z\in\R^n$,
\[
  ({T_\sigma} g)(z)
  \le \sqrt{({T_\sigma} g^2)(x)}\;
      \exp\!\left( \frac{\|z-x\|_2^2}{2{\sigma}^2} \right).
\]
\end{fact}

The inequality formalizes a sense in which overlapping Gaussian neighborhoods share averages.
In our application, $g$ will be the defect of a polynomial, and we will know that ${T_\sigma} g^2$ is small
at a single point $x$ (\Cref{prop:avg} bounds the average of this quantity over $x$).
\Cref{fact:harnack} is what allows us to pass from this one point to every point $z$ near $x$:
it says that ${T_\sigma} g$ cannot be large at $z$ if ${T_\sigma} g^2$ is small at $x$.

\subsection{Zero defect implies function sandwiching (proof of \texorpdfstring{\Cref{prop:comparison}}{Proposition~\ref{prop:comparison}})}

Recall the setting.
The polynomial $f$ satisfies $\defect_f=0$ everywhere,
which by the definition of $\defect_f$ means that $\|\nabla_x f\|_2\le\partial_a f$ at every point:
moving the $a$ variable forward at unit speed increases $f$ at least as fast as moving the $x$ variable in any direction
at unit speed can decrease it.
The points $x,z\in\R^d$ are fixed and $r\ge\|z-x\|_2$.
Our goal is to show $f(x,-r)\le f(z,0)\le f(x,r)$.
We do so by connecting $(x,-r)$ and $(x,r)$ to $(z,0)$ by straight paths along which the $x$ variable moves
at speed at most $1$ and the $a$ variable moves at unit speed towards $0$,
and by showing that $f$ is monotone in the right direction along each path.

The case $r=0$ is immediate, since then $z=x$ and all three quantities equal $f(x,0)$.
Assume henceforth that $r>0$, and set $u = (z-x)/r$, so that $\|u\|_2 = \|z-x\|_2/r \le 1$ and $x+ru = z$.
Consider the two paths
\begin{equation}\label{eq:paths}
  t \mapsto (x+tu,\,-r+t)
  \qquad\text{and}\qquad
  t \mapsto (x+tu,\,r-t),
  \qquad 0\le t\le r.
\end{equation}
The first path runs from $(x,-r)$ to $(z,0)$, and the second from $(x,r)$ to $(z,0)$.

We claim that $f$ is nondecreasing along the first path and nonincreasing along the second.
For $0\le t\le r$, the chain rule gives
\[
  \frac{d}{dt} f(x+tu,-r+t)
  = \langle \nabla_x f, u\rangle + \partial_a f,
  \qquad
  \frac{d}{dt} f(x+tu,r-t)
  = \langle \nabla_x f, u\rangle - \partial_a f,
\]
where in each identity the derivatives on the right are evaluated at the point on the left.
Since $\|u\|_2 \le 1$, Cauchy--Schwarz gives
$|\langle \nabla_x f, u\rangle| \le \|\nabla_x f\|_2\,\|u\|_2 \le \|\nabla_x f\|_2$, and therefore
\begin{equation}\label{eq:monotone}
\begin{aligned}
  \frac{d}{dt} f(x+tu,-r+t)
    &\ge -\|\nabla_x f\|_2 + \partial_a f
     \ge -\defect_f(x+tu,-r+t) = 0, \\
  \frac{d}{dt} f(x+tu,r-t)
    &\le \|\nabla_x f\|_2 - \partial_a f
     \le \defect_f(x+tu,r-t) = 0.
\end{aligned}
\end{equation}
The middle inequality in each line uses $\|\nabla_x f\|_2 - \partial_a f \le \defect_f$,
which holds by the definition of $\defect_f$,
and the last equality uses the assumption $\defect_f=0$
at the point $(x+tu,\mp r\pm t)$ on the corresponding path.

Thus $t\mapsto f(x+tu,-r+t)$ is nondecreasing and $t\mapsto f(x+tu,r-t)$ is nonincreasing on $[0,r]$.
Comparing their values at $t=0$ and $t=r$ gives
\[
  f(x,-r) \le f(z,0)
  \qquad\text{and}\qquad
  f(z,0) \le f(x,r),
\]
which is the claim.
This completes the proof of \Cref{prop:comparison}.

\subsection{Small defect implies function sandwiching (proof of \texorpdfstring{\Cref{prop:robust}}{Proposition~\ref{prop:robust}})}

The setting is as in \Cref{prop:comparison}, except that the condition $\defect_f=0$ is relaxed to
$\defect_f\le\beta$, and only on the segment joining $x$ to $z$ and for $a\in[-r,r]$,
which is the region visited by the paths of \Cref{eq:paths}.
Our goal is $f(x,-r)-\beta r\le f(z,0)\le f(x,r)+\beta r$.
The idea is that along each path, the derivative of $f$ is bounded by the defect rather than by $0$,
so integrating over the parameter interval $[0,r]$, on which the defect is at most $\beta$, costs at most $\beta r$.

The case $r=0$ is immediate, since then $z=x$ and all three quantities equal $f(x,0)$.
Assume henceforth that $r>0$, set $u = (z-x)/r$, and consider the two paths in \Cref{eq:paths}.
We first show the integral bound
\begin{equation}\label{eq:robust-integral}
  f(x,-r) - \int_0^r \defect_f(x+tu,\,-r+t)\,dt
  \;\le\; f(z,0)
  \;\le\; f(x,r) + \int_0^r \defect_f(x+tu,\,r-t)\,dt ,
\end{equation}
which holds without any assumption on $\defect_f$, and then specialize it.
In the proof of \Cref{prop:comparison}, the chain rule, Cauchy--Schwarz, and the definition of $\defect_f$
gave, for $0\le t\le r$,
\begin{equation}\label{eq:deriv-defect}
  \frac{d}{dt} f(x+tu,-r+t) \ge -\defect_f(x+tu,-r+t),
  \qquad
  \frac{d}{dt} f(x+tu,r-t) \le \defect_f(x+tu,r-t);
\end{equation}
these are the first two steps in each line of \Cref{eq:monotone},
and they do not use the assumption of \Cref{prop:comparison}.
Since $f$ is a polynomial,
both $t\mapsto f(x+tu,-r+t)$ and $t\mapsto f(x+tu,r-t)$ are continuously differentiable on $[0,r]$,
and the fundamental theorem of calculus together with \Cref{eq:deriv-defect} gives
\[
\begin{aligned}
  f(z,0) - f(x,-r)
  &= \int_0^r \frac{d}{dt} f(x+tu,-r+t)\,dt
   \ge -\int_0^r \defect_f(x+tu,-r+t)\,dt, \\
  f(z,0) - f(x,r)
  &= \int_0^r \frac{d}{dt} f(x+tu,r-t)\,dt
   \le \int_0^r \defect_f(x+tu,r-t)\,dt .
\end{aligned}
\]
Here we used that the first path runs from $(x,-r)$ to $(z,0)$ and the second from $(x,r)$ to $(z,0)$.
Rearranging the two inequalities yields \Cref{eq:robust-integral}.

Now suppose, as in \Cref{prop:robust}, that $\defect_f(y,a)\le\beta$ for every $y$ on the segment joining $x$ to $z$
and every $a\in[-r,r]$.
For $0\le t\le r$, the point $x+tu$ lies on this segment and both $-r+t$ and $r-t$ lie in $[-r,r]$,
so both integrands in \Cref{eq:robust-integral} are at most $\beta$ on $[0,r]$,
and hence both integrals are at most $\beta r$.
Substituting this into \Cref{eq:robust-integral} gives
\[
  f(x,-r) - \beta r \le f(z,0) \le f(x,r) + \beta r .
\]
This completes the proof of \Cref{prop:robust}.

\subsection{Noise does not increase the defect (proof of \texorpdfstring{\Cref{prop:defect}}{Proposition~\ref{prop:defect}})}

Here ${T_\sigma} f$ is the polynomial obtained by averaging $f$ over Gaussian noise in both variables.
Our goal is to show that the defect of the averaged function is at most the average of the defect,
$\defect_{{T_\sigma} f}(x,a)\le ({T_\sigma}\defect_f)(x,a)$, at every point $(x,a)$.
The proof has two ingredients: the derivatives of ${T_\sigma} f$ are ${\rho_\sigma}$ times the Gaussian averages of the
derivatives of $f$ (\Cref{eq:grad-noise}), and the defect is a jointly convex function of the pair
$(\nabla_x f,\partial_a f)$, so by Jensen's inequality averaging this pair can only decrease the defect.

Fix $(x,a)\in\R^d\times\R$.
Throughout this section, $\bz \sim\gauss$ and $\bb \sim\mathcal N(0,1)$ are independent
and $\E_{\bz,\bb}$ denotes expectation over them,
so that $({T_\sigma} f)(x,a) = \E_{\bz,\bb}[f({\rho_\sigma} x+{\sigma} \bz,{\rho_\sigma} a+{\sigma} \bb)]$.
We compute the partial derivatives of ${T_\sigma} f$:
\begin{equation}\label{eq:grad-noise}
\begin{aligned}
  \nabla_x ({T_\sigma} f)(x,a)
  &= \nabla_x\, \E_{\bz,\bb}\bigl[f({\rho_\sigma} x+{\sigma} \bz,\ {\rho_\sigma} a+{\sigma} \bb)\bigr]
   = \E_{\bz,\bb}\Bigl[\nabla_x\bigl(f({\rho_\sigma} x+{\sigma} \bz,\ {\rho_\sigma} a+{\sigma} \bb)\bigr)\Bigr] \\
  &= {\rho_\sigma}\,\E_{\bz,\bb}\bigl[(\nabla_x f)({\rho_\sigma} x+{\sigma} \bz,{\rho_\sigma} a+{\sigma} \bb)\bigr], \\[4pt]
  \partial_a ({T_\sigma} f)(x,a)
  &= \partial_a\, \E_{\bz,\bb}\bigl[f({\rho_\sigma} x+{\sigma} \bz,\ {\rho_\sigma} a+{\sigma} \bb)\bigr]
   = \E_{\bz,\bb}\Bigl[\partial_a\bigl(f({\rho_\sigma} x+{\sigma} \bz,\ {\rho_\sigma} a+{\sigma} \bb)\bigr)\Bigr] \\
  &= {\rho_\sigma}\,\E_{\bz,\bb}\bigl[(\partial_a f)({\rho_\sigma} x+{\sigma} \bz,{\rho_\sigma} a+{\sigma} \bb)\bigr].
\end{aligned}
\end{equation}
In each case, the first equality interchanges differentiation and expectation,\footnote{This interchange is justified because $f$ is a polynomial:
  $f({\rho_\sigma} x+{\sigma} \bz,{\rho_\sigma} a+{\sigma} \bb)$ and its first partial derivatives in $x$ and in $a$
  are polynomials in $(x,a,\bz,\bb)$, so for $(x,a)$ ranging over any bounded set
  they are dominated by a fixed polynomial in $(\bz,\bb)$, which is integrable with respect to the Gaussian;
  differentiation under the expectation then follows from dominated convergence.}
and the second equality is the chain rule.

Define $\Phi:\R^d\times\R\to[0,\infty)$ by
\[
  \Phi(v,s) := \max\{0,\ \|v\|_2 - s\},
\]
so that, by the definition of the defect, $\defect_h(y,a) = \Phi\bigl(\nabla_x h(y,a),\,\partial_a h(y,a)\bigr)$
for every polynomial $h$ and every $(y,a)$.
The function $\Phi$ is jointly convex in $(v,s)$:
the map $(v,s)\mapsto\|v\|_2-s$ is convex, being the sum of a norm and a linear function,
and $t\mapsto\max\{0,t\}$ is convex and nondecreasing, so their composition is convex.
Moreover, $\Phi$ is positively homogeneous, $\Phi(\lambda v,\lambda s) = \lambda\,\Phi(v,s)$ for all $\lambda\ge0$,
since both $\|v\|_2-s$ and $\max\{0,\cdot\}$ are.

By \Cref{eq:grad-noise}, the pair $\bigl(\nabla_x ({T_\sigma} f)(x,a),\,\partial_a ({T_\sigma} f)(x,a)\bigr)$ equals
${\rho_\sigma}$ times the $\E_{\bz,\bb}$-average of the pair
$\bigl(\nabla_x f({\rho_\sigma} x+{\sigma} \bz,{\rho_\sigma} a+{\sigma} \bb),\,\partial_a f({\rho_\sigma} x+{\sigma} \bz,{\rho_\sigma} a+{\sigma} \bb)\bigr)$.
Using first the homogeneity of $\Phi$, then Jensen's inequality for the jointly convex function $\Phi$,
and finally ${\rho_\sigma}\le1$ together with $\Phi\ge0$, we obtain
\[
\begin{aligned}
  \defect_{{T_\sigma} f}(x,a)
  &= \Phi\Bigl({\rho_\sigma}\,\E_{\bz,\bb}\bigl[\nabla_x f({\rho_\sigma} x+{\sigma} \bz,{\rho_\sigma} a+{\sigma} \bb)\bigr],\
              {\rho_\sigma}\,\E_{\bz,\bb}\bigl[\partial_a f({\rho_\sigma} x+{\sigma} \bz,{\rho_\sigma} a+{\sigma} \bb)\bigr]\Bigr) \\
  &= {\rho_\sigma}\,\Phi\Bigl(\E_{\bz,\bb}\bigl[\nabla_x f({\rho_\sigma} x+{\sigma} \bz,{\rho_\sigma} a+{\sigma} \bb)\bigr],\
                    \E_{\bz,\bb}\bigl[\partial_a f({\rho_\sigma} x+{\sigma} \bz,{\rho_\sigma} a+{\sigma} \bb)\bigr]\Bigr) \\
  &\le {\rho_\sigma}\,\E_{\bz,\bb}\Bigl[\Phi\bigl(\nabla_x f({\rho_\sigma} x+{\sigma} \bz,{\rho_\sigma} a+{\sigma} \bb),\
                                     \partial_a f({\rho_\sigma} x+{\sigma} \bz,{\rho_\sigma} a+{\sigma} \bb)\bigr)\Bigr] \\
  &= {\rho_\sigma}\,\E_{\bz,\bb}\bigl[\defect_f({\rho_\sigma} x+{\sigma} \bz,{\rho_\sigma} a+{\sigma} \bb)\bigr]
   = {\rho_\sigma}\,\bigl({T_\sigma} \defect_f\bigr)(x,a)
   \le \bigl({T_\sigma} \defect_f\bigr)(x,a).
\end{aligned}
\]
This completes the proof of \Cref{prop:defect}.

\subsection{Certifying smallness of the defect on a ball (proof of \texorpdfstring{\Cref{prop:ball}}{Proposition~\ref{prop:ball}})}

Here $x\in\R^d$ and $r\ge0$ are fixed,
and we assume ${T_\sigma}(\defect_f^2)(x,0)\le e^{-2r^2/{\sigma}^2}$,
a bound on the noisy squared defect at the single point $(x,0)$.
Our goal is (i) to show that the defect of ${T_\sigma} f$ is at most $1$ at every point of
$\{z : \|z-x\|_2\le r\}\times[-r,r]$, and (ii) to deduce the comparison inequality of \Cref{prop:robust}
for ${T_\sigma} f$ between $x$ and every $z$ in the ball, namely
\begin{equation}\label{eq:ball-goal}
  ({T_\sigma} f)(x,-r) - r
  \le ({T_\sigma} f)(z,0)
  \le ({T_\sigma} f)(x,r) + r
  \qquad\text{for every } z\in\R^d \text{ with } \|z-x\|_2\le r .
\end{equation}

We first prove (i).
Fix $z\in\R^d$ with $\|z-x\|_2\le r$ and $a\in[-r,r]$.
The function $\defect_f:\R^d\times\R\to[0,\infty)$ is nonnegative and continuous,
so \Cref{fact:harnack} applies to it with $n=d+1$,
the base point $(x,0)$, and the target point $(z,a)$.
Since $\|(z,a)-(x,0)\|_2^2 = \|z-x\|_2^2 + a^2 \le 2r^2$, this gives
\[
  \bigl({T_\sigma} \defect_f\bigr)(z,a)
  \le \sqrt{{T_\sigma}\bigl(\defect_f^2\bigr)(x,0)}\;
      \exp\!\left( \frac{\|z-x\|_2^2 + a^2}{2{\sigma}^2} \right)
  \le e^{-r^2/{\sigma}^2}\, e^{r^2/{\sigma}^2}
  = 1,
\]
where the second inequality uses the assumption of \Cref{prop:ball}.
Applying \Cref{prop:defect} at the point $(z,a)$ then yields
\[
  \defect_{{T_\sigma} f}(z,a) \le \bigl({T_\sigma} \defect_f\bigr)(z,a) \le 1 ,
\]
which is the first part of \Cref{prop:ball}.

We now deduce \Cref{eq:ball-goal} by combining (i) and \Cref{prop:robust}.
Fix $z\in\R^d$ with $\|z-x\|_2\le r$.
Since ${T_\sigma} f$ is a polynomial, \Cref{prop:robust} may be applied to the function ${T_\sigma} f$, the points $x,z$,
the radius $r\ge\|z-x\|_2$, and $\beta := 1$.
Its hypothesis asks that $\defect_{{T_\sigma} f}(y,a)\le 1$ for every $y$ on the line segment joining $x$ to $z$
and every $a\in[-r,r]$.
Every such $y$ satisfies $\|y-x\|_2\le\|z-x\|_2\le r$, so the required bound is exactly (i).
\Cref{prop:robust} therefore gives \Cref{eq:ball-goal}, which is the second part of \Cref{prop:ball}, completing the proof.

\subsection{Certificates exist for most points (proof of \texorpdfstring{\Cref{prop:avg}}{Proposition~\ref{prop:avg}})}

Our goal is to bound the average over $\bx \sim\gauss$ of the noisy squared defect ${T_\sigma}(\defect_f^2)$
at the point $(x,0)$, which is the quantity appearing in the hypothesis of \Cref{prop:ball}.
The bound needs to be in terms of the average of $\defect_f^2$ over $(\bx,\ba)\sim\mathcal N(0,I_{d+1})$.

The difficulty is that the latter averages over $\ba \sim\mathcal N(0,1)$,
whereas the former involves the value $a=0$ only.
We handle this by observing that ${T_\sigma}(\defect_f^2)(x,0)$ averages $\defect_f^2$ over a narrow Gaussian in $a$
centered at $0$, whose density is at most $1/{\sigma}$ times that of the standard Gaussian;
after that, invariance of the Gaussian under ${T_\sigma}$ finishes the proof.

Throughout this section, write\footnote{Since $g = \max\{0,\|\nabla_x f\|_2-\partial_a f\}^2 \le 2\|\nabla_x f\|_2^2 + 2(\partial_a f)^2$
  and $f$ is a polynomial, $g$ is continuous, nonnegative, and bounded by a polynomial,
  so all expectations below are finite.}
$g := \defect_f^2$.
The key step is the inequality
\begin{equation}\label{eq:zero-vs-avg}
  ({T_\sigma} g)(x,0)
  \le \frac{1}{{\sigma}}\,\Ex_{\ba \sim\mathcal N(0,1)}\bigl[({T_\sigma} g)(x,\ba)\bigr]
  \qquad\text{for every } x\in\R^d,
\end{equation}
which replaces the value of ${T_\sigma} g$ at $a=0$ by its average over $\ba \sim\mathcal N(0,1)$
at the price of a factor $1/{\sigma}$;
once it is established, averaging over $\bx$ and the invariance of the Gaussian under ${T_\sigma}$ finish the proof.
\Cref{eq:zero-vs-avg} follows by taking the expectation over $\bz \sim\gauss$, at $\by={\rho_\sigma} x+{\sigma}\bz$,
of the one-dimensional inequality
\begin{equation}\label{eq:one-dim}
  \Ex_{\bb \sim\mathcal N(0,1)}\bigl[g(y,{\sigma} \bb)\bigr]
  \le \frac{1}{{\sigma}}\,\Ex_{\bc \sim\mathcal N(0,1)}\bigl[g(y,\bc)\bigr]
  \qquad\text{for every } y\in\R^d.
\end{equation}
Indeed, by the definition of ${T_\sigma}$, the left-hand side of \Cref{eq:zero-vs-avg} is
$\Ex_{\bz \sim\gauss}$ of the left-hand side of \Cref{eq:one-dim} at $\by={\rho_\sigma} x+{\sigma}\bz$;
and since ${\rho_\sigma} \ba+{\sigma} \bb \sim\mathcal N(0,{\rho_\sigma}^2+{\sigma}^2) = \mathcal N(0,1)$ for independent $\ba,\bb \sim\mathcal N(0,1)$,
the right-hand side of \Cref{eq:zero-vs-avg} is
$\Ex_{\bz \sim\gauss}$ of the right-hand side of \Cref{eq:one-dim} at $\by={\rho_\sigma} x+{\sigma}\bz$.
The one-dimensional inequality \Cref{eq:one-dim} follows by writing out the two Gaussian densities explicitly
and bounding their ratio.\footnote{The density of $\mathcal N(0,{\sigma}^2)$ is at most $1/{\sigma}$ times that of $\mathcal N(0,1)$,
  since their ratio is ${\sigma}^{-1}e^{-t^2({\sigma}^{-2}-1)/2}\le{\sigma}^{-1}$; integrate $g(y,\cdot)\ge0$ against both.}

Taking the expectation of \Cref{eq:zero-vs-avg} over $\bx \sim\gauss$ gives
\begin{equation}\label{eq:avg-both}
  \Ex_{\bx \sim\gauss}\bigl[({T_\sigma} g)(\bx,0)\bigr]
  \le \frac{1}{{\sigma}}\,\Ex_{(\bx,\ba)\sim\mathcal N(0,I_{d+1})}\bigl[({T_\sigma} g)(\bx,\ba)\bigr].
\end{equation}
The standard Gaussian is invariant under ${T_\sigma}$:
if $(\bx,\ba)\sim\mathcal N(0,I_{d+1})$ and $(\bz,\bb)\sim\mathcal N(0,I_{d+1})$ are independent, then
${\rho_\sigma}\,(\bx,\ba)+{\sigma}\,(\bz,\bb)\sim\mathcal N\bigl(0,({\rho_\sigma}^2+{\sigma}^2)I_{d+1}\bigr) = \mathcal N(0,I_{d+1})$,
and therefore
\[
\begin{aligned}
  \Ex_{(\bx,\ba)\sim\mathcal N(0,I_{d+1})}\bigl[({T_\sigma} g)(\bx,\ba)\bigr]
  &= \Ex_{(\bx,\ba),\,(\bz,\bb)\sim\mathcal N(0,I_{d+1})}\bigl[g\bigl({\rho_\sigma}\,(\bx,\ba)+{\sigma}\,(\bz,\bb)\bigr)\bigr] \\
  &= \Ex_{(\bx,\ba)\sim\mathcal N(0,I_{d+1})}\bigl[g(\bx,\ba)\bigr].
\end{aligned}
\]
Substituting this into \Cref{eq:avg-both} yields
\[
  \Ex_{\bx \sim\gauss}\Bigl[{T_\sigma}\bigl(\defect_f^2\bigr)(\bx,0)\Bigr]
  \le \frac{1}{{\sigma}}\,\Ex_{(\bx,\ba)\sim\mathcal N(0,I_{d+1})}\bigl[\defect_f(\bx,\ba)^2\bigr],
\]
which is the claim.
This completes the proof of \Cref{prop:avg}.

\subsection{Small average defect controls the negative part of the signed width (proof of \texorpdfstring{\Cref{prop:width}}{Proposition~\ref{prop:width}})}

Throughout this section, write
\[
  D := \sqrt{\Ex_{(\bx,\ba)\sim\mathcal N(0,I_{d+1})}\bigl[\defect_f(\bx,\ba)^2\bigr]}.
\]
Here $r\in[0,1]$, and $W(x) = ({T_\sigma} f)(x,r)-({T_\sigma} f)(x,-r)$
is the signed width of ${T_\sigma} f$ at $x$ over the interval $[-r,r]$ in the $a$ variable.\footnote{Since ${T_\sigma} f$ is a polynomial, $W$ is a polynomial in $x$, so all expectations below are finite.}
Our goal is to bound the expected negative part of $W$:
\begin{equation}\label{eq:width-goal}
  \Ex_{\bx \sim\gauss}\bigl[\max\{0,-W(\bx)\}\bigr] \le \frac{4r}{{\sigma}}\,D .
\end{equation}
To see why such a bound is plausible, note that if the defect of ${T_\sigma} f$ vanished identically,
then \Cref{prop:comparison} applied to ${T_\sigma} f$ with $z=x$ would give
$({T_\sigma} f)(x,-r)\le({T_\sigma} f)(x,0)\le({T_\sigma} f)(x,r)$, that is, $W(x)\ge0$ for every $x$.
So when the defect is merely small on average, it makes sense that the negative part of $W$ is small on average.

Fix $x\in\R^d$.
For every polynomial $h:\R^d\times\R\to\R$ and every $(y,a)$, the definition of $\defect_h$ gives
$\defect_h(y,a) \ge \|\nabla_x h(y,a)\|_2 - \partial_a h(y,a) \ge -\partial_a h(y,a)$,
that is, $\partial_a h \ge -\defect_h$.
Applying this to $h = {T_\sigma} f$ and integrating over the segment $\{x\}\times[-r,r]$,
the fundamental theorem of calculus gives
\[
  W(x)
  = \int_{-r}^{r} \partial_a ({T_\sigma} f)(x,a)\,da
  \ge -\int_{-r}^{r} \defect_{{T_\sigma} f}(x,a)\,da
  \ge -\int_{-r}^{r} \bigl({T_\sigma} \defect_f\bigr)(x,a)\,da ,
\]
where the last step is \Cref{prop:defect} applied at each point $(x,a)$.
Since the right-hand side is nonpositive, we conclude that
\begin{equation}\label{eq:negpart-pointwise}
  \max\{0,-W(x)\}
  \le \int_{-r}^{r} \bigl({T_\sigma} \defect_f\bigr)(x,a)\,da
  \qquad\text{for every } x\in\R^d.
\end{equation}

Taking the expectation of \Cref{eq:negpart-pointwise} over $\bx \sim\gauss$ and using Tonelli's theorem,
which applies as the integrand is nonnegative, gives
\begin{equation}\label{eq:negpart-avg}
  \Ex_{\bx \sim\gauss}\bigl[\max\{0,-W(\bx)\}\bigr]
  \le \int_{-r}^{r} \Ex_{\bx \sim\gauss}\bigl[\bigl({T_\sigma} \defect_f\bigr)(\bx,a)\bigr]\,da .
\end{equation}
We now bound the expectation inside the integral for a fixed $a\in[-r,r]$.
As in the proof of \Cref{prop:avg}, the bound comes from a one-dimensional inequality in the $a$ variable:
for every $y\in\R^d$,
\begin{equation}\label{eq:one-dim-shifted}
  \Ex_{\bb \sim\mathcal N(0,1)}\bigl[\defect_f(y,\ {\rho_\sigma} a+{\sigma} \bb)\bigr]
  \le \frac{2}{{\sigma}}\,\Ex_{\bc \sim\mathcal N(0,1)}\bigl[\defect_f(y,\bc)\bigr].
\end{equation}
Averaging \Cref{eq:one-dim-shifted} over $\by \sim\gauss$ gives the bound we need.\footnote{Indeed, by the definition of ${T_\sigma}$,
  $\Ex_{\bx \sim\gauss}[({T_\sigma}\defect_f)(\bx,a)] = \Ex_{\bx,\bz \sim\gauss,\ \bb \sim\mathcal N(0,1)}[\defect_f({\rho_\sigma} \bx+{\sigma} \bz,\ {\rho_\sigma} a+{\sigma} \bb)]$
  with $\bx,\bz,\bb$ independent, and ${\rho_\sigma} \bx+{\sigma} \bz \sim\mathcal N(0,({\rho_\sigma}^2+{\sigma}^2)I_d) = \gauss$,
  so this is the expectation over $\by \sim\gauss$ of the left-hand side of \Cref{eq:one-dim-shifted};
  while the expectation over $\by \sim\gauss$ of the right-hand side is
  $\frac{2}{{\sigma}}\Ex_{(\by,\bc)\sim\mathcal N(0,I_{d+1})}[\defect_f(\by,\bc)]$.}
Together with Cauchy--Schwarz, this gives, for every $a\in[-r,r]$,
\begin{equation}\label{eq:shifted-bound}
\begin{aligned}
  \Ex_{\bx \sim\gauss}\bigl[\bigl({T_\sigma} \defect_f\bigr)(\bx,a)\bigr]
  &\le \frac{2}{{\sigma}}\,\Ex_{(\by,\bc)\sim\mathcal N(0,I_{d+1})}\bigl[\defect_f(\by,\bc)\bigr] \\
  &\le \frac{2}{{\sigma}}\,\sqrt{\Ex_{(\by,\bc)\sim\mathcal N(0,I_{d+1})}\bigl[\defect_f(\by,\bc)^2\bigr]}
   = \frac{2}{{\sigma}}\,D .
\end{aligned}
\end{equation}
The one-dimensional inequality \Cref{eq:one-dim-shifted} follows by writing out the densities of
$\mathcal N({\rho_\sigma} a,{\sigma}^2)$ and $\mathcal N(0,1)$ explicitly and comparing them.\footnote{Their ratio is ${\sigma}^{-1}\exp\bigl(\tfrac{t^2}{2}-\tfrac{(t-{\rho_\sigma} a)^2}{2{\sigma}^2}\bigr)$,
  a concave quadratic in $t$ inside the exponential, maximized at $t=a/{\rho_\sigma}$, where it equals
  ${\sigma}^{-1}e^{a^2/2}\le 2/{\sigma}$ as $|a|\le r\le1$; integrate $\defect_f(y,\cdot)\ge0$ against both densities.}

Substituting \Cref{eq:shifted-bound} into \Cref{eq:negpart-avg}
and using that the interval $[-r,r]$ has length $2r$ yields
\[
  \Ex_{\bx \sim\gauss}\bigl[\max\{0,-W(\bx)\}\bigr]
  \le 2r\cdot\frac{2}{{\sigma}}\,D
  = \frac{4r}{{\sigma}}\,D,
\]
which is \Cref{eq:width-goal}, that is, part (i) of \Cref{prop:width}.

Finally, part (ii) follows from part (i): as $\max\{0,W\} = W + \max\{0,-W\}$,
\[
\begin{aligned}
  \Ex_{\bx \sim\gauss}\bigl[\max\{0,W(\bx)\}\bigr]
  &= \Ex_{\bx \sim\gauss}[W(\bx)] + \Ex_{\bx \sim\gauss}\bigl[\max\{0,-W(\bx)\}\bigr] \\
  &\le Kr + \frac{4r}{{\sigma}}\,D ,
\end{aligned}
\]
by the assumption $\Ex_{\bx \sim\gauss}[W(\bx)]\le Kr$ and \Cref{eq:width-goal}.
This completes the proof of \Cref{prop:width}.

\subsection{Proof of \Cref{thm:output-guarantees}}
\label{sec:main}

All the work has been done in \Cref{prop:ball,prop:avg,prop:width};
it remains to put the pieces together and check that the hypothesis of \Cref{thm:output-guarantees}
feeds into each of them with the stated constants.

For (i), \Cref{prop:avg} and the assumption of \Cref{thm:output-guarantees} give
\[
\begin{aligned}
  \Ex_{\bx \sim\gauss}\Bigl[{T_\sigma}\bigl(\defect_f^2\bigr)(\bx,0)\Bigr]
  &\le \frac{1}{{\sigma}}\,\Ex_{(\bx,\ba)\sim\mathcal N(0,I_{d+1})}\bigl[\defect_f(\bx,\ba)^2\bigr] \\
  &\le \frac{1}{{\sigma}}\cdot{\alpha_{\mathrm{Exceptional\ Set}}}\,{\sigma}^2\, e^{-2r^2/{\sigma}^2}\\
  &= {\alpha_{\mathrm{Exceptional\ Set}}}\,{\sigma}\, e^{-2r^2/{\sigma}^2}\\
  &\le {\alpha_{\mathrm{Exceptional\ Set}}}\, e^{-2r^2/{\sigma}^2},
\end{aligned}
\]
using ${\sigma}\le1$ in the last step.

For (ii), fix $x$ with ${T_\sigma}(\defect_f^2)(x,0)\le e^{-2r^2/{\sigma}^2}$.
This is exactly the hypothesis of \Cref{prop:ball} at the point $x$ with the given $r$,
so conclusion (ii) of \Cref{prop:ball} holds for this $x$,
which is precisely the displayed inequality in part (ii) of \Cref{thm:output-guarantees}.

For (iii), the assumption of \Cref{thm:output-guarantees} gives
\[
  \sqrt{\Ex_{(\bx,\ba)\sim\mathcal N(0,I_{d+1})}\bigl[\defect_f(\bx,\ba)^2\bigr]}
  \le \sqrt{{\alpha_{\mathrm{Exceptional\ Set}}}}\,{\sigma}\, e^{-r^2/{\sigma}^2}
  \le {\sigma} ,
\]
using ${\alpha_{\mathrm{Exceptional\ Set}}}\le1$ in the last step.
Since $r\in[0,1]$ and $\Ex_{\bx \sim\gauss}[W(\bx)]\le Kr$, part (ii) of \Cref{prop:width} applies and yields
\[
  \Ex_{\bx \sim\gauss}\bigl[\max\{0,W(\bx)\}\bigr]
  \le Kr + \frac{4r}{{\sigma}}\cdot{\sigma}
  = (K + 4)\, r ,
\]
which is (iii).
This completes the proof of \Cref{thm:output-guarantees}.

\section{Rounding to a Boolean function}
\label{sec:rounding}
In this section, we will show that the output of the previous 
step --- the smoothed polynomial --- can be rounded to a Boolean
function with low error and low boundary volume. 
The goal of this section is to prove the following theorem: 

\begin{theorem}[Learning a Boolean function]
\label{thm:full}
Let $\mathcal D$ be a distribution on $\R^d\times\{0,1\}$
whose first marginal is $\gamma$. Assume $K\ge1$ and $0<r\le1/K$, and write
\[
 \opt_K
 =\inf_{f\in\mathcal C_K}
   \Prx_{(\bx,\by)\sim\mathcal D}[f(\bx)\ne \by].
\]
With probability at least $1-\delta$, \Cref{alg:full} returns
a function $h:\R^d\to\{0,1\}$ satisfying
\[
 \Prx_{(\bx,\by)\sim\mathcal D}[h(\bx)\ne \by]
 \le \opt_K+O(\varepsilon),
 \qquad
 \Bdry_r(h)\le O(Kr).
\]
Its running time, measured in exact real-arithmetic operations, is
\[
 (d+1)^{O(k)}\poly\!\left(\log\frac1\delta\right),
 \qquad
 k=\widetilde O\!\left(
 \frac{K^2}{\varepsilon^2}
 +\frac{K^4r^2}{\varepsilon^4}
 \right),
\]
where the suppressed logarithmic factors may depend on
$K$, $1/\varepsilon$, and $1/(Kr)$.
\end{theorem}

\begin{algorithm}[H]
\caption{\textsc{RobustLearn}$(K,r,\varepsilon,\delta,\mathcal D)$}
\label{alg:full}
\begin{algorithmic}[1]
\Require Boundary profile parameter $K\ge1$; radius $0 < r \le 1/K$;
         accuracy $\varepsilon\in(0,1/4)$;
         failure probability $\delta\in(0,1/4)$;
         sample access to $\mathcal D$.
\Ensure Boolean function $h:\R^d\to\{0,1\}$.
\State Set $\tau=\varepsilon/4$ and $\eta=\min\{\tau,Kr/4\}$.
\State {Set $\sigma=c_\sigma\tau/K$ and}
\[
 {\beta=c_{\beta} Kr\,\sigma^2
             \exp(-C_{\beta} r^2/\sigma^2)},
\]
with constants as required by \Cref{par:parameters}.
\State $P\gets
       \textsc{LearnPolynomial}(K,r,\tau,\delta/3,\mathcal D)$,
using {smoothing standard deviation $\sigma$} and squared-defect budget $\beta$.
\State $(\vec{\theta},\vec{w})\gets
       \textsc{ComputeRoundingThresholds}
       (P,r,\eta,\delta/3,\mathcal D)$,
       using fresh samples.
\State $(A_1,A_2,A_3)\gets
       \textsc{ComputePartition}
       (P,\vec{w},\vec{\theta},r,
        \eta,\delta/3,\mathcal D)$,
       using fresh samples.
\State \Return the function
\[
 h(x)=\sum_{i=1}^3
       \mathbf 1\{x\in A_i\}
       \mathbf 1\{P(x,0)\ge\theta_i\}.
\]
\end{algorithmic}
\end{algorithm}

We may assume $r\le1/K$, since otherwise the allowable boundary
volume is $1$ and the boundary guarantee is automatic.
The restriction $K\ge1$ absorbs the $O(r)$ boundary volume of
the partition seams. For arbitrary $K>0$, replacing $K$ by
$\bar K=\max\{K,1\}$ gives boundary volume $O(\bar K r)$.
At $r=0$ boundary volume is zero for every Boolean function;
one may run the algorithm at working radius $1/K$ to obtain
the clean-error guarantee.

The loss and boundary budgets play different roles. Smoothing uses
$\tau=\varepsilon/4$, whereas threshold discretization, validation,
and empirical partitioning use $\eta\le\min\{\tau,Kr/4\}$.
Thus none of the rounding errors contributes an additive
$O(\varepsilon)$ term to the boundary bound when $Kr\ll\varepsilon$.

The subroutine \textsc{LearnPolynomial} returns the smoothed
polynomial $P$ in $d+1$ dimensions.
Let $\bar p(x)=\max\{0,\min\{1,P(x,0)\}\}$.
The previous section establishes bounds on the expected danger-interval width of $P$ and the $\ell_1$ error of $\bar p$.
As discussed in \Cref{sec:overview-rounding},
these quantities bound the expected boundary volume and error of
a random rounding threshold.

\subsection{Existence of a good small mixture}
Our first step will be to find a mixture of three thresholds
such that both of these quantities are close to their expectations simultaneously. 
This is essentially the threshold-finding step of \cite{LV26},
except that the quantities we want to bound are different.
Here, we want to bound the error and the certified
boundary event (a threshold falling inside the danger
interval).
\begin{algorithm}
\caption{\sc ComputeRoundingThresholds($P, r, \eps, \delta, \mathcal{D}$)}
\label{alg:threshold-compression}
\begin{algorithmic}[1]
\Require Smoothed polynomial $P$; radius $r$;
         accuracy $\varepsilon\in(0,1)$;
         failure probability $\delta\in(0,1)$;
         sample access to $\mathcal D$.
\Ensure Thresholds $\vec{\theta}\in(0,1)^3$
        and weights $\vec{w}\in\Delta_3$.

\State Set $\bar p(x)=\max\{0,\min\{1,P(x,0)\}\}$.
\State Define $g_\theta(x)=\mathbf 1\{\bar p(x)\ge\theta\}$ and
       \[
       I_r(x)=[P(x,-r)-r,\;P(x,r)+r]\cap[0,1],
       \]
       interpreting an interval with reversed endpoints as empty.
\State Set
       $M=\lceil 4/\varepsilon\rceil$,
       $T=\{(j-\tfrac12)/M:j=1,\ldots,M\}$, and
       \[
       N=\left\lceil
          \frac{8}{\varepsilon^2}
          \log\frac{4M}{\delta}
       \right\rceil.
       \]
\State Draw $(\bx_a,\by_a)_{a=1}^N$ independently from $\mathcal D$.
\State Cache $P(\bx_a,0)$, $P(\bx_a,-r)$, and $P(\bx_a,r)$
       for every $a\in[N]$.
\For{$\theta\in T$}
    \State Compute
    \[
    \widehat v(\theta)=
\left(
  \frac1N\sum_{a=1}^N
    \mathbf 1\{g_\theta(\bx_a)\ne \by_a\},
  \frac1N\sum_{a=1}^N
    \mathbf 1\{\theta\in I_r(\bx_a)\}
\right).
    \]
\EndFor
\State Set $\widehat b=M^{-1}\sum_{\theta\in T}\widehat v(\theta)$.
\For{each triple
     $\{\theta_1,\theta_2,\theta_3\}\subseteq T$}
    \State Solve the linear program
    \[
       \sum_{i=1}^{3}w_i\widehat v(\theta_i)=\widehat b,
       \qquad
       \sum_{i=1}^{3}w_i=1, \qquad w_i \ge 0.
    \]
    \If{the system is feasible}
        \State \Return
               $\vec{\theta}=(\theta_1,\theta_2,\theta_3)$,
               $\vec{w}=(w_1,w_2,w_3)$.
    \EndIf
\EndFor
\end{algorithmic}
\end{algorithm}
\begin{lemma}
\label{lem:threshold-compression}
Denote the danger interval 
\[I_r(x)\coloneqq[P(x,-r)-r,\;P(x,r)+r]\cap[0,1].\]
With probability at least $1-\delta$, the output $\vec{\theta}, \vec{w}$ of
\Cref{alg:threshold-compression} satisfies
\[
 \sum_{i=1}^3 w_i
 \Prx_{(\bx,\by)\sim\mathcal D}[g_{\theta_i}(\bx)\ne \by]
 \le
 \mathbb E|\bar p(\bx)-\by|+\varepsilon,
\]
and
\[
 \sum_{i=1}^3 w_i
 \Prx_{\bx \sim\gamma}[\theta_i\in I_r(\bx)]
 \le
 \mathbb E|I_r(\bx)|+\varepsilon.
\]
The algorithm runs in time 
\[
O\!\left(
\varepsilon^{-3}\log\frac1{\varepsilon\delta}\,
T_{\mathrm{eval}}(P)
\right),
\]
where $T_{\mathrm{eval}}(P) \ge 1$ is the time to evaluate $P$ at one point.
\end{lemma}

\begin{proof}
For $\theta\in T$, define the true population vector
\[
 v(\theta)=
 \left(
 \Pr[g_\theta(\bx)\ne \by],\;
 \Pr[\theta\in I_r(\bx)]
 \right),
 \qquad
 b=\frac1M\sum_{\theta\in T}v(\theta).
\]
We first bound the coordinates of $b$ by averaging over thresholds.
For every $u\in[0,1]$ and $y\in\{0,1\}$,
\[
 \int_0^1
 \mathbf 1\{\mathbf 1\{u\ge\theta\}\ne y\}\,d\theta
 =|u-y|.
\]
The algorithm uses a midpoint grid $T\subset(0,1)$ of size
$M=\lceil4/\varepsilon\rceil$, so that the fraction of grid points
in any interval differs from its length by at most $1/M$.
Applying this observation to the error thresholds at each labeled
point, and to the danger interval at each point, gives
\[
 b_1\le \mathbb E|\bar p(\bx)-\by|+\frac1M,
 \qquad
 b_2\le \mathbb E|I_r(\bx)|+\frac1M.
\]

Each coordinate of $\widehat v(\theta)$ is an average of $N$
independent indicators. Hoeffding's inequality and a union bound
over the $2M$ coordinates imply
\[
 \Pr\left[
 \max_{\theta\in T}
 \|\widehat v(\theta)-v(\theta)\|_\infty
 >\frac{\varepsilon}{4}
 \right]
 \le
 4M\exp\left(-\frac{N\varepsilon^2}{8}\right)
 \le\delta.
\]
Condition on the event that all the $\widehat v(\theta)$ estimates are close to the corresponding $v(\theta)$. In particular,
\[
 \|\widehat b-b\|_\infty\le\frac{\varepsilon}{4}.
\]

Since $\widehat b$ lies in the convex hull of the empirical vectors
$\{\widehat v(\theta):\theta\in T\}\subseteq\mathbb R^2$,
Carath\'eodory's theorem expresses it as a convex combination of
at most three such vectors. Thus the search finds
a feasible triple and returns weights satisfying
\[
 \sum_{i=1}^3w_i\widehat v(\theta_i)=\widehat b,
 \qquad
 w_i\ge0,
 \qquad
 \sum_{i=1}^3w_i=1.
\]
Consequently,
\[
 \left\|
 \sum_{i=1}^3w_i v(\theta_i)-b
 \right\|_\infty
 \le
 \sum_{i=1}^3w_i
 \|v(\theta_i)-\widehat v(\theta_i)\|_\infty
 +\|\widehat b-b\|_\infty
 \le\frac{\varepsilon}{2}.
\]
Combining this with the bounds on $b$ and
$1/M\le\varepsilon/4$ proves both claimed inequalities.

Finally, caching the polynomial values takes
$O(NT_{\mathrm{eval}}(P))$ time, and computing all empirical vectors
takes $O(MN)$ additional time. The search examines $O(M^3)$ triples,
each requiring a constant-size linear program.
The total running time is therefore
\[
 O\bigl(NT_{\mathrm{eval}}(P)+MN+M^3\bigr)
 =
 O\left(
 \varepsilon^{-3}\log\frac1{\varepsilon\delta}\,
 T_{\mathrm{eval}}(P)
 \right).
\]
\end{proof}

\subsection{Turning the mixture into a partition}
To output a deterministic function, we now show that we can 
replace these mixing weights by a partition $(A_1,A_2,A_3)$ of 
the domain, such that the ``deterministic mixture'' behaves like
the true mixture.
Each $A_i$ will be a region containing approximately a
$w_i$-share of the error and boundary sets associated
with each threshold $\theta_j$.
The error and boundary sets for $\theta_i$ are
\[
 E_i=\{(x,y):g_i(x)\ne y\},
 \qquad
 C_i=\{(x,y):\theta_i\in I_r(x)\}.
\]
We will cut the domain into a constant number of strips
and assign each strip to one of $A_1$,
$A_2$, or $A_3$, such that \emph{each} of these six
sets (error and boundary sets for each of the three thresholds)
has proportion $w_1$ in $A_1$, $w_2$ in $A_2$, and $w_3$ in $A_3$.
We will then conclude that 
\[\sum_{i=1}^3 \Prx_{(\bx,\by) \sim D}[g_{\theta_i} \ne \by  \land \bx  \in A_i] = \sum_{i=1}^3 w_i \Prx_{(\bx,\by) \sim D}[g_{\theta_i} \ne \by] \pm \eps,\]
and likewise 
\[\sum_{i=1}^3 \Prx_{\bx  \sim \gauss}[\theta_i \in I_r(\bx) \land \bx  \in A_i] = \sum_{i=1}^3 w_i \Prx_{\bx  \sim \gauss}[\theta_i \in I_r(\bx)] \pm \eps.\]
Since the number of these strips
will be constant, and their boundaries are hyperplanes, the increase in boundary volume remains
$O(r)$.

First we will show that such a partition exists: that the real
line can be partitioned into unions of intervals, splitting 
several sets according to an arbitrary weight vector. 
We cite the following result of \cite{StromquistWoodall1985} 
for the case of 2 weights: 

\begin{lemma}[\cite{StromquistWoodall1985}]
\label{lem:2-weights}
For \(q\) nonatomic finite measures \(\nu_1,\ldots,\nu_q\) on the circle $S^1$ and any \(w\in[0,1]\), there is a set \(B\), a union of at most \(q\) intervals, such that
\[
\nu_j(B)=w\,\nu_j(S^1)\qquad(j\in[q]).
\]
\end{lemma}

We generalize this for multiple weights, and take the measures
to be on the real line rather than the circle: 
\begin{corollary}
\label{cor:k-weights}
For $q$ nonatomic finite measures $\nu_1,\ldots,\nu_q$ on $\R$
and any $\vec{w}\in\Delta_k$, there is a partition
$B_1,\ldots,B_k$ of $\R$, determined by at most $2q(k-1)$
cutpoints, such that
\[
 \nu_j(B_i)=w_i\,\nu_j(\R)
 \qquad(j\in[q],\ i\in[k]).
\]
In particular, every part is a finite union of intervals.
\end{corollary}

\begin{proof}
First, map the real line to the circle.
Identify $\R$ with $S^1\setminus\{\infty\}$ in a way that
preserves the order along the circle. Extend each measure to
$S^1$ by setting
\[
 \widetilde\nu_j(A)=\nu_j(A\cap\R)
 \qquad\text{for every Borel set }A\subseteq S^1.
\]
These extensions remain finite and nonatomic, and assign zero
mass to the added point $\infty$.
We construct the
parts one at a time on this circle, then cut it open again.

The proof is by induction.
Initially, let $R_0$ be the whole circle. Suppose that
$B_1,\ldots,B_{i-1}$ have been chosen, leaving the remainder
$R_{i-1}$, and put
\[
 a_i=1-\sum_{h<i}w_h.
\]
Our inductive invariant is
\[
 \nu_j(R_{i-1})=a_i\,\nu_j(S^1)
 \qquad(j\in[q]).
\]
If $w_i=0$, set $B_i=\varnothing$ and leave the remainder
unchanged. Otherwise $a_i\ge w_i>0$. Apply \Cref{lem:2-weights}
to the restricted measures
\[
 \nu_j^{(i)}(A)=\nu_j(A\cap R_{i-1})
\]
with weight $w_i/a_i$. These measures are still finite and
nonatomic. The theorem supplies a union $C_i$ of at most $q$
circular intervals satisfying
\[
 \nu_j(C_i\cap R_{i-1})
 =\frac{w_i}{a_i}\nu_j(R_{i-1})
 =w_i\nu_j(S^1).
\]
Set
\[
 B_i=C_i\cap R_{i-1},
 \qquad
 R_i=R_{i-1}\setminus B_i.
\]
This gives the required mass to $B_i$ and preserves the
inductive invariant for the next step. After constructing the
first $k-1$ parts, take $B_k=R_{k-1}$; its mass is
$w_k\nu_j(S^1)$ under every measure.

It remains to count cutpoints. At step $i$, we assign color $i$
to the points of $C_i$ that remain unassigned. A new boundary
can occur only at an existing boundary or at an endpoint of
$C_i$. Thus each step introduces at most $2q$ new cutpoints.
After $k-1$ steps, there are at most $2q(k-1)$ cutpoints in
total, and the color is constant between consecutive cutpoints.
Endpoints may be assigned to either neighboring part, since
nonatomicity makes every endpoint null under every measure.

Finally, cut the circle open at the added point and remove it.
The resulting parts partition $\R$ and retain their prescribed
masses. No additional finite cutpoint is introduced.
\end{proof}

\subsection{Empirical approximation of the partition}
Now we describe an algorithm for finding an empirical 
approximation of this partition, where the measures to split
are the error and boundary sets corresponding to the three
thresholds. 

\begin{algorithm}
\caption{\sc ComputePartition($P, \vec{w}, \vec{\theta}, r, \eps, \delta, \mathcal{D}$)}
\label{alg:domain-partition}
\begin{algorithmic}[1]
\Require Smoothed polynomial $P$;
         weights $\vec{w}\in\Delta_3$;
         thresholds $\vec{\theta}\in(0,1)^3$;
         radius $r$; accuracy $\varepsilon\in(0,1)$;
         failure probability $\delta\in(0,1)$;
         sample access to $\mathcal D$.
\Ensure A partition $(A_1,A_2,A_3)$ of $\mathbb R^d$.

\State Define
\[
 g_i(x)=\mathbf 1\{P(x,0)\ge\theta_i\},
 \qquad
 I_r(x)=[P(x,-r)-r,\;P(x,r)+r]\cap[0,1],
\]
interpreting an interval with reversed endpoints as empty.
\State Define the six events, for $i\in[3]$,
\[
 E_i=\{(x,y):g_i(x)\ne y\},
 \qquad
 C_i=\{(x,y):\theta_i\in I_r(x)\},
\]
and set $(F_1,\ldots,F_6)=(E_1,E_2,E_3,C_1,C_2,C_3)$.
\State Set ${\alpha_{\mathrm{tolerance}}}=\varepsilon/3$, $S=24$, and
\[
 N=\left\lceil
 C_{\mathrm{VC}} \cdot {\alpha_{\mathrm{tolerance}}}^{-2}
 \left((S+1)\log\frac{2}{{\alpha_{\mathrm{tolerance}}}}
             +\log\frac{36}{\delta}\right)
 +\frac{2S}{{\alpha_{\mathrm{tolerance}}}}
 \right\rceil,
\]
where $C_{\mathrm{VC}}$ is a sufficiently large universal constant.
\State Draw $(\bx_a,\by_a)_{a=1}^N$ independently from $\mathcal D$.
\State Sort the sample so that
       $\bz_1<\cdots<\bz_N$, where $\bz_a=(\bx_a)_1$.
       \Comment{Ties occur with probability zero}
\State Cache $b_{ja}=\mathbf 1\{(\bx_a,\by_a)\in F_j\}$
       for $j\in[6]$, $a\in[N]$.
\State Set $\widehat\mu_j=N^{-1}\sum_{a=1}^N b_{ja}$
       for $j\in[6]$.
\For{each color string $c\in\{1,2,3\}^N$
     with at most $S$ adjacent changes}
    \State Compute, for every $j\in[6]$ and $i\in[3]$,
    \[
       \widehat\mu_{ji}(c)
       =\frac1N\sum_{a=1}^N b_{ja}\mathbf 1\{c_a=i\}.
    \]
    \If{$|\widehat\mu_{ji}(c)-w_i\widehat\mu_j|
          \le \eps/6$ for every $j\in[6]$, $i\in[3]$}
        \State Accept $c$ and exit the loop.
    \EndIf
\EndFor
\State List the change indices as $a_1<\cdots<a_t$, where $t\le S$.
\State Set $d_0=-\infty$, $d_{t+1}=+\infty$, and
       $d_k=(\bz_{a_k}+\bz_{a_k+1})/2$ for $k\in[t]$.
\State Set $\ell_0=c_1$ and $\ell_k=c_{a_k+1}$ for $k\in[t]$.
\State \Return $(A_1,A_2,A_3)$, represented by the seams
       $(d_1,\ldots,d_t)$ and colors $(\ell_0,\ldots,\ell_t)$, where
\[
 A_i=\bigcup_{\substack{0\le k\le t\\ \ell_k=i}}
          \{x\in\mathbb R^d:d_k<x_1\le d_{k+1}\}.
\]
\end{algorithmic}
\end{algorithm}

\begin{lemma}
\label{lem:domain-partition}
With probability at least $1-\delta$, the output 
$(A_1, A_2, A_3)$ of \Cref{alg:domain-partition} satisfies, 
simultaneously for every
$j\in[6]$ and $i\in[3]$,
\[
 \left|
 \Pr[(\bx,\by)\in F_j,\ \bx \in A_i]
 -w_i\Pr[(\bx,\by)\in F_j]
 \right|\le\frac{\varepsilon}{3}.
\]
In particular,
\[
 \sum_{i=1}^3\Pr[(\bx,\by)\in E_i,\ \bx \in A_i]
 \le \sum_{i=1}^3w_i\Pr[(\bx,\by)\in E_i]+\varepsilon,
\]
and the same inequality holds with $C_i$ in place of $E_i$.
The algorithm runs in time 
\[
O\!\left(NT_{\mathrm{eval}}(P)+N^{25}\right),
\qquad
N=O\!\left(\varepsilon^{-2}\log\frac1{\varepsilon\delta}\right).
\]
\end{lemma}

\begin{proof}
Let ${\alpha_{\mathrm{tolerance}}}=\varepsilon/3$ and $S=24$, and define
\[
 \mu_j(A)=\Pr[(\bx,\by)\in F_j,\ \bx \in A],
 \qquad
 \widehat\mu_j(A)=\frac1N\sum_{a=1}^N
 \mathbf 1\{(\bx_a,\by_a)\in F_j,\ \bx_a\in A\}.
\]
Write $\mu_j=\mu_j(\R^d)$ and
$\widehat\mu_j=\widehat\mu_j(\R^d)$.

Let $\mathcal A_S$ be the class of functions of the form $f(x) = \Ind[x_1 \in A]$, where $A$ is a union of at most $S+1$ intervals.
This class has VC dimension at most $2(S+1)$, and intersecting
with any fixed event $F_j$ does not increase its VC dimension.
For the sample size in the algorithm, VC uniform convergence
and a union bound imply that, with probability at least $1-\delta$,
\[
 \sup_{A\in\mathcal A_S}
 |\widehat\mu_j(A)-\mu_j(A)|
 \le\frac{{\alpha_{\mathrm{tolerance}}}}{4}
 \qquad(j\in[6]).
\]
Since $\R^d\in\mathcal A_S$, this also bounds
$|\widehat\mu_j-\mu_j|$ by ${\alpha_{\mathrm{tolerance}}}/4$.
We condition on this event for the remainder of the proof.

To show that the search succeeds, define six measures on $\R$ by
\[
 \nu_j(B)=\Pr[(\bx,\by)\in F_j,\ \bx_1\in B].
\]
These measures are finite and nonatomic, since each is dominated
by the standard Gaussian measure on the first coordinate.
\Cref{cor:k-weights} therefore supplies a partition
$(A_1^\star,A_2^\star,A_3^\star)$ by parallel strips with at most
$S$ seams such that
\[
 \mu_j(A_i^\star)=w_i\mu_j
 \qquad(j\in[6],\ i\in[3]).
\]
Uniform convergence gives
\[
 \left|
 \widehat\mu_j(A_i^\star)-w_i\widehat\mu_j
 \right|
 \le \frac{{\alpha_{\mathrm{tolerance}}}}{4}
       +w_i\frac{{\alpha_{\mathrm{tolerance}}}}{4}
 \le\frac{{\alpha_{\mathrm{tolerance}}}}{2}.
\]
The colors assigned by this partition to the sorted sample form
a string with at most $S$ adjacent changes. Thus the empirical
search considers a string satisfying its acceptance criterion
and must succeed.

The returned partition preserves its accepted sample colors,
so it satisfies
\[
 |\widehat\mu_j(A_i)-w_i\widehat\mu_j|
 \le\frac{{\alpha_{\mathrm{tolerance}}}}{2}.
\]
Applying uniform convergence once more yields
\[
\begin{aligned}
 |\mu_j(A_i)-w_i\mu_j|
 &\le
 |\mu_j(A_i)-\widehat\mu_j(A_i)|
 +|\widehat\mu_j(A_i)-w_i\widehat\mu_j|
 +w_i|\widehat\mu_j-\mu_j|\\
 &\le\frac{{\alpha_{\mathrm{tolerance}}}}{4}+\frac{{\alpha_{\mathrm{tolerance}}}}{2}
       +w_i\frac{{\alpha_{\mathrm{tolerance}}}}{4}
 \le{\alpha_{\mathrm{tolerance}}}.
\end{aligned}
\]
This proves the simultaneous guarantee. Summing the three
inequalities for the error events gives the claimed error bound,
and summing those for the danger events gives the corresponding
danger bound.
\end{proof}

\subsection{Putting the components together}

We now prove \Cref{thm:full}, using
\Cref{lem:empirical-implementation} for the sample implementation
of the learning step and \Cref{thm:output-guarantees}
for the local range certificates and danger-interval width.
\begin{proof}
By \Cref{lem:empirical-implementation},
the learning step, with confidence parameter $\delta/3$, returns
a polynomial $P$ with an auxiliary polynomial $Q$ such that $P={T_\sigma}Q$
satisfying, with probability at least $1-\delta/3$,
\[
 \mathbb E|\bar p(\bx)-\by|\le\opt_K+O(\tau),\qquad
 m_P(r)-m_P(-r)\le C_0Kr,\qquad
 \Ex_{(\bx,\ba)\sim\gamma_{d+1}}
       \defect_Q(\bx,\ba)^2\le\beta,
\]
where $\bar p(x)=\max\{0,\min\{1,P(x,0)\}\}$.
Clipping can only decrease the loss against a Boolean label,
so the first bound follows from the lemma's population loss bound.
Condition on this event.

Set ${\alpha_{\mathrm{Exceptional\ Set}}}=Kr/4\in(0,1/4]$. {The prescribed constants give}
\[
 \beta\le{\alpha_{\mathrm{Exceptional\ Set}}}{\sigma}^2
          \exp\left(-\frac{2r^2}{{\sigma}^2}\right),
\]
so \Cref{thm:output-guarantees} applies at radius $r\le1$.
Define
\[
 B=\left\{x:\bigl({T_\sigma}(\defect_Q^2)\bigr)(x,0)
               >\exp\left(-\frac{2r^2}{{\sigma}^2}\right)
       \right\}.
\]
Part \textup{(i)} and Markov's inequality give
$\gamma(B)\le{\alpha_{\mathrm{Exceptional\ Set}}}$. Part \textup{(ii)} gives, for every
$x\notin B$,
\begin{equation}
\label{eq:rounding-fixed-endpoints}
 P(x,-r)-r\le P(y,0)\le P(x,r)+r
 \qquad\text{for every }\|y-x\|_2\le r.
\end{equation}
Let
\[
 W(x)=P(x,r)-P(x,-r),\qquad
 I_r(x)=[P(x,-r)-r,\;P(x,r)+r]\cap[0,1],
\]
interpreting reversed endpoints as the empty interval.
The expected signed width bound gives $\mathbb EW(\bx)\le C_0Kr$.
Apply part \textup{(iii)} with profile parameter $C_0K$ to obtain
\[
 \mathbb E|I_r(\bx)|
 \le\mathbb E\max\{0,W(\bx)\}+2r
 \le(C_0K+6)r.
\]
Thus the inputs needed for rounding satisfy
\begin{equation}
\label{eq:rounding-input-bounds}
 \mathbb E|\bar p(\bx)-\by|\le\opt_K+O(\tau),\qquad
 \gamma(B)\le Kr/4,\qquad
 \mathbb E|I_r(\bx)|\le O(Kr).
\end{equation}

For each selected threshold, put
\[
 g_i(x)=\mathbf 1\{P(x,0)\ge\theta_i\},
 \qquad
 E_i=\{(x,y):g_i(x)\ne y\},
 \qquad
 C_i=\{x:\theta_i\in I_r(x)\}.
\]
Because $\theta_i\in(0,1)$, thresholding $P(x,0)$ is equivalent
to thresholding $\bar p(x)$.
Apply \Cref{lem:threshold-compression} with accuracy $\eta$
and fresh samples. Except with probability $\delta/3$,
\[
 \sum_{i=1}^3w_i\Pr[E_i]
 \le \mathbb E|\bar p(\bx)-\by|+\eta,
 \qquad
 \sum_{i=1}^3w_i\Pr[C_i]
 \le \mathbb E|I_r(\bx)|+\eta.
\]
Conditional on these outputs, apply \Cref{lem:domain-partition}
with accuracy $\eta$ and another fresh sample. Except with
probability $\delta/3$,
\[
 \sum_{i=1}^3\Pr[E_i,\ \bx \in A_i]
 \le \sum_{i=1}^3w_i\Pr[E_i]+\eta,
 \qquad
 \sum_{i=1}^3\Pr[C_i,\ \bx \in A_i]
 \le \sum_{i=1}^3w_i\Pr[C_i]+\eta.
\]
A union bound shows that all guarantees hold simultaneously
with probability at least $1-\delta$. Since $h=g_i$ on $A_i$,
\[
 \Pr[h(\bx)\ne \by]
 =\sum_{i=1}^3\Pr[E_i,\ \bx \in A_i]
 \le\opt_K+O(\tau)+2\eta
 \le\opt_K+O(\varepsilon).
\]

Let $H$ be the union of the at most $24$ partition seams.
If $x\notin B$, $\operatorname{dist}(x,H)>r$, and
$x\in A_i\setminus C_i$, then the radius-$r$ ball around
$x$ stays in $A_i$. By \eqref{eq:rounding-fixed-endpoints},
all polynomial values on that ball lie in the certified range.
As $\theta_i\notin I_r(x)$, $g_i$ is constant there.
Consequently,
\[
\begin{aligned}
 \Bdry_r(h)
 &\le\gamma(B)
     +\Pr[\operatorname{dist}(\bx,H)\le r]
     +\sum_{i=1}^3\Pr[C_i,\ \bx \in A_i]\\
 &\le O(Kr)+\frac{48r}{\sqrt{2\pi}}
     +\mathbb E|I_r(\bx)|+2\eta\\
 &\le O(Kr),
\end{aligned}
\]
using the bound $1/\sqrt{2\pi}$ on the one-dimensional Gaussian
density, $K\ge1$, and $\eta\le Kr/4$.
This proves the boundary claim.

Finally, the approximation step uses degree
\[
 {k=\left\lceil\frac2{\sigma^2}\log\frac8{\sigma^2\beta}\right\rceil
 =O\!\left(\frac1{\sigma^2}\left[
 1+\frac{r^2}{\sigma^2}
 +\log\frac1{Kr\sigma^4}\right]\right).}
\]
Substituting ${\sigma=c_\sigma\tau/K}$ yields the stated degree bound.
Smoothing only scales the Hermite coefficients, so computing and evaluating $P$ costs at most
$(d+1)^{O(k)}$, up to polynomial factors in $k$.
By \Cref{lem:threshold-compression,lem:domain-partition}, rounding
has overhead polynomial in $1/\eta$ and $\log(1/\delta)$.
The degree choice also gives $\log(1/\eta)=O(k)$: it controls
both $\log(1/\tau)$ through ${\sigma=c_\sigma\tau/K}$ and
$\log(1/(Kr))$ through $\beta\le Kr/4$. Since $d+1\ge2$,
\[
 \poly(1/\eta)\le e^{O(k)}\le(d+1)^{O(k)}.
\]
Together with \Cref{lem:empirical-implementation}, this gives the
displayed runtime. The dependence on $1/(Kr)$ is retained in the
logarithmic factors suppressed in the exponent.

\end{proof}

\bibliographystyle{alpha}
\bibliography{refs}

\appendix
\section{Extension to isotropic strongly log-concave marginals}
\label{apx:log-concave}

{In this section we show that, assuming a recently announced
result in approximation theory, our main theorem (\Cref{thm: main informal})
extends to isotropic strongly log-concave data distributions with the same
running-time exponent up to logarithmic factors.} The learner still
uses only random examples and {additive Gaussian smoothing}.

{To do this, we {first} need to change slightly our polynomial
optimization step. As a result, we need to modify the proof that a good
feasible polynomial exists, to accommodate the adjusted optimization
step, as well as to handle log-concave distributions. In
\Cref{sec:slc-rounding}, we state the required feasibility result and
explain that the rest of our arguments extend to this setting with
minimal adjustments. We prove feasibility in \Cref{sec:slc-witness},
using {some known facts about log-concave distributions that we list} in
\Cref{sec:slc-preliminaries}.}

{Recall that a probability distribution $\mu$ on $\R^d$
is isotropic and log-concave if its mean is zero, its covariance is
$I_d$, and it has density $e^{-{U}}$ for a convex function ${U}$.
Throughout this appendix, we assume that $\mu$ is also
$\alpha$-strongly log-concave, meaning that
${U}(x)-\alpha\|x\|_2^2/2$ is convex for a known constant
$\alpha\in(0,1]$.} Write $\Bdry_{r,\mu}$ and
$\mathcal C_{K,\mu}$ for boundary volume and the boundary-profile
class defined using $\mu$ in place of $\gamma$. Thus
$f\in\mathcal C_{K,\mu}$ means $\Bdry_{u,\mu}(f)\le Ku$ for every
$u>0$. 

{Bizeul and Klartag have announced Gaussian-rate polynomial
approximation results for strongly log-concave probability distributions
\cite{BizeulKlartagAnnouncement2026}. Our proof in this appendix
will depend on their announced result.}
{To state the approximation result, for} any dimension $n$,
define the {additive Gaussian smoothing} operator
\[
 {{H^{\mathrm{smooth}}_{\sigma}}}F(z)=\Ex_{\bg\sim\gamma_n}F(z+{\sigma}\,\bg),\qquad {\sigma>0}.
\]
{
\begin{proposition}[Bizeul and Klartag, based on the announcement~\cite{BizeulKlartagAnnouncement2026}]
\label{prop:slc-polynomial-approximation}
For every $\kappa$-strongly log-concave probability
distribution $\nu$ on $\R^n$, every measurable $F:\R^n\to[-1,1]$,
every ${\sigma>0}$, and every integer $k\ge0$,
\begin{equation}
\label{eq:slc-approximation-assumption}
 \inf_{\deg p\le k}\Ex_{\bz\sim\nu}
 \bigl(p(\bz)-{H_{\sigma}^{\mathrm{smooth}}}F(\bz)\bigr)^2
 \le (1+\kappa {\sigma^2})^{-(k+1)}.
\end{equation}
\end{proposition}
}
{Note that probability distribution $\nu$ in the proposition above
need not be isotropic.}

\begin{theorem}[{Extension of \Cref{thm: main informal} beyond Gaussian marginals}]
\label{thm:slc-extension}
{Assume \Cref{prop:slc-polynomial-approximation}.
Let $\mathcal D$ over $\R^d\times\{0,1\}$ have isotropic $\alpha$-strongly log-concave
first marginal $\mu$.} For $K\ge1$, $0<r\le1/K$, and
$\varepsilon,\delta\in(0,1/4)$, the modification of
\Cref{alg:full} described below returns, with probability at least
$1-\delta$, a Boolean hypothesis $h$ satisfying
\[
 \Prx_{(\bx,\by)\sim\mathcal D}[h(\bx)\ne\by]
 \le\opt_{K,\mu}+\varepsilon,\qquad
 \Bdry_{r,\mu}(h)\le O(Kr),
\]
where $\opt_{K,\mu}=\inf_{f\in\mathcal C_{K,\mu}}
\Prx_{(\bx,\by)\sim\mathcal D}[f(\bx)\ne\by]$.
The sample and real-arithmetic complexity is
\[
 (d+1)^{\widetilde O_\alpha(K^2/\varepsilon^2+K^4r^2/\varepsilon^4)}
 \poly\!\left(\log\frac1\delta\right).
\]
The constants may depend on the fixed curvature bound $\alpha$;
the suppressed logarithmic factors depend on $K$, $1/\varepsilon$,
and $1/(Kr)$, not on $d$.
\end{theorem}

{Note that we give one algorithm that (assuming
\Cref{prop:slc-polynomial-approximation}) works for an arbitrary unknown
isotropic $\alpha$-strongly log-concave data distribution. The algorithm
is only given labeled samples from the distribution, and does not
require other types of additional information about the distribution,
such as a density oracle or a conditional-sample oracle.}

\subsection{The modified polynomial learner}
\label{sec:slc-program}

{Fix a smoothing {standard deviation $\sigma>0$} and
a squared-defect budget $\beta>0$.
The distributions used by the polynomial program can be sampled
using examples from $\mathcal D$ and independent Gaussian variables.}

Our algorithm will learn a polynomial ${\widehat P}={{H^{\mathrm{smooth}}_{\sigma}}}Q$,
{while keeping the defect $\defect_Q=(\|\nabla_xQ\|_2-\partial_aQ)_+$ small.
Note that the notion of defect we use is the same as before.}
{{Specifically, for} degree bound $k\ge1$, set
\[
 M^{\mathrm{monomials}}=\binom{d+1+k}{k},\qquad
 N=d+1+k+{\sigma^{-2}}+\alpha^{-1}+2,\qquad B=N^{Ck},
\]
where $C$ is a sufficiently large universal constant.}
{We write $q_\lambda$ for the monomial coefficients of $Q$,
i.e. with this notation\footnote{{Here
$\lambda=(\lambda_1,\ldots,\lambda_{d+1})$ is a multi-index of nonnegative
integers, $|\lambda|=\sum_{i=1}^{d+1}\lambda_i$, and
$z^\lambda=\prod_{i=1}^{d+1}z_i^{\lambda_i}$ is the corresponding monomial.}}
$Q(z)=\sum_{|\lambda|\le k}q_\lambda z^\lambda$. The population
program is}
\begin{equation}
\label{eq:slc-program}
\begin{aligned}
 \underset{\deg Q\le k}{\operatorname{minimize}}\quad&
 \Ex_{(\bx,\by)\sim\mathcal D}|P(\bx,0)-\by|\\
 \text{subject to}\quad&{\text{(a) }}{\sum_{|\lambda|\le k}q_\lambda^2\le B^2},\\
 &{\text{(b) }}\Ex_{\bx\sim\mu}[P(\bx,r)-P(\bx,-r)]\le C_0Kr,\\
 &{\text{(c) }}\Ex_{{\substack{\bx\sim\mu*\mathcal N(0,{\sigma^2}I_d)\\
                            \ba\sim\mathcal N(0,\alpha^{-1}+{\sigma^2})}}}
   \defect_Q({\bx,\ba})^2\le\beta,\\
 &{P(z)=H^{\mathrm{smooth}}_{\sigma}Q(z)
   \overset{\mathrm{def}}{=}\Ex_{\bg\sim\gamma_{d+1}}Q(z+\sigma\bg).}
\end{aligned}
\end{equation}

{{Compared with Equation~\eqref{eq:polynomial-learning-program},
we {made three changes in the program above}:
\begin{itemize}
\item {Replaced} Ornstein--Uhlenbeck smoothing ${T_{\sigma}}$ by {additive Gaussian smoothing}
${H_{\sigma}^{\mathrm{smooth}}}$, where
\[
\begin{aligned}
 {T_{\sigma}}Q(z)&=\Ex_{\bg\sim\gamma_{d+1}}
 Q({\sqrt{1-\sigma^2}}\,z+{\sigma}\,\bg),\\
 {H_{\sigma}^{\mathrm{smooth}}}Q(z)&=\Ex_{\bg\sim\gamma_{d+1}}Q(z+{\sigma}\,\bg).
\end{aligned}
\]
{In this comparison, $0<\sigma<1$ and $\sigma$ is the standard deviation of the Gaussian noise in both operators.}
\item {Replaced} the bound on the expected square in condition (a) of
Equation~\eqref{eq:polynomial-learning-program} by a ball of monomial coefficients.
\item {Changed} the sampling distribution in condition (c) of
Equation~\eqref{eq:polynomial-learning-program}.
\end{itemize}
The objective and the expected signed-width condition (b) in
Equation~\eqref{eq:polynomial-learning-program} are
still averaged over the original examples.}
The coefficient bound avoids needing an orthogonal basis for an
unknown distribution.}

All constraints are convex. \Cref{lem:slc-empirical-implementation}
implements this program using samples, with slack in the
constraints. It returns ${\widehat P}$ and then uses the threshold and
partition procedures of \Cref{sec:rounding}.
{For a monomial $z^\lambda=\prod_{i=1}^{d+1}z_i^{\lambda_i}$,
{additive Gaussian smoothing} has the explicit formula
\begin{equation}
\label{eq:slc-finite-smoothing}
 {H^{\mathrm{smooth}}_{\sigma}}(z^\lambda)
 =\prod_{i=1}^{d+1}\left(
   \sum_{j=0}^{\lfloor\lambda_i/2\rfloor}
   \frac{\lambda_i!}{(\lambda_i-2j)!\,j!}
   \left(\frac{{\sigma^2}}{2}\right)^j z_i^{\lambda_i-2j}
 \right).
\end{equation}
This formula can be checked by an explicit calculation.
By linearity, $P={H^{\mathrm{smooth}}_{\sigma}}Q$ can be computed exactly
by applying ${H^{\mathrm{smooth}}_{\sigma}}$ to $Q$ monomial-by-monomial.}

{The following algorithm combines this polynomial program
with the rounding steps of \Cref{alg:full}. It essentially implements
\Cref{alg:full} with adjusted parameters and the adjusted optimization program.}
\begin{algorithm}[H]
{
\caption[RobustLearn for strongly log-concave marginals]{{\textsc{RobustLearn} for strongly log-concave marginals}}
\label{alg:slc-full}
\algrenewcommand{\algorithmicensure}{\textbf{Output:}}
\begin{algorithmic}[1]
\Require Curvature bound $\alpha\in(0,1]$; $K\ge1$;
         $0<r\le1/K$; accuracy $\varepsilon\in(0,1/4)$;
         failure probability $\delta\in(0,1/4)$; sample access to
         $\mathcal D$ with isotropic $\alpha$-strongly log-concave
         first marginal.
\Ensure A Boolean function $h:\R^d\to\{0,1\}$.
\State {Set $\tau=c_{\mathrm{acc}}\varepsilon$,
       $\sigma=c\tau/K$, and $\eta=\min\{\tau/4,Kr/4\}$,
       for sufficiently small universal constants
       $c_{\mathrm{acc}},c>0$.}
\State Set
\[
 \beta=\frac{Kr}{16}\frac{\alpha {\sigma^2}}{1+\alpha {\sigma^2}}e^{-2r^2/{\sigma^2}}.
\]
\State Set
\[
 k=\left\lceil\frac{C_1}{\alpha {\sigma^2}}
                 \log\frac{4}{\alpha {\sigma^2}\beta}\right\rceil,
\]
with $C_1$ as in \Cref{prop:slc-polynomial-witness},
and set $N=d+1+k+{\sigma^{-2}}+\alpha^{-1}+2$ and $B=N^{Ck}$,
for a sufficiently large universal constant $C$.
\State Run the sample implementation of \eqref{eq:slc-program}
       from \Cref{lem:slc-empirical-implementation}, with accuracy
       ${\tau}$ and failure probability $\delta/3$, obtaining
       $\widehat P={H_{\sigma}^{\mathrm{smooth}}}Q$ of degree at most $k$.
       Use original examples for loss and signed width. For defect,
       use $(\bx+{\sigma}\,\bg,\ba)$ with
       $\bx\sim\mu$, $\bg\sim{\gamma}$, and
       $\ba\sim\mathcal N(0,\alpha^{-1}+{\sigma^2})$ independent.
\State $(\vec\theta,\vec w)\gets
       \textsc{ComputeRoundingThresholds}
       (\widehat P,r,\eta,\delta/3,\mathcal D)$,
       using fresh samples.
\State $(A_1,A_2,A_3)\gets
       \textsc{ComputePartition}
       (\widehat P,\vec w,\vec\theta,r,\eta,\delta/3,\mathcal D)$,
       using fresh samples.
\State \Return the function
\[
 h(x)=\sum_{i=1}^3
       \mathbf 1\{x\in A_i\}\,
       \mathbf 1\{\widehat P(x,0)\ge\theta_i\}.
\]
\end{algorithmic}
}
\end{algorithm}

\subsection[Analysis of the algorithm]{{Analysis of the algorithm}}
\label{sec:slc-rounding}

{
We begin by stating the proposition saying that there exists a polynomial
satisfying the convex program in Equation~\eqref{eq:slc-program}.

\begin{proposition}[Existence of a polynomial witness]
\label{prop:slc-polynomial-witness}
Assume \Cref{prop:slc-polynomial-approximation}.
Let $f\in\mathcal C_{K,\mu}$ and ${0<\sigma,\beta,r\le1}$.
If $k$ is an integer satisfying
\[
 k\ge\frac{C_1}{\alpha {\sigma^2}}\log\frac{2}{\alpha {\sigma^2}\beta},
\]
for a sufficiently large universal constant $C_1$, then there is
a polynomial ${Q^\star}$ of degree at most $k$ such that,
writing ${P^\star}={H_{\sigma}^{\mathrm{smooth}}}{Q^\star}$,
\[
\begin{aligned}
 &\Ex_{\substack{\bx\sim\mu*\mathcal N(0,{\sigma^2}I_d)\\
                 \ba\sim\mathcal N(0,\alpha^{-1}+{\sigma^2})}}{Q^\star}(\bx,\ba)^2\le1,\\
 &\Ex_{\substack{\bx\sim\mu*\mathcal N(0,{\sigma^2}I_d)\\
                 \ba\sim\mathcal N(0,\alpha^{-1}+{\sigma^2})}}
      \defect_{{Q^\star}}(\bx,\ba)^2\le\beta,\\
 &\Ex_{\bx\sim\mu}|{P^\star}(\bx,0)-f(\bx)|
      \le CK{\sigma}+\beta,\\
 &\Ex_{\bx\sim\mu}[{P^\star}(\bx,r)-{P^\star}(\bx,-r)]
      \le CKr+\beta.
\end{aligned}
\]
Here the samples in the first two expectations are independent,
and $C$ is a universal constant.
The monomial coefficient vector of ${Q^\star}$ also satisfies
$\|{{\vec{q}}^{\,\star}}\|_2\le B$, with $B$ as in \Cref{sec:slc-program}.
\end{proposition}
{We prove \Cref{prop:slc-polynomial-witness} at the end of
\Cref{sec:slc-witness}, using the polynomial-approximation lemma established there.} The analysis below uses
the facts collected in \Cref{sec:slc-preliminaries} only through this proposition;
the sample implementation is supplied by
\Cref{lem:slc-empirical-implementation}.
}

{
\Cref{prop:comparison,prop:robust,prop:defect,prop:ball,prop:avg,prop:width}
extend to the log-concave case as follows.

\begin{proposition}[Output certificates]
\label{prop:slc-output-certificates}
Let $K\ge1$, $0<r\le1/K$, and ${\sigma>0}$.
Suppose a polynomial $Q$, with $P={H_{\sigma}^{\mathrm{smooth}}}Q$,
satisfies {conditions} (b) and (c) {in} Equation~\eqref{eq:slc-program}, where
\[
 0<\beta\le\frac{Kr}{16}\frac{\alpha {\sigma^2}}{1+\alpha {\sigma^2}}e^{-2r^2/{\sigma^2}}.
\]
Then there is a set $B_{\mathrm{exc}}$ with
${\Pr_{\bx\sim\mu}[\bx\in B_{\mathrm{exc}}]}\le Kr/16$ such that, for every
$x\notin B_{\mathrm{exc}}$,
\begin{equation}
\label{eq:slc-sandwich}
 P(x,-r)-r\le P(z,0)\le P(x,r)+r
 \qquad\text{whenever }\|z-x\|_2\le r.
\end{equation}
Moreover, the intervals
\[
 I_r(x)=[P(x,-r)-r,P(x,r)+r]\cap[0,1],
\]
with reversed intervals interpreted as empty, satisfy
{\begin{equation}
\label{eq:slc-interval-length}
 \Ex_{\bx\sim\mu}|I_r(\bx)|\le(C_0+3)Kr.
\end{equation}}
\end{proposition}

\begin{proof}
\Cref{prop:comparison,prop:robust} apply unchanged, since their
statements do not involve a distribution.
\Cref{prop:defect} extends directly with ${T_{\sigma}}$ replaced by
${H_{\sigma}^{\mathrm{smooth}}}$: differentiation now commutes with smoothing
without a contraction factor. The Gaussian Harnack argument in
{\Cref{prop:ball} also applies, with the same noise standard deviation $\sigma$.}
{Consequently, if for some $x$ it holds that
$H_{\sigma}^{\mathrm{smooth}}(\defect_Q^2)(x,0)\le e^{-2r^2/\sigma^2}$,
then \eqref{eq:slc-sandwich} also holds for this $x$.}
We only need to replace the averaging steps in \Cref{prop:avg,prop:width}.

Write ${D_{\sigma}}(a)=\sqrt{(\alpha^{-1}+{\sigma^2})/{\sigma^2}}\,e^{\alpha a^2/2}$.
This bounds the density ratio of $\mathcal N(a,{\sigma^2})$ to
$\mathcal N(0,\alpha^{-1}+{\sigma^2})$.
{Condition (c) in Equation~\eqref{eq:slc-program} therefore gives}
\begin{equation}
\label{eq:slc-average-certificates}
\begin{split}
 \Ex_{\bx\sim\mu}{H^{\mathrm{smooth}}_{\sigma}}(\defect_Q^2)(\bx,0)&\le {D_{\sigma}}(0)\beta,\\
 \Ex_{\bx\sim\mu}\bigl(P(\bx,r)-P(\bx,-r)\bigr)_-
     &\le2r\sqrt{{D_{\sigma}}(r)\beta},
\end{split}
\end{equation}
where $v_-=\max\{0,-v\}$. For the second inequality, integrate
$\partial_aP\ge-{H^{\mathrm{smooth}}_{\sigma}}\defect_Q$ over $[-r,r]$, then use
Cauchy--Schwarz and the same density comparison at each $a$.
Define
\[
 B_{\mathrm{exc}}=
 \{x:{H^{\mathrm{smooth}}_{\sigma}}(\defect_Q^2)(x,0)>e^{-2r^2/{\sigma^2}}\}.
\]
The first inequality and Markov's inequality give
\[
 {\Pr_{\bx\sim\mu}[\bx\in B_{\mathrm{exc}}]}\le {D_{\sigma}}(0)\beta e^{2r^2/{\sigma^2}}\le Kr/16.
\]
Outside this set, \eqref{eq:slc-sandwich} follows as above.

Finally, the bound on $\beta$ gives
${D_{\sigma}}(r)\beta\le(Kr/16)e^{\alpha r^2/2-2r^2/{\sigma^2}}\le e^{1/2}/16$.
Thus \eqref{eq:slc-average-certificates} bounds the expected negative
width by $e^{1/4}r/2<r$.
Writing $w(x)=P(x,r)-P(x,-r)$, we have
$|I_r(x)|\le w(x)+w(x)_-+2r$.
{Condition (b) in Equation~\eqref{eq:slc-program} now gives}
\[
 \Ex_{\bx\sim\mu}|I_r(\bx)|\le C_0Kr+3r\le(C_0+3)Kr.
\]
\end{proof}
}

{
We now combine \Cref{prop:slc-polynomial-witness,prop:slc-output-certificates}
with the sample implementation to obtain the polynomial used for rounding.

\begin{proposition}[Polynomial learner]
\label{prop:slc-polynomial-learner}
Assume \Cref{prop:slc-polynomial-approximation}.
There is an algorithm that, given $\alpha\in(0,1]$, $K\ge1$,
$0<r\le1/K$, $\varepsilon,\delta\in(0,1/4)$, and sample access to
$\mathcal D$ with isotropic $\alpha$-strongly log-concave first marginal
$\mu$, returns a polynomial $\widehat P:\R^{d+1}\to\R$ using
\[
 (d+1)^{\widetilde O_\alpha(K^2/\varepsilon^2+K^4r^2/\varepsilon^4)}
 \poly\!\left(\log\frac1\delta\right)
\]
samples and real-arithmetic operations.
Write
\[
 \bar p(x)=\clip_{[0,1]}\widehat P(x,0),\qquad
 I_r(x)=[\widehat P(x,-r)-r,\widehat P(x,r)+r]\cap[0,1],
\]
with reversed intervals interpreted as empty.
With probability at least $1-\delta$, there is a set $B_{\mathrm{exc}}$
such that
\[
\begin{gathered}
 \Ex_{(\bx,\by)\sim\mathcal D}|\bar p(\bx)-\by|
 \le\opt_{K,\mu}+\frac\varepsilon2,\\
 {\Pr_{\bx\sim\mu}[\bx\in B_{\mathrm{exc}}]}\le Kr/16,\qquad
 \Ex_{\bx\sim\mu}|I_r(\bx)|\le(C_0+3)Kr.
\end{gathered}
\]
For every $x\notin B_{\mathrm{exc}}$ and
$\theta\in(0,1)\setminus I_r(x)$, the classifier
$z\mapsto\Ind[\widehat P(z,0)\ge\theta]$ is constant throughout
the radius-$r$ ball around $x$.
\end{proposition}

\begin{proof}
{{In brief, what} we do in this proof is:
\begin{enumerate}[label=(\alph*)]
\item Pick parameters.
\item Invoke \Cref{prop:slc-polynomial-witness} to conclude that a good
feasible polynomial exists.
\item Use
\Cref{lem:slc-empirical-implementation} to conclude that such a polynomial
$\widehat P$ can be found efficiently.
\item Apply \Cref{prop:slc-output-certificates} to argue that, for
$x\sim\mu$, with high probability
$\widehat P(x,r)-\widehat P(x,-r)$ gives a good bound on how much
$\widehat P(\cdot,0)$ changes in the $r$-neighborhood of $x$
(just as in the case of Gaussian data).
\item Finally, we argue that clipping the output of this polynomial to $[0,1]$
does not increase prediction loss.
\end{enumerate}}

We choose parameters as follows.
{Put $\tau=c_{\mathrm{acc}}\varepsilon$ and choose $\sigma=c\tau/K$,
for sufficiently small universal constants $c_{\mathrm{acc}},c>0$, and set}
\begin{equation}
\label{eq:slc-parameters}
 \beta=\frac{Kr}{16}\frac{\alpha {\sigma^2}}{1+\alpha {\sigma^2}}e^{-2r^2/{\sigma^2}},
 \qquad
 k=\left\lceil\frac{C_1}{\alpha {\sigma^2}}
                 \log\frac{4}{\alpha {\sigma^2}\beta}\right\rceil.
\end{equation}
Here ${\sigma,r\le1}$ and $\beta\le\min\{{\tau},Kr\}/16$.
Apply \Cref{prop:slc-polynomial-witness} with tolerance $\beta/2$.
The witness ${Q^\star}$ has squared defect at most $\beta/2$, and its
smoothed version ${P^\star}$ has expected absolute prediction error
against $f$ at most $CK{\sigma}+\beta/2$.
The triangle inequality therefore bounds its prediction loss by
${\Prx_{(\bx,\by)\sim\mathcal D}[f(\bx)\ne\by]+O({\tau})}$
for every $f\in\mathcal C_{K,\mu}$.
Its expected signed width satisfies
\[
 \Ex_{\bx\sim\mu}[{P^\star}(\bx,r)-{P^\star}(\bx,-r)]
 \le CKr+\frac\beta2\le CKr+\frac{Kr}{32}\le\frac{C_0Kr}{2}.
\]
The last inequality follows when we take $C_0$ large enough.
{The degree choice also gives $k\ge\log(1/{\tau})$, since $\beta\le{\tau}$.}
{Taking the absolute constant $c_{\mathrm{acc}}$ sufficiently small
ensures that the witness loss is at most $3/2$, which is one of the
premises of \Cref{lem:slc-empirical-implementation}.
The bound on the expected square in \Cref{prop:slc-polynomial-witness}
and these inequalities verify its other premises.}
Applying that lemma with accuracy ${\tau}$ and failure probability $\delta$,
and taking the infimum over comparators, gives a learned $\widehat P$
with loss at most $\opt_{K,\mu}+O({\tau})$ and the population constraints
in \eqref{eq:slc-program}.
Taking $c_{\mathrm{acc}}$ small enough makes this loss at most
$\opt_{K,\mu}+\varepsilon/2$.

{Apply \Cref{prop:slc-output-certificates} to $\widehat P$ to obtain
$\Pr_{\bx\sim\mu}[\bx\in B_{\mathrm{exc}}]\le Kr/16$
and the expected interval length bound in Equation~\eqref{eq:slc-interval-length}.}
For $x\notin B_{\mathrm{exc}}$, \eqref{eq:slc-sandwich} ensures that
every threshold in $(0,1)$ outside $I_r(x)$ gives a classifier constant
throughout the radius-$r$ ball around $x$.
Clipping can only decrease prediction loss.

Finally, substituting the choice of $\beta$ into the degree bound gives
\begin{equation}
\label{eq:slc-degree}
 k=O\!\left(\frac1{\alpha {\sigma^2}}
       \left[\frac{r^2}{{\sigma^2}}+\log\frac1{\alpha {\sigma^2}{\tau} Kr}\right]\right)
 =\widetilde O_\alpha\!\left(
       \frac{K^2}{\varepsilon^2}+\frac{K^4r^2}{\varepsilon^4}\right).
\end{equation}
The sample implementation in \Cref{lem:slc-empirical-implementation}
costs $N^{O_\alpha(k)}\poly(\log(1/\delta))$.
For the prescribed parameters,
$N$ {equals} $d+1+\poly_\alpha(K,\varepsilon^{-1},\log(1/(Kr)))$,
so its extra parameter factors can be absorbed into logarithmic
factors in the exponent of $(d+1)$.
Together with \eqref{eq:slc-degree}, this proves the stated
sample and real-arithmetic complexity.
\end{proof}
}

{Now, it remains to show that this function
$\bar p(x)=\clip_{[0,1]}\widehat P(x,0)$ can be rounded to a Boolean function
using the tools we developed for doing this under the Gaussian distribution.}
\begin{proof}[{Proof of \Cref{thm:slc-extension}}]
{Apply \Cref{prop:slc-polynomial-learner} with failure
probability $\delta/3$, and use its output $\widehat P$.
{Put $\tau=c_{\mathrm{acc}}\varepsilon$ and
$\eta=\min\{\tau/4,Kr/4\}$, as in \Cref{alg:slc-full}.}
The proposition gives the loss, coverage, and expected-width
guarantees required for rounding.}

{The proofs of
\Cref{lem:threshold-compression,cor:k-weights,lem:domain-partition}
use only the error and danger-interval bounds and nonatomicity.
Apply their procedures to $\widehat P$ with accuracy $\eta$,
fresh examples from $\mathcal D$, and failure probability $\delta/3$
for each of threshold selection and partitioning.}
{The first-coordinate marginal of $\mu$ has density at most
$1$.\footnote{{This is a standard property of isotropic
log-concave distributions; see, e.g.,
\cite[Lemma~5.5(a)]{LovaszVempala2007}.}}}
Thus the radius-$r$ neighborhoods of the at most $24$ seams have
$\mu$-mass at most $48r$.
{The proof of \Cref{thm:full}, using the threshold guarantee in
\Cref{prop:slc-polynomial-learner}, now gives error
$\opt_{K,\mu}+\varepsilon/2+O(\eta)$ and boundary volume $O(Kr)$.
Taking $c_{\mathrm{acc}}$ sufficiently small makes the error at most
$\opt_{K,\mu}+\varepsilon$.
The degree bound \eqref{eq:slc-degree} gives
$\log(1/\eta)=O_\alpha(k)$, so the polynomial factors in $1/\eta$
from rounding are absorbed by $(d+1)^{O_\alpha(k)}$.
Together with the complexity of \Cref{prop:slc-polynomial-learner},
this proves the complexity in \Cref{thm:slc-extension}.
A union bound over learning, threshold selection, and partitioning
completes the proof.}
\end{proof}

\subsection[Facts about strongly log-concave distributions]{{Facts about strongly log-concave distributions}}
\label{sec:slc-preliminaries}

{
{The rest of the section is dedicated to proving the feasibility of the
optimization program, i.e.\ \Cref{prop:slc-polynomial-witness}.
To do this, we will use some known properties of strongly log-concave
distributions we give in this subsection.}

\paragraph[Basic Facts]{{Basic Facts.}}
For a probability distribution on $\R^n$ with density proportional to $e^{-U}$,
we call $U$ its \emph{potential}. We say that the distribution has
\emph{curvature at least $\kappa$} when
\[
 U(x)-\frac\kappa2\|x\|_2^2\quad\text{is convex}.
\]
Thus curvature at least $\kappa>0$ means exactly
$\kappa$-strong log-concavity. If $U$ is twice differentiable, this
condition is equivalent to $\nabla^2U(x)\succeq\kappa I_n$ for every $x$.
For example, $\mathcal N(m,{{\sigma^2}}I_n)$ has curvature $1/{{\sigma^2}}$ for ${{\sigma}}>0$.
For ${\sigma>0}$, multiplying a density of curvature at least $\kappa$ by
$e^{-\|x-z\|_2^2/(2{\sigma^2})}$ and normalizing gives curvature at least
$\kappa+{\sigma^{-2}}$: the Gaussian factor adds
$\|x-z\|_2^2/(2{\sigma^2})$ to the potential.

For probability distributions $\rho,{\xi}$ on $\R^n$, their convolution
$\rho*{\xi}$ is the law of $\bx+\by$ for independent
$\bx\sim\rho$ and $\by\sim{\xi}$.
Thus $\rho*\mathcal N(0,{{\sigma^2}}I_n)$ is obtained by adding independent
Gaussian noise with covariance ${{\sigma^2}}I_n$ to a sample from $\rho$.
For probability distributions on possibly different Euclidean spaces, $\rho\otimes{\xi}$
denotes their product distribution: the law of $(\bx,\by)$ for independent
$\bx\sim\rho$ and $\by\sim{\xi}$.

\noindent\begin{minipage}{\linewidth}
This fact lets us apply polynomial approximation after adding Gaussian noise.
\par\nopagebreak[4]

\begin{proposition}[{Additive Gaussian smoothing}, see e.g.~{\cite[Lemma~1.7]{KhudiakovaMaasPedrotti2025}}]
\label{prop:slc-convolution}
Let $\kappa>0$. If $\rho$ is a $\kappa$-strongly log-concave probability distribution
on $\R^n$ and ${{\sigma}}>0$, then $\rho*\mathcal N(0,{{\sigma^2}}I_n)$ is
$\kappa/(1+\kappa{{\sigma^2}})$-strongly log-concave.
A product of two strongly log-concave probability distributions
has curvature at least the minimum of their curvature bounds.
Consequently,
\begin{equation}
\label{eq:slc-smoothed-curvature}
 \begin{gathered}
 (\mu*\mathcal N(0,2{\sigma^2}I_d))\otimes\mathcal N(0,\alpha^{-1}+2{\sigma^2})
 \quad\text{is $\kappa_2$-strongly log-concave},\\
 \kappa_2=\frac{\alpha}{1+2\alpha {\sigma^2}}\qquad({\sigma>0}).
 \end{gathered}
\end{equation}
\end{proposition}
\end{minipage}
\par\smallskip

\paragraph[Posterior Smoothing]{{Posterior Smoothing.}}
Recall that in the Gaussian case we used Ornstein--Uhlenbeck smoothing
(\Cref{sec:OU-smoothing}). The convenient
thing about it was that, if $\bx,\bg\sim{\gamma}$ are independent and
\[
{{\bx'=\sqrt{1-\sigma^2}\,\bx+\sigma\,\bg,}}
\]
then both coupled random variables $\bx$ and ${\bx'}$ have the standard
Gaussian distribution. {{Here $\sigma\in(0,1)$ is the standard deviation of the Gaussian noise.}}

If we simply did the same for $\mu$, the distribution would not be
preserved as it was in the Gaussian case. {This property is called
\emph{stationarity} and was used to prove
\Cref{lem:smoothed-hard-lift,lem:smoothing-signed-width}.
Because of this, we define the following noise process for $\mu$ that will
have the same stationarity property.}

\begin{itemize}
\item Sample $\bx\sim\mu$ and add independent Gaussian
noise to obtain $\bz=\bx+{\sigma}\,\bg$, where $\bg\sim{\gamma}$
and {$\sigma>0$ is the standard deviation of the added Gaussian noise.}
\item Draw a fresh sample ${\bx'}$
from the conditional distribution of the original point $\bx$ given
the noisy observation $\bz$.
\end{itemize}
This is the \emph{posterior distribution}:
given $\bz$, the new sample ${\bx'}$ is independent of $\bx$ and has
the same conditional distribution as $\bx$.

{For fixed ${\sigma}$, write} $\pi_z$ for this posterior distribution when the noisy observation
is $z$. Its density is
\[
 \pi_z(x)\propto
 \exp\!\left(-{U}(x)-\frac{\|x-z\|_2^2}{2{\sigma^2}}\right),
\]
where proportionality means normalization to total probability one.
Averaging over the noisy observation, ${\bx'}$ again has distribution
$\mu$, just as $\bx$ does: for every measurable set $A$,
\[
 {\Ex_{\bz\sim\mu*\mathcal N(0,\sigma^2I_d)}
 \Pr_{\bx'\sim\pi_{\bz}}[\bx'\in A]
 =\Pr_{\bx\sim\mu}[\bx\in A].}
\]

\noindent\begin{minipage}{\linewidth}
This noise process is well studied in log-concave sampling, where it
is known as the proximal sampler
\cite{LeeShenTian2021,ChenChewiSalimWibisono2022}.
{The following proposition tells us that one can couple the posterior samples
coming from points $z$ and $z'$ that are nearby.}
\par\nopagebreak[4]

\begin{proposition}[Posterior coupling]
\label{prop:slc-posterior-coupling}
For every ${\sigma>0}$ and $z,z'\in\R^d$, the posterior distributions
$\pi_z$ and $\pi_{z'}$ defined above admit a coupling
$\by_z\sim\pi_z$, $\by_{z'}\sim\pi_{z'}$ satisfying
\begin{equation}
\label{eq:slc-posterior-coupling}
{ \|\by_z-\by_{z'}\|_2\le\frac{\|z-z'\|_2}{1+\alpha {\sigma^2}}
 \quad\text{almost surely}.}
\end{equation}
\end{proposition}
\end{minipage}
\par\smallskip

\noindent\begin{minipage}{\linewidth}
This is a corollary of Khudiakova, Maas, and Pedrotti's more general transport bound.
\par\nopagebreak[4]

\begin{proposition}[Khudiakova, Maas, and Pedrotti~{\cite[Proposition~1.1]{KhudiakovaMaasPedrotti2025}}]
\label{prop:slc-transport}
Let $\kappa>0$ and let $\rho$ be a $\kappa$-strongly log-concave probability density
on $\R^n$, and let $H:\R^n\to\R$ be $L$-Lipschitz.
If $d{\xi}=Z^{-1}e^{-H}\,d\rho$, then there is a coupling
$(\bx,\by)$ of $\rho$ and ${\xi}$ such that
\[
 \|\bx-\by\|_2\le L/\kappa\qquad\text{almost surely}.
\]
\end{proposition}
\end{minipage}
\par\smallskip

\begin{proof}[Proof of \Cref{prop:slc-posterior-coupling}]
By the Gaussian reweighting fact above, both $\pi_z$ and $\pi_{z'}$
have curvature at least $\alpha+{\sigma^{-2}}$. Up to an additive constant,
the potential of $\pi_{z'}$ minus that of $\pi_z$ is
$\langle z-z',x\rangle/{\sigma^2}$, whose Lipschitz constant in $x$ is
$\|z-z'\|_2/{\sigma^2}$. Apply \Cref{prop:slc-transport} with
$\rho=\pi_z$, ${\xi}=\pi_{z'}$, $\kappa=\alpha+{\sigma^{-2}}$,
and $L=\|z-z'\|_2/{\sigma^2}$. The resulting coupling satisfies the claimed
bound because
\[
 \frac L\kappa
 =\frac{\|z-z'\|_2/{\sigma^2}}{\alpha+{\sigma^{-2}}}
 =\frac{\|z-z'\|_2}{1+\alpha {\sigma^2}}.
\]
\end{proof}

We will use the following Poincar\'e inequality to bound prediction error caused by noise and resampling.
\par\nopagebreak[4]

\begin{proposition}[Poincar\'e inequality~{\cite[Section~3.1 of the arXiv version]{ChafaiLehec2020}}]
\label{prop:slc-poincare}
Let $\kappa>0$. If $\rho$ is a $\kappa$-strongly log-concave probability distribution
on $\R^n$, then, for every locally Lipschitz $g:\R^n\to\R$ with
$\Ex_{\bx\sim\rho}g(\bx)^2<\infty$,
\begin{equation}
\label{eq:slc-poincare}
 \operatorname{Var}_{\rho}(g)
 \le\frac1\kappa\Ex_{\bx\sim\rho}\|\nabla g(\bx)\|_2^2.
\end{equation}
{In particular, the posterior distribution $\pi_z$ defined above is
$(\alpha+\sigma^{-2})$-strongly log-concave, so it satisfies}
\begin{equation}
\label{eq:slc-posterior-poincare}
{\operatorname{Var}_{\pi_z}(g)
 \le\frac{\sigma^2}{1+\alpha\sigma^2}
 \Ex_{\bx\sim\pi_z}\|\nabla g(\bx)\|_2^2.}
\end{equation}
\end{proposition}

}

\subsection[There exists a polynomial satisfying the constraints]{{There exists a polynomial satisfying the constraints}}
\label{sec:slc-witness}

\begingroup
\makeatletter
\renewcommand\paragraph{\@startsection{paragraph}{4}{\z@}{1.3ex \@plus.4ex \@minus.08ex}{-1em}{\normalfont\normalsize\bfseries}}
\makeatother

{
We now prove \Cref{prop:slc-polynomial-witness}, which supplies the
feasible polynomial used in \Cref{sec:slc-rounding}.
Recall the offset lift from \Cref{sec:offset-lift}:
\[
 B_f(x,a)=\Ind[\sdist_f(x)\ge-a].
\]
At $a=0$ it agrees with $f$ almost surely; positive values of $a$
enlarge the set where it equals one, and negative values shrink that set.
{We seek a polynomial $Q^\star$ of degree at most $k$ such that,
writing $P^\star=H_{\sigma}^{\mathrm{smooth}}Q^\star$,
\[
\begin{aligned}
 &\Ex_{\bx\sim\mu}|P^\star(\bx,0)-f(\bx)|\le CK\sigma+\beta,\\
 &\Ex_{\bx\sim\mu}[P^\star(\bx,r)-P^\star(\bx,-r)]\le CKr+\beta,\\
 &\Ex_{\substack{\bx\sim\mu*\mathcal N(0,\sigma^2I_d)\\
                 \ba\sim\mathcal N(0,\alpha^{-1}+\sigma^2)}}
       \defect_{Q^\star}(\bx,\ba)^2\le\beta,\\
 &\sum_{|\lambda|\le k}(q_\lambda^\star)^2\le B^2,
\end{aligned}
\]
where $q_\lambda^\star$ are the monomial coefficients of $Q^\star$,
$C$ is a universal constant, and $B$ is as in \eqref{eq:slc-program}.}
As in the Gaussian proof, we first smooth
$B_f$ and establish prediction, signed-width, and zero-defect bounds.
We then use this {smoothed} function to build the polynomial.

\paragraph{Posterior smoothing.}
In the Gaussian proof, Ornstein--Uhlenbeck smoothing couples two points
that both have the Gaussian distribution. {Here we use the noise process
of \Cref{sec:slc-preliminaries} on the pair $(x,a)$.}

{\begin{definition}[Posterior smoothing]
\label{def:slc-posterior-smoothing}
{
{Use the prior distribution $\mu\otimes\mathcal N(0,\alpha^{-1})$.}
For a fixed starting pair $(x,a)$ and {noise standard deviation $\sigma>0$,
the process maps $(x,a)$ to a random pair $(\bx',\ba')$ through the following steps}:
\begin{itemize}
\item Add independent Gaussian noise to obtain
$({\bx_{\mathrm{perturbed}},\ba_{\mathrm{perturbed}}})=(x+{\sigma}\,\bg,a+{\sigma}\,\bg_0)$, where
$\bg\sim\gamma$ and $\bg_0\sim\gamma_1$.
\item Given $({\bx_{\mathrm{perturbed}},\ba_{\mathrm{perturbed}}})$, draw a fresh pair $(\bx',\ba')$ from the
posterior for this prior and observation. In particular,
$\bx'\sim\pi_{{\bx_{\mathrm{perturbed}}}}$, with the posterior parameter set to ${\sigma}$.
\end{itemize}
For a bounded measurable function $F$, define ${T_{\sigma}^\mu} F(x,a)$ as
the expected value of $F(\bx',\ba')$ under this procedure.
We call ${T_{\sigma}^\mu}$ \emph{posterior smoothing with respect to $\mu$};
the fixed prior on the last coordinate is implicit in the notation.
{The parameter $\sigma$ is the standard deviation of the added Gaussian noise.}
}
\end{definition}}

The complete process preserves the prior distribution, as explained in
\Cref{sec:slc-preliminaries}. In particular, starting with $\bx\sim\mu$
makes the returned point $\bx'$ have distribution $\mu$, even when
the starting value of $a$ is fixed.

Set ${V_{\sigma}}={T_{\sigma}^\mu} B_f$. The following proposition plays the role of
\Cref{lem:smoothed-hard-lift,lem:smoothing-signed-width}.
Prediction error controls the objective of \eqref{eq:slc-program}, and
signed width controls condition (b). Zero defect gives a starting point
from which we will obtain the small defect required by condition (c).

\begin{proposition}[Bounds for the smoothed witness]
\label{prop:slc-witness-bounds}
Let $f\in\mathcal C_{K,\mu}$ and ${\sigma,r>0}$. Then ${V_{\sigma}}={T_{\sigma}^\mu} B_f$
is smooth and satisfies
\begin{equation}
\label{eq:slc-witness-bounds}
\begin{aligned}
 &0\le {V_{\sigma}}\le1,\qquad \defect_{{V_{\sigma}}}=0,\\
 &\Ex_{\bx\sim\mu}|{V_{\sigma}}(\bx,0)-f(\bx)|\le CK{\sigma},\\
 &\Ex_{\bx\sim\mu}[{V_{\sigma}}(\bx,r)-{V_{\sigma}}(\bx,-r)]\le CKr,
\end{aligned}
\end{equation}
for a universal constant $C$.
\end{proposition}

We prove the proposition in three steps: preserve the inequality used
in the proof of \Cref{lem:smoothed-hard-lift}, bound how often resampling
the point changes its label, and compute the random value that replaces
$a$. The last calculation will be used for both prediction error and
signed width.

\begin{claim}[Increasing the last coordinate compensates for moving the point]
\label{claim:slc-noise-offset}
For ${\sigma>0}$, $x,y\in\R^d$, and $a,v\in\R$,
\begin{equation}
\label{eq:slc-noise-offset}
 {V_{\sigma}}(x,a)\le {V_{\sigma}}(y,a+v)
 \qquad\text{whenever }v\ge\|x-y\|_2.
\end{equation}
{For noisy observations $(z,a)$ and $(z',a+v)$ with
$v\ge\|z-z'\|_2$, we also have
\begin{equation}
\label{eq:slc-posterior-expectation-comparison}
 \Ex_{\substack{\bx\sim\pi_z\\
   \ba\sim\mathcal N\!\left(\frac{a}{1+\alpha\sigma^2},\frac{\sigma^2}{1+\alpha\sigma^2}\right)}}
 B_f(\bx,\ba)
 \le
 \Ex_{\substack{\bx\sim\pi_{z'}\\
   \ba\sim\mathcal N\!\left(\frac{a+v}{1+\alpha\sigma^2},\frac{\sigma^2}{1+\alpha\sigma^2}\right)}}
 B_f(\bx,\ba).
\end{equation}}
\end{claim}

\begin{proof}
Recall from \Cref{sec:offset-lift} that
\begin{equation}
\label{eq:slc-lift-comparison}
 B_f(x,a)\le B_f(y,a+v)
 \qquad\text{whenever }v\ge\|x-y\|_2.
\end{equation}
This follows from the signed distance being Lipschitz and does not
depend on the underlying distribution.

First consider posterior sampling from two noisy observations
$(z,a)$ and $(z',a+v)$, where $v\ge\|z-z'\|_2$.
By \Cref{prop:slc-posterior-coupling}, we can couple
$\bx_z\sim\pi_z$ and $\bx_{z'}\sim\pi_{z'}$ so that
\begin{equation}
\label{eq:slc-coupled-spatial-distance}
 \|\bx_z-\bx_{z'}\|_2\le\frac{\|z-z'\|_2}{1+\alpha {\sigma^2}}
 \quad\text{almost surely}.
\end{equation}
For the last coordinate, Gaussian conditioning gives the posterior
$\mathcal N(a/(1+\alpha {\sigma^2}),{\sigma^2}/(1+\alpha {\sigma^2}))$ at observation $a$.
{To couple the two posterior samples, draw $\bg_0\sim\gamma_1$ and set
\[
\begin{aligned}
 \ba&=\frac{a}{1+\alpha\sigma^2}
      +\frac{\sigma}{\sqrt{1+\alpha\sigma^2}}\,\bg_0,\\
 \ba'&=\frac{a+v}{1+\alpha\sigma^2}
      +\frac{\sigma}{\sqrt{1+\alpha\sigma^2}}\,\bg_0.
\end{aligned}
\]
The same Gaussian draw appears in both samples, so
\begin{equation}
\label{eq:slc-coupled-offset-difference}
 \ba'-\ba=\frac{v}{1+\alpha\sigma^2}
 \qquad\text{almost surely}.
\end{equation}
Thus, combining Equations~\eqref{eq:slc-lift-comparison},
\eqref{eq:slc-coupled-spatial-distance}, and
\eqref{eq:slc-coupled-offset-difference} gives us}
\[
 B_f(\bx_z,\ba)
 \le B_f\!\left(\bx_{z'},\ba+\frac{v}{1+\alpha {\sigma^2}}\right)
 \quad\text{almost surely}.
\]
{Taking expectations proves Equation~\eqref{eq:slc-posterior-expectation-comparison}.}\footnote{{This also explains our choice of the prior variance
$\alpha^{-1}$: increasing the last coordinate by $v$ still suffices
after the two points are resampled.}}

For the complete process {of \Cref{def:slc-posterior-smoothing}}
{(i.e. Equation~\eqref{eq:slc-noise-offset})}, use the same added Gaussian noise for the
starting pairs $(x,a)$ and $(y,a+v)$. The two noisy spatial observations
remain $\|x-y\|_2$ apart, and their last coordinates differ by $v$.
Apply the preceding posterior comparison at each observation and
average over the added noise to obtain \eqref{eq:slc-noise-offset}.
\end{proof}

{The function $V_{\sigma}$ is smooth because it is the Gaussian smoothing of a bounded function.} Differentiating
\eqref{eq:slc-noise-offset}, exactly as in the proof of
\Cref{lem:smoothed-hard-lift}, gives
$\|\nabla_x{V_{\sigma}}\|_2\le\partial_a{V_{\sigma}}$, and hence zero defect.

We next show that the noise process rarely changes the prediction.
The Gaussian proof used a Gaussian noise-stability bound here;
the following argument replaces that input with the Poincar\'e inequality.

\begin{claim}[Resampling rarely changes the label]
\label{claim:slc-spatial-error}
Draw $\bx\sim\mu$ and $\bg\sim\gamma$ independently, put
$\bz=\bx+{\sigma}\,\bg$, and draw $\bx'\sim\pi_{\bz}$ independently
of $\bx$ given $\bz$.
{This is the spatial part of the process in \Cref{def:slc-posterior-smoothing}.}
Then
\[
 \Ex_{\bz\sim\mu*\mathcal N(0,{\sigma^2}I_d)}
 \Pr_{\bx,\bx'\sim\pi_{\bz}}[f(\bx)\ne f(\bx')]
 \le CK{\sigma}.
\]
\end{claim}

\begin{proof}
For $\ell>0$, replace $f$ by the Lipschitz ramp
\[
 g_\ell(x)=\clip_{[0,1]}\!\left(\frac12+
                         \frac{\sdist_f(x)}{2\ell}\right).
\]
{Since $f$ is Boolean, we have
\[
\begin{aligned}
 &\Pr_{\substack{\bz\sim\mu*\mathcal N(0,\sigma^2I_d)\\
                  \bx,\bx'\sim\pi_{\bz}}}
       [f(\bx)\ne f(\bx')]
 =\Ex_{\substack{\bz\sim\mu*\mathcal N(0,\sigma^2I_d)\\
                  \bx,\bx'\sim\pi_{\bz}}}
       \bigl(f(\bx)-f(\bx')\bigr)^2\\
 &=\Ex_{\substack{\bz\sim\mu*\mathcal N(0,\sigma^2I_d)\\
                  \bx,\bx'\sim\pi_{\bz}}}
       \bigl(f(\bx)-g_\ell(\bx)+g_\ell(\bx)-g_\ell(\bx')
             +g_\ell(\bx')-f(\bx')\bigr)^2\\
 &\le3\Ex_{\substack{\bz\sim\mu*\mathcal N(0,\sigma^2I_d)\\
                     \bx,\bx'\sim\pi_{\bz}}}
       \bigl(f(\bx)-g_\ell(\bx)\bigr)^2
   +3\Ex_{\substack{\bz\sim\mu*\mathcal N(0,\sigma^2I_d)\\
                        \bx,\bx'\sim\pi_{\bz}}}
       \bigl(g_\ell(\bx)-g_\ell(\bx')\bigr)^2\\
 &\quad+3\Ex_{\substack{\bz\sim\mu*\mathcal N(0,\sigma^2I_d)\\
                        \bx,\bx'\sim\pi_{\bz}}}
       \bigl(g_\ell(\bx')-f(\bx')\bigr)^2.
\end{aligned}
\]
We will show bounds on each of these three error terms.}

The ramp agrees with $f$ outside the distance-$\ell$ boundary strip,
and its gradient has length at most $1/(2\ell)$ and vanishes outside
that strip. {Therefore, the bounded boundary ratio of $f$ and
the $1/(2\ell)$-Lipschitzness of $g_\ell$ give us, respectively, that}
\[
 \Ex_{\bx\sim\mu}(f(\bx)-g_\ell(\bx))^2\le K\ell,
 \qquad
 \Ex_{\bx\sim\mu}\|\nabla g_\ell(\bx)\|_2^2\le\frac{K}{4\ell}.
\]
{Both $\bx$ and $\bx'$ have distribution $\mu$, so the first and
third expectations in the decomposition are each at most $K\ell$.}

{For the middle error term, recall that, given $\bz$, the two
points are independent samples from $\pi_{\bz}$.}
Apply \Cref{prop:slc-poincare} and average over the observation:
\[
\begin{aligned}
 &{\Ex_{\substack{\bz\sim\mu*\mathcal N(0,\sigma^2I_d)\\
                            \bx,\bx'\sim\pi_{\bz}}}}
       (g_\ell(\bx)-g_\ell(\bx'))^2\\
 &=2\Ex_{\bz\sim\mu*\mathcal N(0,{\sigma^2}I_d)}
       \operatorname{Var}_{\bx\sim\pi_{\bz}}(g_\ell(\bx))\\
 &\le\frac{2{\sigma^2}}{1+\alpha {\sigma^2}}
       \Ex_{\bx\sim\mu}\|\nabla g_\ell(\bx)\|_2^2
 \le\frac{K{\sigma^2}}{2\ell}.
\end{aligned}
\]
Here averaging the posterior recovers $\mu$.
{The three estimates bound the probability
in the claim by $CK(\ell+\sigma^2/\ell)$.}
Choose $\ell={\sigma}$.
\end{proof}

So far, we have bounded the change from $f(\bx)$ to $f(\bx')$.
The process evaluates $B_f(\bx',\ba')$, so we must also account for
the random value $\ba'$ that replaces the starting value $a$.

\begin{claim}[The process replaces $a$ by $a/(1+\alpha {\sigma^2})$ plus independent Gaussian noise]
\label{claim:slc-last-coordinate}
For fixed $a$, draw $\bx\sim\mu$ and apply the posterior-smoothing
process to $(\bx,a)$. {(See \Cref{def:slc-posterior-smoothing}.)}
The returned pair $(\bx',\ba')$ has independent
coordinates with
\begin{equation}
\label{eq:slc-last-coordinate}
\begin{gathered}
 \bx'\sim\mu,\qquad
 \ba'\sim\mathcal N\!\left(\frac{a}{1+\alpha {\sigma^2}},{v_{\sigma}}\right),
 \\
 {v_{\sigma}}=\frac{{\sigma^2}(2+\alpha {\sigma^2})}{(1+\alpha {\sigma^2})^2}\le2{\sigma^2}.
\end{gathered}
\end{equation}
\end{claim}

\begin{proof}
The returned point $\bx'$ has distribution $\mu$ by stationarity.
The first step adds ${\sigma}\,\bg_1$ to $a$. Given this observation,
the Gaussian posterior has mean $(a+{\sigma}\,\bg_1)/(1+\alpha {\sigma^2})$
and variance ${\sigma^2}/(1+\alpha {\sigma^2})$. Thus
\[
 \ba'=\frac{a}{1+\alpha {\sigma^2}}
       +\frac{{\sigma}}{1+\alpha {\sigma^2}}\bg_1
       +\sqrt{\frac{{\sigma^2}}{1+\alpha {\sigma^2}}}\,\bg_2,
 \qquad \bg_1,\bg_2\sim\gamma_1\text{ independently}.
\]
The two noise variances add to ${v_{\sigma}}$. The prior is a product
distribution, and the Gaussian noises in the spatial and last
coordinates are independent, which proves the independence assertion.
\end{proof}

{Now, we have all the ingredients for proving
\Cref{prop:slc-witness-bounds}, i.e. that the ideal witness
$V_\sigma=T^\mu_\sigma B_f$ has the prediction, signed-width, and defect
bounds needed for the optimization program~\eqref{eq:slc-program}.
Below we finish proving \Cref{prop:slc-witness-bounds} and afterwards
proceed to showing that there is a low-degree polynomial that can stand
in for the ideal witness $V_\sigma$.}

\begin{proof}[Proof of \Cref{prop:slc-witness-bounds}]
Boundedness follows {by} averaging the Boolean function $B_f$;
\Cref{claim:slc-noise-offset} and the differentiation argument following
it give smoothness and zero defect.

\paragraph{Prediction error.}
\Cref{claim:slc-spatial-error} bounds the change in the label caused
by resampling the point. We now use \Cref{claim:slc-last-coordinate}
to finish the comparison with ${V_{\sigma}}(\bx,0)$.
When the starting last coordinate is zero, $\ba'\sim\mathcal N(0,{v_{\sigma}})$
independently of $\bx'\sim\mu$. For a fixed value $a'$,
$B_f(\bx',a')$ can differ from $f(\bx')$ only within distance $|a'|$
of the boundary. This is the same boundary-strip argument used in the
proof of \Cref{lem:smoothed-hard-lift}, and it gives
\[
 \Ex_{\ba'\sim\mathcal N(0,{v_{\sigma}})}\Ex_{\bx'\sim\mu}
       |B_f(\bx',\ba')-f(\bx')|
 \le K\Ex_{\ba'\sim\mathcal N(0,{v_{\sigma}})}|\ba'|
 \le CK{\sigma}.
\]
Combining this with \Cref{claim:slc-spatial-error}, the triangle
inequality and Jensen's inequality give
\[
 \Ex_{\bx\sim\mu}|{V_{\sigma}}(\bx,0)-f(\bx)|\le CK{\sigma}.
\]

\paragraph{Signed width.}
Prediction error is now established. The same description of $\ba'$
lets us compare the outputs starting at $r$ and $-r$.
By definition, ${V_{\sigma}}(x,a)$ is the probability that the returned pair
$(\bx',\ba')$ satisfies $B_f(\bx',\ba')=1$, or equivalently
$\sdist_f(\bx')+\ba'\ge0$. After averaging the starting point over
$\bx\sim\mu$, \Cref{claim:slc-last-coordinate} lets us describe the
returned pair by drawing $\bx'\sim\mu$ and $\bg\sim\gamma_1$
independently and setting
\[
 \ba'=\frac{a}{1+\alpha {\sigma^2}}+\sqrt{{v_{\sigma}}}\,\bg.
\]
{Recalling that $\sdist_f(x)$ is $(2f(x)-1)$ times the distance
from $x$ to the decision boundary (see \Cref{sec:offset-lift}), we see
that, for a fixed returned point $x'$,}
\[
\begin{aligned}
 &{\Pr_{\ba'\sim\mathcal N(a/(1+\alpha\sigma^2),v_\sigma)}
     [B_f(x',\ba')=1]}\\
 &{=}\Pr_{\bg\sim\gamma_1}\!\left[
     \sdist_f(x')+\frac{a}{1+\alpha {\sigma^2}}+\sqrt{{v_{\sigma}}}\,\bg\ge0\right]\\
 &\quad=\Pr_{\bg\sim\gamma_1}\!\left[
     \bg\ge-\frac{\sdist_f(x')+a/(1+\alpha {\sigma^2})}{\sqrt{{v_{\sigma}}}}\right]
 =\Phi\!\left(\frac{\sdist_f(x')+a/(1+\alpha {\sigma^2})}{\sqrt{{v_{\sigma}}}}\right).
\end{aligned}
\]
Here $\Phi$ is the standard Gaussian cumulative distribution function;
the last equality uses Gaussian symmetry, $\Pr_{\bg\sim\gamma_1}[\bg\ge-z]=\Phi(z)$.
Averaging this probability over the returned point gives
\[
 \Ex_{\bx\sim\mu}{V_{\sigma}}(\bx,a)
 =\Ex_{\bx'\sim\mu}
   \Phi\!\left(\frac{\sdist_f(\bx')+a/(1+\alpha {\sigma^2})}{\sqrt{{v_{\sigma}}}}\right).
\]

To compare starting at $r$ and $-r$, use the same returned point
$\bx'\sim\mu$ and the same $\bg\sim\gamma_1$, independently sampled.
The two Boolean outputs are
\[
 B_f\!\left(\bx',\frac{r}{1+\alpha {\sigma^2}}+\sqrt{{v_{\sigma}}}\,\bg\right)
 \quad\text{and}\quad
 B_f\!\left(\bx',-\frac{r}{1+\alpha {\sigma^2}}+\sqrt{{v_{\sigma}}}\,\bg\right).
\]
The first is one and the second is zero exactly when
\[
 -\frac{r}{1+\alpha {\sigma^2}}
 \le\sdist_f(\bx')+\sqrt{{v_{\sigma}}}\,\bg
 <\frac{r}{1+\alpha {\sigma^2}}.
\]
The expected signed width is therefore the probability of this event.
We can now use the proof of \Cref{lem:smoothing-signed-width}, with
$\mu$ in place of $\gamma$ and
$((1+\alpha {\sigma^2})^{-1},\sqrt{{v_{\sigma}}})$ in place of $({\rho_{\sigma}},{\sigma})$.
That argument uses only the boundary-profile bound and ${\rho_{\sigma}}\le1$.
For $r>\sqrt{{v_{\sigma}}}$ it applies the boundary-profile bound directly;
for $r\le\sqrt{{v_{\sigma}}}$ it differentiates the Gaussian average and bounds
the derivative by $CK$. The same argument therefore gives
\[
 \Ex_{\bx\sim\mu}[{V_{\sigma}}(\bx,r)-{V_{\sigma}}(\bx,-r)]\le CKr.
\]
\end{proof}

}

{
\paragraph{Polynomial approximation.}
The ideal witness has the required prediction, signed-width, and zero-defect
bounds, but it is not a polynomial. We now use the ideal witness to build a
polynomial that satisfies the constraints in \eqref{eq:slc-program} and has
small prediction error. The following lemma supplies the approximation,
defect, and coefficient bounds; we then combine it with
\Cref{prop:slc-witness-bounds} to prove \Cref{prop:slc-polynomial-witness}.
The approximation guarantee concerns the output $P^\star={H_{\sigma}^{\mathrm{smooth}}}Q^\star$,
whereas the program constrains the defect and coefficients of $Q^\star$.

\begin{lemma}[Polynomial approximation with small defect]
\label{lem:slc-polynomial-transfer}
Assume \Cref{prop:slc-polynomial-approximation}. Let $f\in\mathcal C_{K,\mu}$
and ${0<\sigma,\beta,r\le1}$. If
\[
 k\ge\frac{C_1}{\alpha {\sigma^2}}\log\frac{2}{\alpha {\sigma^2}\beta}
\]
is an integer, for a sufficiently large universal constant $C_1$, then
there is a polynomial $Q^\star$ of degree at most $k$ such that, writing
$P^\star={H_{\sigma}^{\mathrm{smooth}}}Q^\star$,
\[
\begin{aligned}
 &\Ex_{\bx\sim\mu}
   |P^\star(\bx,a)-{T_{\sqrt3\,\sigma}^{\mu}}B_f(\bx,a)|\le\frac\beta2
   \qquad(a\in\{0,-r,r\}),\\
 &\Ex_{\substack{\bx\sim\mu*\mathcal N(0,{\sigma^2}I_d)\\
                  \ba\sim\mathcal N(0,\alpha^{-1}+{\sigma^2})}}
   \defect_{Q^\star}(\bx,\ba)^2\le\beta,\\
 &\Ex_{\substack{\bx\sim\mu*\mathcal N(0,{\sigma^2}I_d)\\
                  \ba\sim\mathcal N(0,\alpha^{-1}+{\sigma^2})}}
   Q^\star(\bx,\ba)^2\le1.
\end{aligned}
\]
The samples in the last two expectations are independent, and the
monomial coefficient vector satisfies $\|{\vec{q}}^{\,\star}\|_2\le B$, with
$B$ as in \Cref{sec:slc-program}.
\end{lemma}

\begin{proof}
In the Gaussian proof, \Cref{lem:truncation-defect,lem:witness-slices}
use Hermite truncation to obtain a polynomial and control its defect and
prediction error. Here we use \Cref{prop:slc-polynomial-approximation},
and retain {additive Gaussian smoothing} to turn approximation error into derivative
control. This is the point where we separate the two steps of ${T_{\sigma}^\mu}$.

{Write $J_{\sigma}$ for posterior averaging alone, i.e.}
\[
 {J_{\sigma}}B_f(y,a)=
 \Ex_{\substack{\bx\sim\pi_y\\
   \bv\sim\mathcal N(a/(1+\alpha {\sigma^2}),\,{\sigma^2}/(1+\alpha {\sigma^2}))}}
 B_f(\bx,\bv),
\]
where $\pi_y$ is the spatial posterior at noise variance ${\sigma^2}$, and the two
samples are independent. {By \Cref{def:slc-posterior-smoothing},}
\[
 {T_{\sigma}^\mu} B_f={H_{\sigma}^{\mathrm{smooth}}}{J_{\sigma}}B_f.
\]
Since $0\le B_f\le1$, also $0\le {J_{\sigma}}B_f\le1$. The proof of
\Cref{claim:slc-noise-offset} gives
\begin{equation}
\label{eq:slc-posterior-average-comparison}
 {J_{\sigma}}B_f(x,a)\le {J_{\sigma}}B_f(y,a+v)
 \qquad\text{whenever }v\ge\|x-y\|_2.
\end{equation}
Gaussian averaging preserves \eqref{eq:slc-posterior-average-comparison}.

{Before giving our construction of the polynomial witness formally,
we give a short summary. Specifically, our construction follows the following blueprint:}
\[
\begin{gathered}
 R\approx {H_{\sigma}^{\mathrm{smooth}}}{J_{\sqrt3\,\sigma}}B_f,\qquad
 Q^\star={H_{\sigma}^{\mathrm{smooth}}}R,\\
 P^\star={H_{\sigma}^{\mathrm{smooth}}}Q^\star={H_{\sqrt2\,\sigma}^{\mathrm{smooth}}}R.
\end{gathered}
\]
The first {additive Gaussian smoothing} puts the bounded function ${J_{\sqrt3\,\sigma}}B_f$ in
the form required by \Cref{prop:slc-polynomial-approximation}.
The next smoothing controls derivative error, and the last is the
learner's output map. Their variances add to $3{\sigma^2}$, so the target for
$P^\star$ is
\[
 {H_{\sqrt3\,\sigma}^{\mathrm{smooth}}}{J_{\sqrt3\,\sigma}}B_f={T_{\sqrt3\,\sigma}^{\mu}}B_f.
\]
{This explains the parameter $\sqrt3\,\sigma$ in the posterior average: its squared value matches the total Gaussian variance.}
{Now we give this construction formally and with more detail.}

\paragraph{Approximating the bounded smoothed function.}
Choose $R$ of degree at most $k$ to minimize the expected squared error
from ${H_{\sigma}^{\mathrm{smooth}}}{J_{\sqrt3\,\sigma}}B_f$ under independent samples
$\bx\sim\mu*\mathcal N(0,2{\sigma^2}I_d)$ and
$\ba\sim\mathcal N(0,\alpha^{-1}+2{\sigma^2})$.
This is the distribution reached by the two Gaussian smoothings that
remain after choosing $R$, when the starting pair has distribution
$\mu\otimes\mathcal N(0,\alpha^{-1})$.
By \Cref{prop:slc-convolution}, its curvature is at least
$\alpha/(1+2\alpha {\sigma^2})$. Thus \Cref{prop:slc-polynomial-approximation}
gives
\begin{equation}
\label{eq:slc-polynomial-error}
 \Delta^2:=\Ex_{\substack{\bx\sim\mu*\mathcal N(0,2{\sigma^2}I_d)\\
                   \ba\sim\mathcal N(0,\alpha^{-1}+2{\sigma^2})}}
 \bigl(R(\bx,\ba)-{H_{\sigma}^{\mathrm{smooth}}}{J_{\sqrt3\,\sigma}}B_f(\bx,\ba)\bigr)^2
 \le\left(\frac{1+2\alpha {\sigma^2}}{1+3\alpha {\sigma^2}}\right)^{k+1}.
\end{equation}
{Since $R$ is the orthogonal projection of
$H_{\sigma}^{\mathrm{smooth}}J_{\sqrt3\,\sigma}B_f$ onto the space of
degree-$\le k$ polynomials,}
\[
\begin{aligned}
 &\Ex_{\substack{\bx\sim\mu*\mathcal N(0,2{\sigma^2}I_d)\\
                  \ba\sim\mathcal N(0,\alpha^{-1}+2{\sigma^2})}}R(\bx,\ba)^2\\
 &\quad\le
 \Ex_{\substack{\bx\sim\mu*\mathcal N(0,2{\sigma^2}I_d)\\
                  \ba\sim\mathcal N(0,\alpha^{-1}+2{\sigma^2})}}
 \bigl({H_{\sigma}^{\mathrm{smooth}}}{J_{\sqrt3\,\sigma}}B_f(\bx,\ba)\bigr)^2\le1.
\end{aligned}
\]
Jensen's inequality then gives
\[
 \Ex_{\substack{\bx\sim\mu*\mathcal N(0,{\sigma^2}I_d)\\
                  \ba\sim\mathcal N(0,\alpha^{-1}+{\sigma^2})}}
 Q^\star(\bx,\ba)^2\le1.
\]
The coefficient estimate in the proof of
\Cref{lem:slc-empirical-implementation} now gives
$\|{\vec{q}}^{\,\star}\|_2\le B$. It remains to control the defect and the
output approximation error.

\paragraph{Controlling the defect before the final smoothing.}
In the proof of \Cref{lem:truncation-defect}, a Hermite estimate bounds
the derivative error. Here, differentiating the Gaussian expectation
in ${H_{\sigma}^{\mathrm{smooth}}}$ gives\footnote{{Writing the expectation as an
integral and differentiating the Gaussian density yields
$\nabla {H_{\sigma}^{\mathrm{smooth}}}F(z)={\sigma^{-1}}
\Ex_{\bg\sim\gamma_{d+1}}[\bg F(z+{\sigma}\,\bg)]$.
For any unit vector $v$, apply Cauchy--Schwarz to $\langle v,\bg\rangle$
and $F(z+{\sigma}\,\bg)$, using
$\Ex_{\bg\sim\gamma_{d+1}}\langle v,\bg\rangle^2=1$.
Taking the supremum over $v$ gives the displayed inequality.}}
\[
 \|\nabla {H_{\sigma}^{\mathrm{smooth}}}F(z)\|_2^2
 =\frac1{{\sigma^2}}\left\|\Ex_{\bg\sim\gamma_{d+1}}
       [\bg F(z+{\sigma}\,\bg)]\right\|_2^2
 \le\frac1{{\sigma^2}} {H_{\sigma}^{\mathrm{smooth}}}(F^2)(z).
\]
Equation~\eqref{eq:slc-posterior-average-comparison} is preserved by
${H_{\sqrt2\,\sigma}^{\mathrm{smooth}}}$, so ${H_{\sqrt2\,\sigma}^{\mathrm{smooth}}}{J_{\sqrt3\,\sigma}}B_f$
has zero defect. The pointwise comparison in the proof of
\Cref{lem:truncation-defect} therefore gives
\[
\begin{aligned}
 \defect_{Q^\star}(z)^2
 &\le2\bigl\|\nabla\bigl(Q^\star-{H_{\sqrt2\,\sigma}^{\mathrm{smooth}}}{J_{\sqrt3\,\sigma}}B_f\bigr)(z)\bigr\|_2^2\\
 &\le\frac2{{\sigma^2}} {H_{\sigma}^{\mathrm{smooth}}}
       \bigl((R-{H_{\sigma}^{\mathrm{smooth}}}{J_{\sqrt3\,\sigma}}B_f)^2\bigr)(z).
\end{aligned}
\]
Averaging over the distribution in condition (c) of
\eqref{eq:slc-program} adds another variance-${\sigma^2}$ Gaussian and reaches
the distribution used in \eqref{eq:slc-polynomial-error}. Hence
\begin{equation}
\label{eq:slc-witness-defect}
 \Ex_{\substack{\bx\sim\mu*\mathcal N(0,{\sigma^2}I_d)\\
                  \ba\sim\mathcal N(0,\alpha^{-1}+{\sigma^2})}}
 \defect_{Q^\star}(\bx,\ba)^2\le\frac{2\Delta^2}{{\sigma^2}}.
\end{equation}

\paragraph{Controlling the output on the three required slices.}
The remaining task is to compare $P^\star$ with the ideal witness at
$a=0,-r,r$. As in the proof of \Cref{lem:witness-slices}, smoothing
allows us to obtain these slice bounds from an expected squared error.
Here the error is
\[
 P^\star-{T_{\sqrt3\,\sigma}^{\mu}}B_f
 ={H_{\sqrt2\,\sigma}^{\mathrm{smooth}}}(R-{H_{\sigma}^{\mathrm{smooth}}}{J_{\sqrt3\,\sigma}}B_f).
\]
Starting with $\bx\sim\mu$ at a fixed offset $a$, the right-hand side
averages over independent samples from $\mu*\mathcal N(0,2{\sigma^2}I_d)$ and
$\mathcal N(a,2{\sigma^2})$. The spatial distribution already matches
\eqref{eq:slc-polynomial-error}; only the last coordinate requires a
density comparison. The Gaussian density ratio satisfies
\[
 \frac{d\mathcal N(a,2{\sigma^2})}{d\mathcal N(0,\alpha^{-1}+2{\sigma^2})}(v)
 \le\sqrt{\frac{\alpha^{-1}+2{\sigma^2}}{2{\sigma^2}}}\,e^{\alpha a^2/2}
 \qquad(v\in\R).
\]
Jensen's and Cauchy--Schwarz inequalities, as in the proof of
\Cref{lem:witness-slices}, yield
\begin{equation}
\label{eq:slc-slices}
 \Ex_{\bx\sim\mu}|P^\star(\bx,a)-{T_{\sqrt3\,\sigma}^{\mu}}B_f(\bx,a)|
 \le\left(\frac{\alpha^{-1}+2{\sigma^2}}{2{\sigma^2}}\right)^{1/4}
       e^{\alpha a^2/4}\Delta\qquad(|a|\le r).
\end{equation}

\paragraph{Choosing the degree.}
We now make both error bounds small enough. Since
${0<\alpha,\sigma,\beta,r\le1}$,
\[
 \frac{1+2\alpha {\sigma^2}}{1+3\alpha {\sigma^2}}\le e^{-\alpha {\sigma^2}/4},
 \qquad
 \left(\frac{\alpha^{-1}+2{\sigma^2}}{2{\sigma^2}}\right)^{1/4}
       e^{\alpha r^2/4}
 \le\left(\frac{3e}{2\alpha {\sigma^2}}\right)^{1/4}.
\]
For $C_1\ge16$, the degree assumption and
\eqref{eq:slc-polynomial-error} imply
$\Delta\le(\alpha {\sigma^2}\beta/2)^2$. Consequently,
\[
 \frac{2\Delta^2}{{\sigma^2}}\le\beta,
 \qquad
 \left(\frac{3e}{2\alpha {\sigma^2}}\right)^{1/4}\Delta\le\frac\beta2.
\]
Together with \eqref{eq:slc-witness-defect} and \eqref{eq:slc-slices}, and the size
bounds already proved, these are the conclusions of the lemma.
\end{proof}

\begin{proof}[Proof of \Cref{prop:slc-polynomial-witness}]
We can now combine the ideal witness with its polynomial approximation.
Apply \Cref{lem:slc-polynomial-transfer}, and use
\Cref{prop:slc-witness-bounds} {at standard deviation $\sqrt3\,\sigma$.}
Since ${V_{\sqrt3\,\sigma}}={T_{\sqrt3\,\sigma}^{\mu}}B_f$, the slice bound at $a=0$ gives
\[
\begin{aligned}
 \Ex_{\bx\sim\mu}|P^\star(\bx,0)-f(\bx)|
 &\le\Ex_{\bx\sim\mu}|{V_{\sqrt3\,\sigma}}(\bx,0)-f(\bx)|
     +\frac\beta2\\
 &\le CK{\sigma}+\frac\beta2,
\end{aligned}
\]
after enlarging the universal constant $C$. At $a=-r$ and $a=r$,
the two approximation errors add to at most $\beta$, so
\[
 \Ex_{\bx\sim\mu}[P^\star(\bx,r)-P^\star(\bx,-r)]
 \le\Ex_{\bx\sim\mu}[{V_{\sqrt3\,\sigma}}(\bx,r)-{V_{\sqrt3\,\sigma}}(\bx,-r)]+\beta
 \le CKr+\beta.
\]
The expected-square, squared-defect, degree, and coefficient bounds
are supplied by \Cref{lem:slc-polynomial-transfer}. These prove
\Cref{prop:slc-polynomial-witness}, which is the feasibility statement
used by the algorithm analysis in \Cref{sec:slc-rounding}.
\end{proof}

}
\endgroup

\section{Empirical approximation}
\label{apx:empirical}

The polynomial programs use exact population statistics. We implement
both the Gaussian learner and the distribution extension using the
same uniform-estimation and convex-optimization argument. Their
only different requirements here are bounds on the coefficients
and on the contribution of large samples. We prove the shared
argument first, then verify these bounds for each distribution.
Throughout, samples have exact real coordinates and complexity
counts samples and exact real-arithmetic operations.

\subsection{Uniform estimation and convex optimization}
\label{sec:empirical-shared}

Let $Q_{\vec q}$ and $P_{\vec q}$ be degree-$\le k$ polynomials
that depend linearly on $N$ coefficients $\vec q$, with
$\|\vec q\|_2\le1$. The map from $Q$ to $P$ will be Gaussian
smoothing in both applications. For a labeled distribution
$\mathcal D$ with first marginal $\mu$, and a defect-sampling
law $\nu$ on $\R^{d+1}$, write
\[
\begin{split}
 \mathcal L(\vec q)&=
   \Ex_{(\bx,\by)\sim\mathcal D}|P_{\vec q}(\bx,0)-\by|,\\
 \mathcal V(\vec q)&=
   \Ex_{\bz\sim\nu}\defect_{Q_{\vec q}}(\bz)^2,\\
 \mathcal W(\vec q)&=
   \Ex_{\bx\sim\mu}[P_{\vec q}(\bx,r)-P_{\vec q}(\bx,-r)].
\end{split}
\]
The first two functions are convex in $\vec q$ and the third
is linear. Our task is to minimize loss while keeping squared
defect and expected signed width below their prescribed bounds.

We estimate these expectations after discarding points outside a
fixed ball. Discarded samples contribute zero, and averages are
still divided by the original sample count. Because the cutoff
does not depend on $\vec q$, the empirical loss and squared
defect remain convex and the empirical signed width remains
linear. In the Gaussian application, we can instead compute
$\mathcal W$ exactly.

\begin{lemma}[From uniform estimates to a population guarantee]
\label{lem:empirical-shared}
Let $K\ge1$, $0<r\le1/K$, $C_0\ge1$, and
$\tau,\delta\in(0,1/4)$. For a sufficiently small universal
constant $c>0$, put $\zeta=c\min\{\tau,\beta,Kr\}$, where
$\beta>0$. Suppose fixed norm cutoffs have the following properties
uniformly over $\|\vec q\|_2\le1$: each discarded expectation
has absolute value at most $\zeta/2$, and the retained integrands
have absolute value and coefficient Lipschitz constant at most
$H\ge1$. If $\mathcal W$ is computed exactly, only its Lipschitz
constant must be at most $H$.

Suppose there is a coefficient vector ${{\vec{q}}^{\,\star}}$ in the unit
ball with
\[
 \mathcal L({{\vec{q}}^{\,\star}})\le3/2,\qquad
 \mathcal V({{\vec{q}}^{\,\star}})\le\beta/2,\qquad
 \mathcal W({{\vec{q}}^{\,\star}})\le C_0Kr/2.
\]
With probability at least $1-\delta$, an empirical optimization
returns $\vec q$ satisfying, simultaneously for every such witness,
\begin{equation}
\label{eq:empirical-reserved-slack}
 \mathcal L(\vec q)\le\mathcal L({{\vec{q}}^{\,\star}})+O(\tau),\qquad
 \mathcal V(\vec q)\le3\beta/4,\qquad
 \mathcal W(\vec q)\le3C_0Kr/4.
\end{equation}
It uses
\[
 O\!\left(\frac{H^2}{\zeta^2}
       \left[N\log\frac{CH}{\zeta}+\log\frac1\delta\right]\right)
\]
samples of each required kind. After forming their feature values
for $P,Q$, and $\nabla Q$, the number of real-arithmetic operations
is polynomial in $N,d,H,\beta^{-1},(Kr)^{-1},\tau^{-1}$, and
$\log(1/\delta)$.
\end{lemma}

We prove the lemma in two steps. Uniform convergence lets us use
the same samples to estimate the objectives and to choose the
polynomial. A penalty objective then converts approximate
optimization into the population constraints.

\begin{definition}[Pseudo-dimension; see {\cite[Section~2]{MorgensternRoughgarden2016}}]
\label{def:pseudo-dimension}
Let $\mathcal F$ be a class of functions from a set $\mathcal X$
to $\R$. The \emph{subgraph} of $f\in\mathcal F$ is the set
\[
 \{(x,t)\in\mathcal X\times\R:t<f(x)\}.
\]
The \emph{pseudo-dimension} of $\mathcal F$ is the VC dimension
of the set class
\[
 \bigl\{\{(x,t)\in\mathcal X\times\R:t<f(x)\}:f\in\mathcal F\bigr\}.
\]
Thus the points being classified are pairs $(x,t)$, and membership
records whether the function value at $x$ exceeds $t$.
\end{definition}

\begin{proof}[Proof of \Cref{lem:empirical-shared}]
At a fixed input, each loss threshold test is a disjunction of two
linear inequalities in $\vec q$:
$P_{\vec q}(x,0)-y>t$ or $P_{\vec q}(x,0)-y<-t$, for $t\ge0$.
For squared defect, put $c=\partial_aQ_{\vec q}(z)+\sqrt t$.
Then
\[
 \defect_{Q_{\vec q}}(z)^2>t
 \quad\Longleftrightarrow\quad
 c<0\quad\text{or}\quad
 \bigl(c\ge0\ \text{and}\ \|\nabla_xQ_{\vec q}(z)\|_2^2>c^2\bigr).
\]
Thus its threshold tests use a constant number of linear and
quadratic inequalities in the $N$ coefficients. Negative
thresholds are automatic for these two nonnegative integrands.
Signed width may be negative, but its threshold test is a single
linear inequality for every real threshold. Polynomial sign-pattern
bounds \cite{GoldbergJerrum1995} therefore give pseudo-dimension
$O(N)$ for all three classes. The cutoff does not increase this
complexity: at a discarded point the function value is always
zero, so the comparison outcome is independent of $\vec q$.

Let $\widehat{\mathcal L},\widehat{\mathcal V}$, and
$\widehat{\mathcal W}$ denote the truncated sample averages;
set $\widehat{\mathcal W}=\mathcal W$ if width is known exactly.
Uniform convergence for bounded function classes
\cite[Theorem~1]{MorgensternRoughgarden2016}, with a union bound
over the three estimates, gives sampling error at most $\zeta/2$
using the stated sample count. For signed width, shift the range
$[-H,H]$ before applying the bounded-function result. Adding the
discarded-tail bias gives
\begin{equation}
\label{eq:empirical-uniform-estimates}
 \sup_{\|\vec q\|_2\le1}
 |\widehat{\mathcal J}(\vec q)-\mathcal J(\vec q)|\le\zeta
 \qquad(\mathcal J=\mathcal L,\mathcal V,\mathcal W).
\end{equation}
We work on this simultaneous event for the rest of the proof.

Minimize the following convex objective over the unit ball:
\begin{equation}
\label{eq:empirical-penalized-program}
 \widehat{\mathcal L}(\vec q)
 +\frac{64}{\beta}
       \bigl(\widehat{\mathcal V}(\vec q)-5\beta/8\bigr)_+
 +\frac{64}{C_0Kr}
       \bigl(\widehat{\mathcal W}(\vec q)-5C_0Kr/8\bigr)_+.
\end{equation}
Every witness incurs neither penalty, since its empirical defect
is at most $\beta/2+\zeta<5\beta/8$ and its empirical signed
width is at most $C_0Kr/2+\zeta<5C_0Kr/8$. An additive $\tau$
approximate minimizer therefore has population loss at most
$\mathcal L({{\vec{q}}^{\,\star}})+\tau+2\zeta$. Its objective is also
at most $3/2+\zeta+\tau<2$. Each term in the objective is
nonnegative, so
\[
 \widehat{\mathcal V}(\vec q)\le5\beta/8+\beta/32,
 \qquad
 \widehat{\mathcal W}(\vec q)\le5C_0Kr/8+C_0Kr/32.
\]
Transferring these bounds to the population using
\eqref{eq:empirical-uniform-estimates} proves
\eqref{eq:empirical-reserved-slack} for sufficiently small $c$.

For completeness, projected subgradient descent finds the required
approximate minimizer in polynomially many operations. At each
step, move against a subgradient and project back to the unit
ball. Subgradients of \eqref{eq:empirical-penalized-program}
can be chosen with norm at most
$G=H(1+64/\beta+64/(C_0Kr))$. With step size
$a=1/(G\sqrt T)$, projection gives
\[
 \|\vec q_{j+1}-{{\vec{q}}^{\,\star}}\|_2^2
 \le\|\vec q_j-{{\vec{q}}^{\,\star}}\|_2^2
       -2a\langle\vec g_j,\vec q_j-{{\vec{q}}^{\,\star}}\rangle
       +a^2G^2.
\]
Here $\vec g_j$ is a subgradient at $\vec q_j$ and, for this
calculation, ${{\vec{q}}^{\,\star}}$ can be an exact minimizer of the
empirical objective. Summing cancels the intermediate squared
distances. The subgradient inequality and convexity then bound
the objective gap of the average iterate by $O(G/\sqrt T)$.
Taking $T=O(G^2/\tau^2)$ suffices. The feature values make each
iteration a polynomial number of arithmetic operations.
\end{proof}

\subsection{Gaussian marginals}
\label{sec:empirical-gaussian}

We now verify the hypotheses of \Cref{lem:empirical-shared} for
\textsc{LearnPolynomial}. Here the coefficient ball is supplied
by Hermite orthogonality, and Mehler's formula bounds both the
retained samples and the discarded tails.

\begin{lemma}[Sample implementation of the population learner]
\label{lem:empirical-implementation}
Let $K\ge1$, $0<r\le1/K$, and $\tau,\delta\in(0,1/4)$.
Choose ${\sigma},\beta,k$ as in \Cref{par:parameters}, with $\tau$ in
place of $\varepsilon$.
For a sufficiently large universal choice of $C_0$, there is an
implementation of \textsc{LearnPolynomial} using samples that,
with probability at least $1-\delta$, returns a polynomial $P$
of degree at most $k$. There is a polynomial $Q$ such that
$P={T_\sigma}Q$ and
\begin{equation}
\label{eq:empirical-population-guarantees}
\begin{split}
 {\Ex_{(\bx,\ba)\sim\gamma_{d+1}}Q(\bx,\ba)^2}&\le1,\\
 \Ex_{(\bx,\by)\sim\mathcal D}|P(\bx,0)-\by|
     &\le\opt_K+O(\tau),\\
 m_P(r)-m_P(-r)&\le C_0Kr,\\
 \Ex_{(\bx,\ba)\sim\gamma_{d+1}}
       \defect_Q(\bx,\ba)^2&\le\beta.
\end{split}
\end{equation}
The sample and real-arithmetic complexity is
\[
 (d+1)^{O(k)}\poly\!\left(\log\frac1\delta\right).
\]
\end{lemma}

Clipping $P(x,0)$ to $[0,1]$ can only reduce its loss against a
Boolean label. Thus \eqref{eq:empirical-population-guarantees}
supplies the population bounds used in \Cref{sec:rounding}; the
threshold and partition procedures are analyzed there.

\begin{proof}[Proof of \Cref{lem:empirical-implementation}]
Put $\eta=\min\{\tau,Kr/4\}$ and $N_k=\binom{d+1+k}{k}$.
Using the Hermite basis from \Cref{sec:Hermite}, parameterize
\[
 Q_{\vec q}=\sum_{|\nu|\le k}q_\nu H_\nu,
 \qquad P_{\vec q}={T_\sigma}Q_{\vec q},
 \qquad \|\vec q\|_2\le1.
\]
The expected signed width requires no sampling: spatial
averaging removes every Hermite term with a nonconstant spatial
factor, giving
\begin{equation}
\label{eq:empirical-exact-width}
 \mathcal W(\vec q)
 =\sum_{j=0}^k q_{(0,\ldots,0,j)}{(1-\sigma^2)^{j/2}}
       \bigl(H_j(r)-H_j(-r)\bigr).
\end{equation}

To bound the loss and squared defect, write their coefficient
maps explicitly. For $z=(x,a)$, define
\[
 \phi(x)=\bigl({(1-\sigma^2)^{|\nu|/2}}H_\nu(x,0)\bigr)_\nu,
 \qquad
 D(z)=\bigl(\nabla H_\nu(z)\bigr)_\nu.
\]
Thus $P_{\vec q}(x,0)=\langle\vec q,\phi(x)\rangle$ and
$\nabla Q_{\vec q}(z)=D(z)\vec q$. The matrix $D(z)$ has one
column for each basis polynomial.

\paragraph{Norm cutoff.}
The cutoff bounds follow from a pointwise estimate on the
Hermite basis.
For this calculation write $m=d+1$ and choose
$t=\min\{1/4,k/(4m)\}$. Mehler's formula, in our normalization
and evaluated at equal arguments, gives
\cite[Eq.~18.18.28]{DLMFMehler}
\[
 \sum_{\nu\in\mathbb Z_{\ge0}^m}t^{|\nu|}H_\nu(z)^2
 =(1-t^2)^{-m/2}
       \exp\!\left(\frac{t\|z\|_2^2}{1+t}\right).
\]
The multivariate identity is the product of the univariate
identities. Since $t^{|\nu|}\ge t^k$ for $|\nu|\le k$,
\[
 \sum_{|\nu|\le k}H_\nu(z)^2
 \le t^{-k}(1-t^2)^{-m/2}e^{t\|z\|_2^2}.
\]
The derivative identity from \Cref{sec:Hermite} also gives
\[
 \|D(z)\|_F^2
 =\sum_{|\mu|\le k-1}(|\mu|+m)H_\mu(z)^2
 \le(k+m)\sum_{|\nu|\le k}H_\nu(z)^2.
\]
Moreover, $\|\phi(x)\|_2^2\le\sum_{|\nu|\le k}H_\nu(x,0)^2$.
Consequently, setting
\[
 A=4(k+m)t^{-k}(1-t^2)^{-m/2},
\]
we obtain the coefficient-independent bounds
\begin{equation}
\label{eq:empirical-growth-bound}
 |P_{\vec q}(x,0)-y|\le Ae^{t\|x\|_2^2},\qquad
 \defect_{Q_{\vec q}}(z)^2\le Ae^{t\|z\|_2^2}.
\end{equation}
The loss is at most $1+\|\phi(x)\|_2$, and the squared
defect is at most $2\|D(z)\|_F^2$. The same bounds with a factor
of $2$ also bound their Lipschitz constants in $\vec q$ on the
unit ball. The choice of $t$ makes these bounds useful even in
large dimension: $tm\le k/4$, so the dimension-dependent factor
in $A$ costs only $(d+1)^{O(k)}$.

To bound the omitted expectations, let $\bg\sim\gamma_j$ with
$j=d$ or $d+1$. A direct Gaussian integral gives
$\mathbb E e^{v\|\bg\|_2^2}=(1-2v)^{-j/2}$ for $v<1/2$.
Since $t\le1/4$, for any $u\ge0$,
\[
 \mathbb E\bigl[e^{t\|\bg\|_2^2}\Ind[\|\bg\|_2>u]\bigr]
 \le e^{-u^2/8}\mathbb E e^{(t+1/8)\|\bg\|_2^2}
 \le2^m e^{-u^2/8}.
\]
Choose $\zeta=c\min\{\tau,\beta,Kr\}$ and a radius
$R\ge2$ with
\[
 R^2=\Theta\!\left(m+\log\frac A\zeta\right),
 \qquad A2^m e^{-R^2/8}\le\zeta/2.
\]
The implied constant is a sufficiently large universal constant.
By \eqref{eq:empirical-growth-bound}, discarding only samples
outside radius $R$ changes either expectation by at most
$\zeta/2$. Samples retained inside radius $R$ have loss and
squared defect at most $Ae^{tR^2}$, and coefficient Lipschitz constants at most
$2Ae^{tR^2}$. {The coefficient vector of the exact signed-width
functional \eqref{eq:empirical-exact-width} has Euclidean norm at most
$2e^{1/4}/\sqrt{{\sigma}}$ by the slice estimate
\eqref{eq:slice-estimate}.} Thus we may take
\[
 H=\max\{2Ae^{tR^2},2e^{1/4}/\sqrt{{\sigma}},1\}.
\]

\paragraph{Applying the shared lemma.}
For a sufficiently large $C_0$, the witnesses from
\Cref{lem:truncation-defect,lem:witness-slices} satisfy
\[
 \|{{\vec{q}}^{\,\star}}\|_2\le1,\qquad
 \mathcal V({{\vec{q}}^{\,\star}})\le\beta/2,\qquad
 \mathcal W({{\vec{q}}^{\,\star}})\le C_0Kr/2,
\]
and have loss at most ${\Prx_{(\bx,\by)\sim\mathcal D}[f(\bx)\ne\by]}+O(\tau)$ for every
$f\in\mathcal C_K$. The internal accuracy constants can be
chosen so that this loss is at most $3/2$.
\Cref{lem:empirical-shared} therefore returns a polynomial with
the required population constraints. Taking the infimum over
$f$ gives loss at most $\opt_K+O(\tau)$; an optimal comparator
need not exist. Return $P={T_\sigma}Q_{\vec q}$. Hermite orthogonality
gives ${\Ex_{(\bx,\ba)\sim\gamma_{d+1}}
Q_{\vec q}(\bx,\ba)^2\le1}$ throughout optimization.

\paragraph{Complexity.}
The estimates and optimization above
have complexity polynomial in
\[
 N_k,\ H,\ k,\ {\sigma}^{-1},\ \beta^{-1},\
 \eta^{-1},\ \log(1/\delta).
\]
Although $\beta$ can be exponentially small in ${r^2/\sigma^2}$, the chosen
degree satisfies
\[
 {\log(1/\beta)\le\sigma^2 k\le k,\qquad
 \sigma^{-2}=O(k).}
\]
The norm cutoff has the same permissible cost. In fact,
$\log A=O(k\log(d+2))$ and $t(d+1)\le k/4$.
The parameter choices give $\beta\le\min\{\tau,Kr\}$, so
$\zeta=c\beta$ and $\log(1/\zeta)=O(k)$.
Consequently
\[
 \log H=O(k\log(d+2)).
\]
Also $\log(1/\eta)=O(k)$, since $\beta\le\eta$, so every
polynomial factor in $1/\eta$ is absorbed by $(d+1)^{O(k)}$.
Together with $N_k\le(d+2)^k$, these bounds give complexity
$(d+1)^{O(k)}\poly(\log(1/\delta))$.
\end{proof}

\subsection{Strongly log-concave marginals}
\label{sec:slc-samples}

For the learner in \Cref{sec:slc-program}, the signed width must
also be estimated. We use monomial coefficients because the
sampling distribution is unknown. The following result supplies
its sample implementation; the polynomial approximation
{result} is needed only to construct the witness in
\Cref{sec:slc-witness}, not for the estimation argument here.

\begin{lemma}[Sample implementation for the distribution extension]
\label{lem:slc-empirical-implementation}
{Fix the distributions and parameters ${\sigma},\beta,k,N,B$ from
\Cref{sec:slc-program,eq:slc-parameters}.
Let $\tau,\delta\in(0,1/4)$ be the optimization accuracy and failure
probability, with $k\ge\log(1/\tau)$.}
{Suppose there exists a degree-$\le k$ polynomial $Q^\star$
over $\R^{d+1}$ satisfying}
\[
\begin{aligned}
 &{\Ex_{\substack{\bx\sim\mu*\mathcal N(0,{\sigma^2}I_d)\\
                            \ba\sim\mathcal N(0,\alpha^{-1}+{\sigma^2})}}{Q^\star}(\bx,\ba)^2\le1},\\
 &{\Ex_{(\bx,\by)\sim\mathcal D}
     \bigl|H^{\mathrm{smooth}}_\sigma Q^\star(\bx,0)-\by\bigr|\le3/2,}\\
 &{\Ex_{\substack{\bx\sim\mu*\mathcal N(0,\sigma^2I_d)\\
                            \ba\sim\mathcal N(0,\alpha^{-1}+\sigma^2)}}
     \defect_{Q^\star}(\bx,\ba)^2\le\beta/2,}\\
 &{\Ex_{\bx\sim\mu}
     \bigl[H^{\mathrm{smooth}}_\sigma Q^\star(\bx,r)
           -H^{\mathrm{smooth}}_\sigma Q^\star(\bx,-r)\bigr]\le C_0Kr/2.}
\end{aligned}
\]
With probability at least $1-\delta$, the sample implementation
returns ${\widehat P}={{H^{\mathrm{smooth}}_\sigma}}Q$ satisfying the population constraints of
\eqref{eq:slc-program} and, simultaneously for every such witness,
\[
 \Ex_{(\bx,\by)\sim\mathcal D}|{\widehat P}(\bx,0)-\by|
 \le{\Ex_{(\bx,\by)\sim\mathcal D}
       \bigl|H^{\mathrm{smooth}}_\sigma Q^\star(\bx,0)-\by\bigr|}
       +O({\tau}).
\]
Its sample and real-arithmetic complexity is
\[
 {N}^{O_\alpha(k)}\poly\!\left(\log\frac1\delta\right).
\]
\end{lemma}

\begin{proof}
We first show that the known coefficient ball contains the witness.

Write ${(\mu*\mathcal N(0,{\sigma^2}I_d))\otimes\mathcal N(0,\alpha^{-1}+{\sigma^2})}$
as the law of $\bu+{\sigma}\,\bg$, where
$\bu\sim{\mu\otimes\mathcal N(0,\alpha^{-1})}$ and
$\bg\sim{\gamma_{d+1}}$ are independent.
We have $\mathbb E\|\bu\|_2^2=d+{\alpha^{-1}}\le{(d+1)/\alpha}$.
For any $Q$ with
\[
{\Ex_{\substack{\bx\sim\mu*\mathcal N(0,{\sigma^2}I_d)\\
                            \ba\sim\mathcal N(0,\alpha^{-1}+{\sigma^2})}}Q(\bx,\ba)^2\le1},
\]
Markov's inequality
applied to $\|\bu\|_2^2$ and to
$\mathbb E_{\bg}Q(\bu+{\sigma}\,\bg)^2$ gives a point $u_0$ with
\[
 \|u_0\|_2\le2\sqrt{{(d+1)/\alpha}},\qquad
 {\Ex_{\bg\sim\gamma_{d+1}}Q(u_0+{\sigma}\,\bg)^2\le4}.
\]
Indeed, each bad event has probability at most $1/4$.
The translated polynomial has Hermite coefficient norm at most $2$.
The coefficient formula for the Hermite polynomials and the
binomial formula for undoing translation show that its monomial
coefficient norm, after rescaling by ${\sigma^{-1}}$, is at most
\[
 {M^{\mathrm{monomials}}}\,[C(k+1)(1+\|u_0\|_2)(1+{\sigma^{-1}})]^{Ck}
 \le {N}^{C'k}.
\]
{Enlarging the universal constant in $B$ therefore includes
every witness ${Q^\star}$ satisfying the lemma's bound on its expected square.
Other polynomials in this ball need not satisfy that bound:
the output certificates use only their loss, signed width, and
squared defect, so this enlargement preserves the proof.}

\paragraph{Cutoff bounds and the shared lemma.}
Normalize the coefficient ball by writing $\vec q=B\vec v$,
$\|\vec v\|_2\le1$. Monomial evaluation, differentiation, and
{additive Gaussian smoothing} show that the loss, absolute signed-width
integrand, squared defect, and their Lipschitz constants in
$\vec v$ have a common upper bound
\[
 {N}^{Ck}(1+\|z\|_2)^{2k}.
\]
For the loss and signed width, $z=x\sim\mu$; for the defect,
{$z=(x,a)$, where $x\sim\mu*\mathcal N(0,{\sigma^2}I_d)$ and
$a\sim\mathcal N(0,\alpha^{-1}+{\sigma^2})$ are independent}.
{For either sampling distribution $\rho$, strong log-concavity gives Gaussian concentration
of the norm, and hence
$\Pr_{\bz\sim\rho}[\|\bz\|_2>v]\le C_\alpha e^{-c_\alpha v^2}$ for
$v\ge C_\alpha\sqrt{d+1}$
\cite{saumard2014log}.} Integrating this tail against the polynomial
envelope bounds the omitted expectation outside radius $R$ by
\[
 {N}^{C_\alpha k}\exp(C_\alpha{(d+1)}-c_\alpha R^2).
\]
Thus, with $\zeta=c\min\{{\tau},\beta,Kr\}$, choosing
\[
 R^2=C_\alpha\bigl({d+1}+k\log {N}+\log(1/\zeta)\bigr)
\]
with a sufficiently large constant makes every omitted expectation
at most $\zeta/2$. {The degree choice and $k\ge\log(1/\tau)$ give
$\log(1/\zeta)=O_\alpha(k)$, so the retained values and coefficient
Lipschitz constants are bounded by $H=N^{O_\alpha(k)}$.}
These are precisely the cutoff hypotheses of
\Cref{lem:empirical-shared}. {Apply that lemma
in coefficient dimension $M^{\mathrm{monomials}}$, with accuracy $\tau$
and empirical signed width.} Its guarantees imply the
population constraints and loss bound claimed here.

The samples and feature values required by the lemma can be
formed explicitly. Loss and width use original examples, while
defect uses $(\bx+{\sigma}\,\bg,\ba)$ with $\bx\sim\mu$,
$\bg\sim\gamma$, and $\ba\sim\mathcal N(0,{\alpha^{-1}}+{\sigma^2})$ independent.
Monomials and their derivatives are evaluated directly, and the
finite expansion \eqref{eq:slc-finite-smoothing} computes ${\widehat P}$.
Their cost is polynomial in ${M^{\mathrm{monomials}}},k,d$.
{Finally, $M^{\mathrm{monomials}}\le N^k$ and
$\log(1/\beta)=O_\alpha(k)$ by the degree choice.
Since $\beta\le Kr$, $\log(1/\tau)\le k$, and $N\ge2$,
every polynomial factor in $\beta^{-1},(Kr)^{-1},\tau^{-1}$
is absorbed by $N^{O_\alpha(k)}$.}
Substituting these bounds and $H={N}^{O_\alpha(k)}$ in
\Cref{lem:empirical-shared} proves the stated complexity.
\end{proof}

\section{\texorpdfstring{Proofs of bounds from \Cref{tab:boundary-profiles}}{Proofs of the table bounds}}
\label{apx:boundary-profile-bounds}
The boundary-ratio bounds below, substituted into \Cref{thm:full},
give the running times in \Cref{tab:boundary-profiles}.
Indeed, for $r\le1/K$ and $0<\eps<1$,
\[
 \frac{K^2}{\eps^2}+\frac{K^4r^2}{\eps^4}
 \le \frac{2K^2}{\eps^4}.
\]
Thus the table uses the general bound $d^{\widetilde O(K^2/\eps^4)}$
at constant success probability. Its lower bounds concern agnostic
learning without a robustness requirement.

\subsection{SQ lower bounds}
We measure SQ cost by $Q/\tau^2$, where $Q$ is the number of
queries and $\tau\le1$ is their minimum tolerance.
All lower bounds are for constant success probability and excess
error below a sufficiently small absolute constant.
The bounds inherited from \cite{DKPZ2021,HSSV2022} require
$\eps\ge d^{-c}$ for a sufficiently small constant $c>0$.
These sources give an alternative between exponentially many
queries and a small query tolerance. As in the table, we restrict
to parameter ranges where the exponential-query alternative is
at least as large as the displayed cost bound.

\paragraph{Boundary ratio $K$ and Boolean combinations of halfspaces.}
{Theorem~20 of \cite{DKPZ2021} constructs a univariate degree-$k$
polynomial threshold function $f=\operatorname{sign}(p)$ such that
every polynomial $q$ satisfying
$\Ex_{\bg\sim\gamma_1}|q(\bg)-f(\bg)|\le2\eps$
has degree $\Omega(k^2/\eps^2)$.}
Their Theorem~4 converts this approximation lower bound into an
SQ lower bound for the class containing all rotations
\[
 x\mapsto\operatorname{sign}(p(\langle v,x\rangle)),
 \qquad \|v\|=1,
\]
of this function. Together, the two theorems give cost
$d^{\Omega(k^2/\eps^2)}$ in the parameter range above.

These functions change sign at at most $k$ thresholds, so they
are Boolean combinations of at most $k$ parallel halfspaces.
Taking $k=m$ proves the table's lower bound for Boolean functions
of $m$ halfspaces. The boundary calculation below also shows that
their boundary ratio is at most $2k/\sqrt{2\pi}$.
Taking $k=\lfloor K\rfloor=\Theta(K)$ for $K\ge1$ therefore gives
the lower bound $d^{\Omega(K^2/\eps^2)}$ for $\mathcal C_K$.
The case $k=1$ gives the halfspace lower bound used below.

\paragraph{Balls and Boolean combinations of balls.}
We transfer the halfspace bounds by approximating halfspaces
arbitrarily well with balls. For a unit vector $v$ and
$t\in\mathbb R$, let
\[
 H=\{x:\langle v,x\rangle\ge t\},
 \qquad
 B_R=\{x:\|x-Rv\|\le R-t\},
 \quad R>\max\{t,0\}.
\]
Expanding the defining inequality gives
\[
 x\in B_R
 \quad\Longleftrightarrow\quad
 \langle v,x\rangle
 \ge t+\frac{\|x\|^2-t^2}{2R}.
\]
Consequently $\mathbf1_{B_R}\to\mathbf1_H$ almost everywhere as
$R\to\infty$, and dominated convergence gives
${\Pr_{\bx\sim\gamma}[\bx\in B_R\triangle H]}\to0$.
For any Boolean function $F:\{0,1\}^m\to\{0,1\}$,
\[
 {\begin{aligned}
 &\Pr_{\bx\sim\gamma}\!\left[
 \begin{aligned}
 &F(\mathbf1_{H_1}(\bx),\ldots,\mathbf1_{H_m}(\bx))\\
 &\qquad\ne F(\mathbf1_{B_1}(\bx),\ldots,\mathbf1_{B_m}(\bx))
 \end{aligned}\right]\\
 &\qquad\le \sum_{i=1}^m\Pr_{\bx\sim\gamma}[\bx\in H_i\triangle B_i].
 \end{aligned}}
\]
Thus the same approximation holds for Boolean combinations.
For every labeling distribution {$\mathcal D$} with Gaussian marginal,
\[
 {\left|\Prx_{(\bx,\by)\sim\mathcal D}[f(\bx)\ne\by]
 -\Prx_{(\bx,\by)\sim\mathcal D}[g(\bx)\ne\by]\right|}
 \le {\Pr_{\bx\sim\gamma}[f(\bx)\ne g(\bx)]}.
\]
Taking infima, the optimal error over the ball class is no larger
than that over the corresponding halfspace class. A learner for
the former therefore also competes with the latter.
This gives costs $d^{\Omega(1/\eps^2)}$ for balls and
$d^{\Omega(m^2/\eps^2)}$ for Boolean functions of $m$ balls.

\paragraph{Intersections and unions of halfspaces.}
\cite[Theorem~18]{HSSV2022} gives the table's bound
\[
 d^{\Omega(\log m/\log\log m+\eps^{-2})}
\]
for intersections of $m$ halfspaces, with
$m=2^{O(d^{0.245})}$ and $\eps\ge d^{-c}$ for a sufficiently
small constant $c>0$, subject to the query-cost qualification above.
Complementing both the labels and the output hypothesis preserves
classification error and SQ cost, so the bound also holds for unions.

\paragraph{Axis-aligned boxes.}
Let $\bz_i=\operatorname{sign}(\bx_i)$ for $\bx\sim\gamma$.
Then $\bz$ is uniform on $\{-1,1\}^d$.
For $S\subseteq[d]$ of size $s\ge1$, define
\[
 \chi_S(\bz)=\prod_{i\in S}\bz_i,
 \qquad
 g_S(\bx)=2\mathbf1\{\bz_i=1\text{ for all }i\in S\}-1.
\]
Here $g_S$ is an orthant indicator in $\{-1,1\}$ notation, and
\[
 \mathbb E[g_S(\bx)\chi_S(\bz)]=2^{1-s}.
\]
Under labels $\by=\chi_S(\bz)$, its classification error is therefore
$1/2-2^{-s}$.
Choose $s=\Theta(\log(1/\eps))$ with $2^{-s}\ge4\eps$.
An agnostic box learner must then output a hypothesis with error
at most $1/2-3\eps$, hence correlation at least $6\eps$ with the
unknown parity.

{The $\binom ds$ parity functions satisfy
\[
 \Ex_{\bz\sim\mathrm{Unif}(\{-1,1\}^d)}
       [\chi_S(\bz)\chi_T(\bz)]=\mathbf1\{S=T\}.
\]}
By Bessel's inequality, a bounded SQ query of
tolerance $\tau$ can distinguish only $O(1/\tau^2)$ of these
labelings from independent fair labels.
The learner's output can be tested with one additional correlation
query of tolerance $\eps$. Thus learning requires cost
$\Omega(\binom ds)-O(\eps^{-2})$.
For $s=\Theta(\log(1/\eps))=O(\log d)$ and $\eps$ below a
sufficiently small absolute constant, the first term dominates and gives
$d^{\Omega(\log(1/\eps))}$.
This is the SQ version of the conjunction--sparse-parity reduction
in \cite[Section~5]{FeldmanKothari2015}.
Orthants can be approximated by bounded boxes with arbitrarily
small Gaussian disagreement, so the bound also applies when
boxes are required to be bounded.

\paragraph{Convex sets.}
The sample lower bound of \cite{KOS2008} requires
$2^{\Omega(\sqrt d)}$ examples even for realizable learning to
constant accuracy.
An SQ algorithm making $Q$ queries of tolerance at least $\tau$
can be simulated using
$O(Q\tau^{-2}\log(Q/\delta))$ fresh examples.
The sample lower bound therefore implies SQ cost
$2^{\Omega(\sqrt d)}$; the logarithmic simulation overhead is
absorbed in the constant in the exponent.
Convex sets also contain halfspaces, giving the independent
$d^{\Omega(1/\eps^2)}$ bound from
\cite[Theorems~4 and~20]{DKPZ2021}.
Taking the maximum of these two lower bounds gives the table's
\[
 2^{\Omega(\sqrt d+\log(d)/\eps^2)},
\]
since the maximum of two nonnegative exponents is at least half
their sum.

\subsection{Boundary-ratio calculations}

\paragraph{Halfspaces and balls.}
For $H=\{x:\langle v,x\rangle\le b\}$ with $\|v\|=1$,
\[
 \Bdry_r(\mathbf1_H)
 \le {\Pr_{\bx\sim\gamma}}\bigl[|\langle v,\bx \rangle-b|\le r\bigr]
 \le \frac{2r}{\sqrt{2\pi}}.
\]
Euclidean balls have uniformly bounded Gaussian surface area,
independently of their centers, radii, and dimension
\cite{KOS2008}.
Writing $\operatorname{Per}_{\gamma}$ for Gaussian perimeter,
the coarea formula therefore gives, for $R>0$,
\[
 \begin{aligned}
 \Bdry_r(\mathbf1_{B(c,R)})
 &\le
 {\Pr_{\bx\sim\gamma}}\bigl[\max\{0,R-r\}\le\|\bx-c\|\le R+r\bigr]\\
 &=\int_{\max\{0,R-r\}}^{R+r}
       \operatorname{Per}_{\gamma}(B(c,t))\,dt
 \le Cr.
 \end{aligned}
\]
Degenerate balls satisfy the same bound by a limiting argument.

\paragraph{Intersections, unions, and boxes.}
Let
\[
 P=\bigcap_{i=1}^m
       \{x:\langle v_i,x\rangle\le b_i\},
 \qquad \|v_i\|=1,
 \qquad
 M(x)=\max_{i\le m}(\langle v_i,x\rangle-b_i).
\]
If $x\in P$ can leave $P$ within distance $r$, then
$-r\le M(x)\le0$.
If $x\notin P$ can enter $P$ within distance $r$, then
$0<M(x)\le r$.
Consequently,
\[
 \Bdry_r(\mathbf1_P)
 \le {\Pr_{\bx\sim\gamma}}[-r\le M(\bx)\le r].
\]
Nazarov's Gaussian anti-concentration inequality
\cite{ChernozhukovChetverikovKato2017} bounds the right-hand side by
\[
 Cr\bigl(1+\sqrt{\log m}\bigr)
 =O\!\left(r\sqrt{\log(m+1)}\right).
\]
The inequality allows arbitrary thresholds and correlations among
the unit-variance Gaussian coordinates $\langle v_i,\bx \rangle$.
Complementation gives the same bound for unions.
An axis-aligned box is an intersection of at most $2d$ halfspaces,
giving $K=O(\sqrt{\log(d+1)})$.

\paragraph{Boolean combinations and univariate PTFs.}
For any Boolean function $F$ and classifiers $f_1,\ldots,f_m$,
a perturbation that changes $F(f_1,\ldots,f_m)$ must change at
least one $f_i$. Thus
\[
 \Bdry_r(F(f_1,\ldots,f_m))
 \le \sum_{i=1}^m\Bdry_r(f_i).
\]
The halfspace and ball bounds imply $K=O(m)$ for their arbitrary
Boolean combinations.
In particular, a univariate degree-$k$ PTF embedded along a unit
direction has at most $k$ parallel boundary hyperplanes, so
$K=O(k)$.

\paragraph{Convex sets.}
Let $C$ be a closed convex set with nonempty interior, and let
$\rho_C$ be signed distance, positive inside $C$.
For $t\ge0$, the superlevel set $\{\rho_C\ge t\}$ is an inner
parallel body of $C$; for $t<0$, it is an outer parallel body.
Both are convex.
Since $|\nabla\rho_C|=1$ almost everywhere away from the boundary,
the coarea formula and Ball's bound
\cite{Ball1993} imply
\[
 \begin{aligned}
 \Bdry_r(\mathbf1_C)
 &\le {\Pr_{\bx\sim\gamma}[|\rho_C(\bx)|\le r]}\\
 &=\int_{-r}^{r}
       \operatorname{Per}_{\gamma}(\{\rho_C\ge t\})\,dt\\
 &\le Crd^{1/4}.
 \end{aligned}
\]
The perimeter bound is needed only for almost every level.
Lower-dimensional convex sets have Gaussian measure zero, and
the same estimate follows by applying coarea only to their outer
parallel bodies. Empty and full sets are immediate.

\section{Harnack inequality for the Gaussian noise operator (proof of \texorpdfstring{\Cref{fact:harnack}}{Fact~\ref{fact:harnack}})}
\label{app:harnack}
Here, we include the proof of Fact~\ref{fact:harnack}. This is a special case of a  well-known inequality, but we include the proof for the reader's convenience.

\begin{proof}
Let $\mu_x := \mathcal N({\rho_\sigma} x,{\sigma}^2 I_n)$ and $\mu_z := \mathcal N({\rho_\sigma} z,{\sigma}^2 I_n)$,
so that $({T_\sigma} g)(z) = \int g\,d\mu_z$ and $({T_\sigma} g^2)(x) = \int g^2\,d\mu_x$.
The two Gaussians have the same covariance, so $\mu_z$ has a density with respect to $\mu_x$, namely
\[
  \frac{d\mu_z}{d\mu_x}(y)
  = \exp\!\left( \frac{\|y-{\rho_\sigma} x\|_2^2 - \|y-{\rho_\sigma} z\|_2^2}{2{\sigma}^2} \right).
\]
By Cauchy--Schwarz,
\begin{equation}\label{eq:harnack-cs}
\begin{aligned}
  ({T_\sigma} g)(z)
  = \int g\,\frac{d\mu_z}{d\mu_x}\,d\mu_x
  &\le \left( \int g^2\,d\mu_x \right)^{1/2}
       \left( \int \Bigl(\frac{d\mu_z}{d\mu_x}\Bigr)^2 d\mu_x \right)^{1/2} \\
  &= \sqrt{({T_\sigma} g^2)(x)}\,
     \left( \int \Bigl(\frac{d\mu_z}{d\mu_x}\Bigr)^2 d\mu_x \right)^{1/2}.
\end{aligned}
\end{equation}
It remains to compute the last integral.
Writing $w := y-{\rho_\sigma} z$ and $\delta := {\rho_\sigma}(z-x)$, so that $y-{\rho_\sigma} x = w+\delta$, we have
\[
  \|y-{\rho_\sigma} x\|_2^2 - 2\|y-{\rho_\sigma} z\|_2^2
  = \|w+\delta\|_2^2 - 2\|w\|_2^2
  = -\|w-\delta\|_2^2 + 2\|\delta\|_2^2 .
\]
Hence the function $y\mapsto \bigl(\frac{d\mu_z}{d\mu_x}(y)\bigr)^2$ times the density of $\mu_x$ equals
\[
  (2\pi{\sigma}^2)^{-n/2}
  \exp\!\left( \frac{-\|y-{\rho_\sigma} z-\delta\|_2^2}{2{\sigma}^2} \right)
  \exp\!\left( \frac{\|\delta\|_2^2}{{\sigma}^2} \right),
\]
which is $e^{\|\delta\|_2^2/{\sigma}^2}$ times the density of $\mathcal N({\rho_\sigma} z+\delta,{\sigma}^2 I_n)$.
Integrating over $y\in\R^n$ gives
\[
  \int \Bigl(\frac{d\mu_z}{d\mu_x}\Bigr)^2 d\mu_x
  = \exp\!\left( \frac{\|\delta\|_2^2}{{\sigma}^2} \right)
  = \exp\!\left( \frac{{\rho_\sigma}^2\|z-x\|_2^2}{{\sigma}^2} \right)
  \le \exp\!\left( \frac{\|z-x\|_2^2}{{\sigma}^2} \right),
\]
where the last step uses ${\rho_\sigma}\le1$.
Substituting this into \Cref{eq:harnack-cs} and taking the square root of the exponential proves the claim.
\end{proof}

\end{document}